\documentclass[conference]{IEEEtran}
\IEEEoverridecommandlockouts

\usepackage{graphicx}

\usepackage{float}

\usepackage{hyperref}
\usepackage[ruled,vlined,noend]{algorithm2e}

\usepackage{tikz}
\usetikzlibrary{decorations.pathmorphing,shapes,arrows,automata,chains,calc}
\usepackage{listings}
\usepackage{pifont}
\usepackage{hhline}
\usepackage{xspace}
\usepackage{todonotes}
\usepackage{enumitem}

\lstdefinestyle{gcore}{
  belowcaptionskip=1\baselineskip,
  breaklines=true,
  xleftmargin=\parindent,
  language=SQL,
  showstringspaces=false,
  basicstyle=\bfseries\ttfamily,
  keywordstyle=\color{green!40!black},
  commentstyle=\itshape\color{purple!40!black},
  identifierstyle=\color{blue},
  stringstyle=\color{black},
  morekeywords={FWD,BWD,PREV,NEXT,ON,MATCH}
}
\usepackage{cite}
\usepackage{amsmath,amssymb,amsfonts}
\usepackage{graphicx}
\usepackage{textcomp}
\usepackage{xcolor}

\usepackage[T1]{fontenc}
\usepackage{amsmath}
\usepackage{tikz}
\usetikzlibrary{automata}
\usetikzlibrary{chains}
\usetikzlibrary{arrows}
\usetikzlibrary{calc}
\usepackage{url}
\usepackage{amsmath} 
\usepackage{subfigure} 
\usepackage{upquote}
\usepackage{multirow}
\usepackage{apxproof}
\usepackage{bbm}

\usepackage{fancyvrb}
\DefineVerbatimEnvironment{verbatim}{Verbatim}{xleftmargin=-1cm}

\newcommand{\covuri}[1]{\text{\footnotesize {\tt\vphantom{yI}#1}}}

\newcommand{\covurie}[1]{\covuri{#1\,=\,}}

\newcommand{\covurieb}[1]{\covuri{{\bf #1}\,=\,}}

\newcommand{\covuriq}[1]{\covuri{#1}}

\newcommand{\etal}{et al.\xspace}

\newcommand{\alt}[2]{
\addtolength{\jot}{-1ex} 
\begin{minipage}{#1} 
\vspace{2pt}
\begin{center}
$\begin{aligned}
#2
\end{aligned}$
\end{center}
\end{minipage}
}

\usetikzlibrary{arrows,positioning,backgrounds} 
\tikzset{
    rt/.style={
		rectangle,
		fill = white,
		draw=black, 
		text centered,
		inner sep=0.5ex
		},
    rtt/.style={ 
    	rt,
    	inner sep=0.1ex
    	},
    ert/.style={ 
     	rt,
     	dashed
     	}, 
    ertt/.style={ 
        rtt,
        dashed
        }, 
    rect/.style={ 
        rectangle,
        fill = white,
        rounded corners,
        draw=black, 
        text centered,
        inner sep=0.8ex
        },
    rectw/.style={
        rect,
        draw=white
        },
    erect/.style={ 
    	rect,
    	dashed
    	},
    erectw/.style={ 
     	rectw,
     	dashed
     	},
    arrout/.style={
           ->,
           -latex,
           },
    arrin/.style={
           <-,
           latex-,
           },
    arrb/.style={
           <->,
           >=latex,
           }
}

\newcommand{\tpg}{\text{TPG}\xspace}
\newcommand{\tpgs}{\text{TPGs}\xspace}
\newcommand{\eg}{e.g.,\xspace}

\def\Val{\textit{Val}}
\def\Prop{\textit{Prop}}
\def\Lab{\textit{Lab}}
\def\G{G}

\newcommand{\N}{\mathbb{N}}
\newcommand{\src}{\mathsf{src}}
\newcommand{\tgt}{\mathsf{tgt}}

\newcommand{\ignore}[1]{}

\newcommand{\Lname}{{\rm {\sf {\small NavL}}}}
\newcommand{\PC}{{\rm {\sf {\small PC}}}}
\newcommand{\NOI}{{\rm {\sf {\small NOI}}}}
\newcommand{\ANOI}{{\rm {\small {\sf ANOI}}}}

\newcommand{\Ltwo}{\Lname[\PC]\xspace}
\newcommand{\Lthree}{\Lname{\rm [}\PC,\NOI{\rm ]}\xspace}
\newcommand{\Lthreep}{\Lname{\rm [}\PC,\ANOI{\rm ]}\xspace}
\newcommand{\Lonep}{\Lname{\rm [}\NOI{\rm ]}}
\newcommand{\Lonepp}{\Lname{\rm [}\ANOI{\rm ]}}
\newcommand{\pt}{\mathsf{path}}
\newcommand{\test}{\mathsf{test}}
\newcommand{\axis}{\mathsf{axis}}
\newcommand{\mt}{k}
\newcommand{\ex}{\exists}
\newcommand{\vd}{\mathbf{Node}}
\newcommand{\ed}{\mathbf{Edge}}
\newcommand{\propt}[2]{#1 \mapsto #2}
\newcommand{\bw}{\mathbf{B}}
\newcommand{\fw}{\mathbf{F}}
\newcommand{\nxt}{\mathbf{N}}
\newcommand{\prv}{\mathbf{P}}
\newcommand{\sem}[1]{\llbracket #1 \rrbracket_\G}
\newcommand{\semp}[2]{\llbracket #1 \rrbracket_{#2}}
\newcommand{\Pt}{\mathsf{PTO}}
\newcommand{\past}{\mathbf{P}}
\newcommand{\ctpg}{\text{ITPG}\xspace}

\newcommand{\vtab}{\mathbf{Nodes}}
\newcommand{\etab}{\mathbf{Edges}}

\newcommand{\F}{\mathcal{F}}
\newcommand{\C}{C}
\newcommand{\coal}{coalesced\xspace}
\newcommand{\fnc}{\mathsf{FC}}
\newcommand{\vfnc}{\mathsf{vFC}}
\newcommand{\can}{\mathsf{can}}
\newcommand{\tupleeval}{\text{\rm {\sf {\small Eval}}}}

\newcommand{\qbf}{\textsf{QBF}}
\newcommand{\tqbf}{\textsf{TQBF}}

\newcommand{\bit}{\textsf{bit}}
\newcommand{\gsss}{\textsf{G-SUBSET-SUM}}
\newcommand{\sss}{\textsf{SUBSET-SUM}}

\newcommand{\ptime}{\textsc{PTIME}\xspace}
\newcommand{\pspace}{\text{\rm {\sc Pspace}}\xspace}

\newcommand{\np}{\text{\rm {\sc NP}}\xspace}
\newcommand{\conp}{\text{\rm {\sc coNP}}\xspace}
\newcommand{\sigmatwop}{$\Sigma_2^p$}

\newcommand{\insql}[1]{{\small \lstinline{#1}}}

\def\BibTeX{{\rm B\kern-.05em{\sc i\kern-.025em b}\kern-.08em
    T\kern-.1667em\lower.7ex\hbox{E}\kern-.125emX}}
    
\allowdisplaybreaks

\usepackage{stmaryrd}
\usepackage{amsmath}
\usepackage{mathtools}
\usepackage{etoolbox}
\usepackage{dsfont}
\usepackage{color}
\usepackage{colortbl}
\usepackage{multirow}
\usepackage{xcolor,colortbl}
\usepackage{cite}
\usepackage{balance}
\usepackage{url}
\usepackage{xspace}
\usepackage{todonotes}
\usepackage{bbm}
\usepackage{listings}
\usepackage{graphicx}

\definecolor{Gray}{gray}{0.5}
\definecolor{LightCyan}{rgb}{0.88,1,1}
\newcolumntype{a}{>{\columncolor{Gray}}c}

\newcommand*{\tgs}{\textsf{TGraphs}\xspace}

\def\true{\textit{true}}
\def\false{\textit{false}}

\def\Val{\textit{Val}}
\def\Prop{\textit{Prop}}
\def\Lab{\textit{Lab}}

\def\G{G}

\def\charSet[#1][#2]{\boldsymbol{\chi}_{#1}^{#2}}

\newcommand{\eat}[1]{}

\newcommand{\algname}[1]{{\sf #1}}
\def\myrulewidth{3.25in}
\def\therule{\rule{\myrulewidth}{0.2pt}}

\def\myrulewidthwide{4in}
\def\therulewide{\rule{\myrulewidthwide}{0.2pt}}

\newenvironment{insidecode}[3]
{
	\begin{tabular}{p{\myrulewidth}}
		\multicolumn{1}{c}{\rule{0mm}{3mm}{\bf #3} $\algname{#1}(\mbox{#2})$\vspace{-0.6em}}\\
		\therule\vskip-0.8em\therule
		\vspace{-1em}
		\begin{algorithmic}[1]}
		{\end{algorithmic}
		\vskip-0.3em\therule
	\end{tabular}}
	
\newenvironment{insidecodewide}[3]
{
	\begin{tabular}{p{\myrulewidthwide}}
		\multicolumn{1}{c}{\rule{0mm}{3mm}{\bf #3} $\algname{#1}(\mbox{#2})$\vspace{-0.6em}}\\
		\therulewide\vskip-0.8em\therulewide
		\vspace{-1em}
		\begin{algorithmic}[1]}
		{\end{algorithmic}
		\vskip-0.3em\therulewide
	\end{tabular}}

\newenvironment{repeatresult}[2]
{\vskip0.5em\par\textsc{#1} #2.\em}
{\vskip1em}

\renewcommand{\C}{C}

\newtheorem{theorem}{Theorem}[section]

\newtheorem{definition}[theorem]{Definition}

\renewcommand{\thealgocf}{}

\newcommand\rev[1]{\textcolor{black}{#1}}

\begin{document}

\title{Temporal Regular Path Queries} 

\author{\IEEEauthorblockN{Marcelo Arenas, Pedro Bahamondes}
\IEEEauthorblockA{\textit{Universidad Católica \& IMFD, Chile} \\
marenas@ing.puc.cl, pibahamondes@uc.cl}
\and
\IEEEauthorblockN{Amir Aghasadeghi, Julia Stoyanovich}
\IEEEauthorblockA{\textit{New York University, USA} \\
apa374@nyu.edu, stoyanovich@nyu.edu}
}

\maketitle

\begin{abstract}

In the last decade, substantial progress has been made towards standardizing the syntax of graph query languages, and towards understanding their semantics and complexity of evaluation. In this paper, we consider temporal property graphs (\tpgs) and propose temporal regular path queries (TRPQs) that incorporate time into \tpg navigation.  Starting with design principles, we propose a natural syntactic extension of the MATCH clause of popular graph query languages.  We then formally present the semantics of TRPQs, and study the complexity of their evaluation.  We show that TRPQs can be evaluated in polynomial time if \tpgs are time-stamped with time points, and identify  fragments of the TRPQ language that admit efficient evaluation over a more succinct interval-annotated representation.  Finally, we implement a fragment of the language in a state-of-the-art dataflow framework, and experimentally demonstrate that TRPQ can be evaluated efficiently. 
\end{abstract}


\section{Introduction}
\label{sec-intro}

The importance of networks in scientific and commercial domains is undeniable. Networks are represented by graphs, and we will use the terms {\em network} and {\em graph} interchangeably.   Considerable research and engineering effort is devoted to the development of effective and efficient graph representations and query languages. Property graphs have emerged as the de facto standard, and have been studied extensively, with efforts underway to unify the semantics of query languages for these graphs~\cite{DBLP:journals/csur/AnglesABHRV17,GCore18}.  
Many interesting questions about graphs are related to their evolution rather than to their static state~\cite{DBLP:conf/icwsm/GoetzLMF09,DBLP:journals/tweb/LeskovecAH07,DBLP:conf/kdd/LeskovecBKT08,DBLP:conf/icml/SarkarCJ12,DBLP:journals/tkdd/AsurPU09,DBLP:journals/tcsb/BeyerTLSF10,Stuart2003,Chan2008,DBLP:journals/jisa/PapadimitriouDG10}.   
Consequently, several recent proposals seek to extend query languages for property graphs with time~\cite{DBLP:journals/tkde/ByunWK20,Debrouvier21,DBLP:conf/sigmod/JohnsonKLS16,Labouseur2015,DBLP:conf/dbpl/MoffittS17}.

Our focus in this paper is on incorporating time into path queries. More precisely, we (a) outline the design principles for a temporal extension of Regular Path Queries (RPQs) with time; (b) propose a natural syntactic extension of state of the art query languages for conventional (non-temporal) property graphs, which supports temporal RPQs (TRPQs);
(c) formally present the semantics of this language;  (d) study the complexity of evaluation of several variants of this language; (e) implement a practical fragment of this language in a dataflow framework; and (f) empirically demonstrate that TRPQs can be evaluated efficiently. We show that, by adhering to the design principles that draw on decades of work on graph databases and on temporal relational databases, we are able to achieve polynomial-time complexity of evaluation, paving the way to implementations that are both usable and practical, as supported by our implementation and experiments.

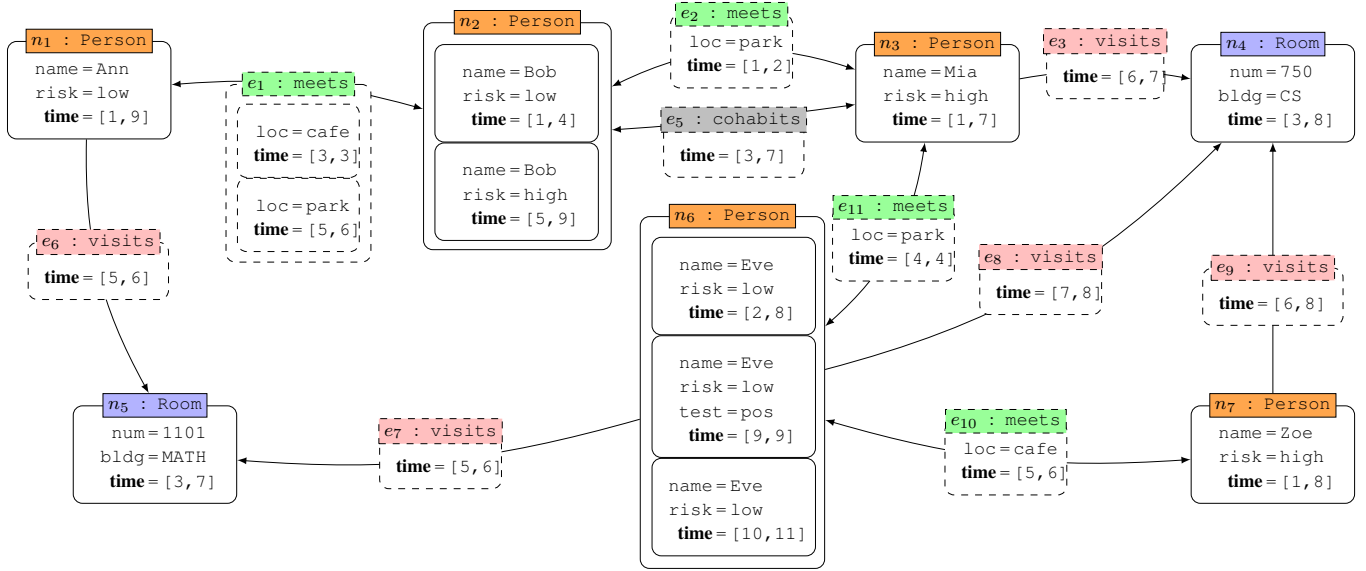
\begin{figure*}[t]
    \begin{center}
        \resizebox{\textwidth}{!}{
        \begin{tikzpicture} 
            \node[rect] (n1) {
                \alt{60pt}{ 
                    \covurie{name} & 
                    \covuriq{Ann}\\
                    \covurie{risk} & \covuriq{low}\\
                    \covurieb{time} & \covuriq{[1,9]}
                }
            };

            \node[rt, fill=orange!70] (ln1) at (n1.north) {
                \covuri{$n_1$ $:$ Person}
            };
         
            \node[rect, text width=26mm, text height=32mm] (bn2) at (65.1mm,-7.0mm) {}; 
             
            \node[rect, below right= -1.45cm and 4cm of n1] (n2) {
                \alt{60pt}{ 
                    \covurie{name} & \covuriq{Bob}\\
                    \covurie{risk} & \covuriq{low}\\
                    \covurieb{time} & \covuriq{[1,4]}
                }
            };
            \node[rect, below=0cm of n2] (n22) {
                \alt{60pt}{
                    \covurie{name} & \covuriq{Bob}\\
                    \covurie{risk} & \covuriq{high}\\
                    \covurieb{time} & \covuriq{[5,9]}
                }
            };
    
            \node[rt, fill=orange!70] (ln2) at (bn2.north) {
                \covuri{$n_2$ $:$ Person}
            };
        
            \draw[arrb, bend left=10] (n1) to node[below, erect, text width=19.5mm, text height=24.5mm] (be1) {}
            (bn2);
            
            
            \node[erect] (e1) at (31.8mm,-7.7mm) {
                \alt{42pt}{
                    \covurie{loc} & \covuriq{cafe}\\
                    \covurieb{time} & \covuriq{[3,3]}
                }
            };
            \node[below=0cm of e1, erect] (e12) {
                \alt{42pt}{
                    \covurie{loc} & \covuriq{park}\\
                    \covurieb{time} & \covuriq{[5,6]}
                }
            };
        
            \node[ert, fill=green!40] (le1) at (be1.north) {
                \covuri{$e_1$ $:$ meets}
            };
        
            \node[rect, right=3.9cm of n2] (n3) {
                \alt{60pt}{
                    \covurie{name} & \covuriq{Mia}\\
                    \covurie{risk} & \covuriq{high}\\
                    \covurieb{time} & \covuriq{[1,7]}
                }
            };
        
            \node[rt, fill=orange!70] (ln3) at (n3.north) {
                \covuri{$n_3$ $:$ Person}
            };

            \node[rect, right=2.6cm of n3] (n4) {
                \alt{60pt}{
                    \covurie{num} & \covuriq{750}\\
                    \covurie{bldg} & \covuriq{CS}\\
                    \covurieb{time} & \covuriq{[3,8]}
                }
            };
        
            \node[rt, fill=blue!30] (ln4) at (n4.north) {
                \covuri{$n_4$ $:$ Room}
            };
            
            \node[rect, below left=4cm and 3cm of n2] (n5) {
                \alt{60pt}{
                    \covurie{num} & \covuriq{1101}\\
                    \covurie{bldg} & \covuriq{MATH}\\
                    \covurieb{time} & \covuriq{[3,7]}
                }
            };
            
            \node[rt, fill=blue!30] (ln5) at (n5.north) {
                \covuri{$n_5$ $:$ Room}
            };

            \draw[arrb, bend right=2] (bn2) to node[below, erect] (e5) {
                \alt{50pt}{
                  \covurieb{time} & \covuriq{[3,7]}
                }
            }
            (n3);
            
            \node[ert, fill=gray!50] (le5) at (e5.north) {
                \covuri{$e_5$ $:$ cohabits}
            };

            \node[rect, text width=25.5mm, text height=51mm] (bn6) at (97.8mm,-46mm) {}; 
            
            \node[rect, below=0.9cm of e5] (n6) {
                \alt{60pt}{
                    \covurie{name} & \covuriq{Eve}\\
                    \covurie{risk} & \covuriq{low}\\
                    \covurieb{time} & \covuriq{[2,8]}
                }
            };
            \node[rect, below=0cm of n6] (n62) {
                \alt{60pt}{
                    \covurie{name} & \covuriq{Eve}\\
                    \covurie{risk} & \covuriq{low}\\
                    \covurie{test} & \covuriq{pos}\\
                    \covurieb{time} & \covuriq{[9,9]}
                }
            };

            \node[rect, below=0cm of n62] (n63) {
                \alt{60pt}{
                    \covurie{name} & \covuriq{Eve}\\
                    \covurie{risk} & \covuriq{low}\\
                    \covurieb{time} & \covuriq{[10,11]}
                }
            };
            
            \node[rt, fill=orange!70] (ln6) at (bn6.north) {
                \covuri{$n_6$ $:$ Person}
            };
            
            \node[rect, below=4cm of n4] (n7) {
                \alt{60pt}{
                    \covurie{name} & \covuriq{Zoe}\\
                    \covurie{risk} & \covuriq{high}\\
                    \covurieb{time} & \covuriq{[1,8]}
                }
            };
        
            \node[rt, fill=orange!70] (ln7) at (n7.north) {
                \covuri{$n_7$ $:$ Person}
            };

            \draw[arrb, bend left=22] (bn2) to node[erect] (e2) {
                \alt{42pt}{
                 \covurie{loc} & \covuriq{park}\\
                    \covurieb{time} & \covuriq{[1,2]}
                }
            }
            (n3);
            
            \node[ert, fill=green!40] (le2) at (e2.north) {
                \covuri{$e_2$ $:$ meets}
            };
            
            \draw[arrout, bend left=10] (n3) to node[erect] (e3) {
                \alt{40pt}{
                    \covurieb{time} & \covuriq{[6,7]}
                }
            }
            (n4);
            
            \node[ert, fill=pink] (le3) at (e3.north) {
                \covuri{$e_3$ $:$ visits}
            };

            \draw[arrout, bend right=15] (n1) to node[erect] (e6) {
                \alt{50pt}{
                    \covurieb{time} & \covuriq{[5,6]}
                }
            }
            (ln5);
            
            \node[ert, fill=pink] (le6) at (e6.north) {
                \covuri{$e_6$ $:$ visits}
            };
            
            \draw[arrin, bend right=10] (n5) to node[erect] (e7) {
                \alt{42pt}{
                  \covurieb{time} & \covuriq{[5,6]}
                }
            }
            (bn6);
            
            \node[ert, fill=pink] (le7) at (e7.north) {
                \covuri{$e_7$ $:$ visits}
            };
            
            \draw[arrout, bend right=15] (bn6) to node[erect] (e8) {
                \alt{42pt}{
                    \covurieb{time} & \covuriq{[7,8]}
                }
            }
            (n4);
            
            \node[ert, fill=pink] (le8a) at (e8.north) {
                \covuri{$e_8$ $:$ visits}
            };

            \draw[arrout] (ln7) to node[below, erect] (e9) {
                \alt{50pt}{
                    \covurieb{time} & \covuriq{[6,8]}
                }
            }
            (n4);
            
            \node[ert, fill=pink] (le8a) at (e9.north) {
                \covuri{$e_9$ $:$ visits}
            };
            
            \draw[arrb, bend right=10] (bn6) to node[erect] (e10) {
                \alt{42pt}{
                 \covurie{loc} & \covuriq{cafe}\\
                    \covurieb{time} & \covuriq{[5,6]}
                }
            }
            (n7);
            
            \node[ert, fill=green!40] (le10) at (e10.north) {
                \covuri{$e_{10}$ $:$ meets}
            };
            
            \draw[arrb, bend right=20] (bn6) to node[erect] (e11) {
                \alt{42pt}{
                 \covurie{loc} & \covuriq{park}\\
                    \covurieb{time} & \covuriq{[4,4]}
                }
            }
            (n3);
            
            \node[ert, fill=green!40] (le11) at (e11.north) {
                \covuri{$e_{11}$ $:$ meets}
            };
        \end{tikzpicture}
    }
    \end{center}
    \caption{\small A \tpg used for contact tracing. The graph contains two types of nodes, \insql{Person} and \insql{Room}, and three types of edges: bi-directional edges \insql{meets} and \insql{cohabits}, and directed edge \insql{visits}. \insql{Person} nodes have properties \insql{name}, \insql{risk} (\insql{'high'} or \insql{'low'}) of complications, and \insql{test} (\insql{'pos'} or \insql{'neg'}) of disease status. Eve (node $n_6$) tested positive for a communicable disease at time 9.
    \label{fig:covid}}
\end{figure*}

\subsection{Running example}
\label{sec-intro-example}

As a preview of our proposed methods, consider Figure~\ref{fig:covid} that depicts a contact tracing network for a communicable disease with airborne transmission between people in enclosed locations on a university campus.  In this network, different actors and their interactions are presented as a \emph{temporal property graph} or \tpg for short. (We will define temporal property graphs formally in Section~\ref{sec-tgraph}).

As in conventional property graphs~\cite{DBLP:journals/csur/AnglesABHRV17}, nodes and edges in a \tpg are labeled.  The graph in Figure~\ref{fig:covid} contains two types of nodes, \insql{Person} and \insql{Room} (representing a classroom), and three types of edges: bi-directional edges \mbox{\insql{meets}} and \insql{cohabits} (lives together), and directed edge \mbox{\insql{visits}.} Nodes and edges have optional properties that are associated with values.  For example, node $n_1$ of type \insql{Person} has properties \insql{name} with value \insql{'Ann'} and \insql{risk} with value \insql{'low'}.  As another example, edge $e_2$ of type \insql{meets} has property \insql{loc} with value \insql{'park'}.
 
The purpose of the graph in Figure~\ref{fig:covid} is to allow identification of individuals who may have been exposed to the disease.  In particular, we are interested in identifying potentially infected individuals who are considered high risk, due to age or pre-existing conditions. These types of questions can be naturally phrased as \emph{temporal regular path queries} (TRPQs) that interrogate reachability over time.  We will give an example of a TRPQ momentarily.

To support TRPQs, all nodes and edges in a \tpg are associated with  \emph{time intervals of validity} (or \emph{intervals} for short) that represent consecutive time points during which no change  occurred for a node or an edge, in terms of its existence or property values. For example, node $n_1$ (Ann) is associated with the interval \insql{[1, 9]}, signifying that $n_1$ was present in the graph and took on the specified property values during 9 consecutive time points. 
As another example, node $n_2$ (Bob) exists during the same interval as $n_1$, but undergoes a change in the  value of the property \insql{risk} at time 4, when it changes from \insql{'low'} to \insql{'high'}.  We represent a change in the state of an entity (a node or an edge) with nested boxes inside an outer box that denotes the entity in Figure~\ref{fig:covid}. 

Now, consider an example of a TRPQ that extends the syntax of Cypher 
to retrieve the list of high-risk people (\insql{x}) who met someone (\insql{y}), who subsequently tested positive for an infectious disease:

{\footnotesize
\begin{lstlisting}
MATCH (x:Person {risk = 'high'})-
  /FWD/:meets/FWD/NEXT*/-(y:Person {test = 'pos'})
ON contact_tracing
\end{lstlisting}
}

This contact tracing query  produces the following \emph{temporal binding table} when evaluated over the \tpg in Figure~\ref{fig:covid}:
\begin{center}
{\footnotesize
    \begin{tabular}{|cccc|}
        \hline
        {\color{blue}{\bfseries\ttfamily x}} & {\color{blue}{\bfseries\ttfamily x\_time}} & 
        {\color{blue}{\bfseries\ttfamily y}} & {\color{blue}{\bfseries\ttfamily y\_time}} \\
        \hline
        \hline
        $n_7$ & 5 & $n_6$ & 9\\
        $n_7$ & 6 & $n_6$ & 9\\
        $n_3$ & 4 & $n_6$ & 9\\
        \hline
    \end{tabular}
}
\end{center}

\subsection{Summary of our approach}
\label{sec-intro-summary}

In the remainder of this paper, we formally develop the concepts that are necessary to evaluate this and other useful TRPQs over \tpgs. We adopt a conceptual \tpg model that naturally extends property graphs with time, and is both simple and sufficiently flexible to support the evolution of graph topology and of the properties of its nodes and edges. We evaluate TRPQs on \tpgs under \emph{point-based semantics}~\cite{Bohlen2000}, in which operators adhere to two principles: snapshot reducibility and extended snapshot reducibility, discussed in Section~\ref{sec-related}. Our conceptual \tpg model admits two logical representations that differ 
in the kind of time-stamping they use~\cite{Montanari2009}. One associates objects with time points, while the other associates them with time intervals, for a more compact representation.
\textit{Design principles.} 
We carefully designed our TRPQ language based on the following 
principles:

    \textbf{Navigability}: 
    Include operators that refer to the dynamics of navigating through  
    the \tpg: temporal navigation refers to movements on the graph over time, and structural navigation refers to movements across locations in its topology.
    
    \textbf{Navigation orthogonality}: Temporal and structural navigation operators must be orthogonal, allowing non-simultaneous single-step temporal and structural movement.

    
    \textbf{Node-edge symmetry}: The language should treat nodes and edges as first-class citizens, supporting equivalent operations.

    \textbf{Static testability}: Testing is independent of navigation. 


    \textbf{Snapshot reducibility}: 
    When time is removed from a query, pairs of temporal objects satisfying the query should correspond to pairs of objects in a single snapshot of the graph, and every pair satisfying the query in the snapshot of the \tpg should correspond to a path satisfying it in the \tpg. 

By adhering to these principles, we achieved polynomial-time complexity of evaluation for \tpgs that are time-stamped with time points, and also identified a significant fragment of the language that can be efficiently evaluated for interval time-stamped \tpgs. In addition to theoretical results, these principles also allowed us to efficiently implement TRPQs by decoupling non-temporal and temporal processing.  


\paragraph*{Paper organization} 
We first give some background on temporal graph models and path query languages in Section~\ref{sec-related}. We then formally define temporal graphs in Section~\ref{sec-tgraph}. We go on to propose a syntax for adding time to a practical graph query language in Section~\ref{sec-standard}. Next, in Section~\ref{sec-t-rpq}, we give the precise syntax and semantics of the language, and study the complexity of evaluating it.
We describe an implementation of our language over an interval-based \tpg in  Section~\ref{sec-system}, and present results of an experimental evaluation in Section~\ref{sec-experiments}. We conclude in Section~\ref{sec-conc}. Additional complexity results and proofs, and supplementary experiments are available in the Appendix. 
System implementation and experimental evaluation are available at \url{https://github.com/amirpouya/tpath}.








\section{Background and Related Work}
\label{sec-related}
Substantial research has been undertaken in the area of {\em temporal relational databases} since the 1980s, producing a significant body of work~\cite{Liu:2009:EDS:1804422},
which includes representation of
time~\cite{Clifford1985,JENSEN1994513,Snodgrass:1985:TTD:318898.318921}, semantics of temporal models~\cite{Bohlen1998}, temporal algebras~\cite{Dignos2012}, and access methods~\cite{Salzberg1999}. Results of some of this work are part of the SQL:2011 standard~\cite{DBLP:journals/sigmod/KulkarniM12}.

\paragraph{Temporal graph models} Temporal graph models differ in what temporal semantics they encode, what time representation they use (time point, interval, or implicitly with a sequence), what entities they time-stamp (graphs, nodes, edges, or attribute-value assignments), and whether they represent evolution of topology only or also of the attributes. With a few exceptions, discussed next, the current de facto standard representation of temporal graphs is the \emph{snapshot sequence}, where a state of a graph is associated with either a time point or an interval during which the graph was in that state~\cite{Borgwardt2006,Fard2012,Ferreira2004,Kan2009,Khurana2013,Khurana2016,Lahiri2008,Ren2011,Semertzidis2015,Sricharan2014,DBLP:conf/cikm/YangQZGL07}.  This representation supports operations within each snapshot under the principle of \emph{snapshot reducibility}, namely, that applying a temporal operator to a database is equivalent to applying the non-temporal variant of the operator to each database state~\cite{Bohlen2000}.
For example, the G* system~\cite{Labouseur2015}
stores a temporal graph as a snapshot sequence and provides two query languages, the procedural PGQL and the declarative DGQL. PGQL includes operators such as retrieving graph vertices and their edges at a given time point, along with non-graph operators like aggregation, union, projection, and join. Neither PGQL nor DGQL support temporal path queries. 

The fundamental disadvantage of using the snapshot sequence as the conceptual representation of a temporal graph is that it does not support operations that explicitly reference temporal information. 
Semantics of operations that make explicit references to time are formalized as the principle of \emph{extended snapshot reducibility}, where timestamps are made available to operators by propagating time as data~\cite{Bohlen2000}. Considering that our goal in this work is to support temporal regular path queries, having access to temporal information during navigation is crucial.
 
In response to this important limitation of the snapshot sequence representation, proposals have been made to annotate graph nodes, edges, or attributes with time.  Moffitt and Stoyanovich~\cite{DBLP:conf/dbpl/MoffittS17} proposed to model property graph evolution by associating intervals of validity with nodes, edges, and property values. They also developed a compositional temporal graph algebra that provides a temporal generalization of common graph operations including subgraph, node creation, union, and join, but does not include reachability or path constructs. In our work, we adopt a similar representation of temporal graphs, but focus on temporal regular path queries.

\paragraph{Paths in temporal graphs} Specific kinds of path queries over temporal graphs have been considered in the literature. Wu~\etal~\cite{DBLP:journals/pvldb/WuCHKLX14,DBLP:journals/tkde/WuCKHHW16,DBLP:conf/icde/WuHCLK16} studied path query variants over temporal graphs, in which nodes are time-invariant and edges are associated with a starting time and an ending time.  (Nodes and edges do not have type labels or attributes.) The authors introduced four types of ``minimum temporal path'' queries, including the earliest-arriving path and the fastest path, which can be seen as generalizations of the shortest path query for temporal graphs. They proposed  algorithms and indexing methods to process minimum temporal path and temporal reachability queries efficiently.

Byun~\etal~\cite{DBLP:journals/tkde/ByunWK20} introduced ChronoGraph, a temporal graph traversal system in which edges are traversed time-forward.
The authors show three use cases: temporal breadth-first search, temporal depth-first search, and temporal single-source shortest-path, instantiated over Apache Tinkerpop. 
Johnson~\etal~\cite{DBLP:conf/sigmod/JohnsonKLS16} introduced Nepal, a query language that has SQL-like syntax and supports regular path queries over temporal multi-layer communication networks, represented by temporal graphs that associate a sequence of intervals of validity with each node and edge. The key novelty of this work are time-travel path queries to retrieve past network states. 
Finally, Debrouvier~\etal~\cite{Debrouvier21} introduced T-GQL, a query language for \tpgs with Cypher-like syntax~\cite{Cypher18}. T-GQL operates over graphs in which (a) nodes persists  but their attributes (with values) can change over time, 
and so are associated with periods of validity; and (b) edges are associated with periods of validity but their attributes are time-invariant.  This asymmetry in the handling of nodes and edges is due to the authors' commitment to a specific (lower-level) representation of such \tpgs in a conventional property graph system.  Specifically, they assume that Objects (representing nodes), Attributes, and Values are stored as conventional property graph nodes, whereas time intervals are stored as properties of these nodes. Temporal edges are, in turn, stored as conventional edges, with time interval as one of their properties. T-GQL supports three types of path queries over such graphs, syntactically specified with the help of named functions:  (1) ``Continuous path'' queries retrieve paths valid during each time point---snapshot semantics. (2) ``Pairwise continuous paths'' require that the incoming and the outgoing edge for a node being traversed must exist during some overlapping time period. 
\rev{(3) ``Consecutive paths'' encode temporal journeys; for example,  to indicate a way to fly from Tokyo to Buenos Aires with a couple of stopovers in a temporal graph for flight scheduling. Consecutive paths are used in T-GQL for encoding earliest arrival, latest departure, fastest, and shortest path queries.}

\rev{A more detailed comparison of our proposal with other temporal query languages is given in Section \ref{sec-comp-other}.} 
In summary, our proposal differs from prior work in that we develop a general-purpose query language for temporal paths, which works over a simple conceptual definition of temporal property graphs and  is nonetheless general enough to represent different kinds of temporal and structural evolution of such
graphs.  Our language is syntactically simple: it directly, and minimally, extends the MATCH clause of popular graph query languages, and does not rely on custom functions. In fact, as we show in Section~\ref{sec:sem-comp-navl}, there is a simple way to define its formal semantics, which allows us to develop efficient algorithms for query evaluation.

\section{A Temporal Graph Model}
\label{sec-tgraph}

In this section, we formalize the notion of temporal property graph, which extends the widely used notion of property graph \cite{DBLP:journals/csur/AnglesABHRV17,Cypher18,GCore18} to include explicit access to time. In this way, we can model the evolution of the topology of such a  graph, as well as the changes in node and edge properties.

A temporal property graph defines a point-based representation of the evolution of a property graph, which is a simple and suitable framework to represent and reason about this evolution. However, time-stamping objects with time points may be impractical in terms of space overhead.   This motivates the development of interval-based representations, which are common for temporal models  for both relations (\eg ~\cite{DBLP:conf/sigmod/DignosBG12,Montanari2009}) and graphs (\eg ~\cite{DBLP:journals/tkde/ByunWK20,Debrouvier21,DBLP:conf/dbpl/MoffittS17}). In this section, we also define a succinct representation of temporal property graphs that uses interval time-stamping.
Notice that point-based temporal semantics requires this succinct representation to be temporally coalesced: a pair of value-equivalent temporally adjacent intervals should be stored as a single interval, and this property should be maintained through operations~\cite{DBLP:conf/vldb/BohlenSS96}.



\subsection{Temporal property graphs}
Assume $\Lab$, $\Prop$ and $\Val$ to be 
sets of label names, property names and actual values, respectively. 
We define temporal property graphs over finite sets of time points. Time points can take on values that correspond to the units of time as appropriate for the application domain, and may represent seconds, weeks, or years. For the sake of presentation, we 
represent the universe of time points by $\N$:
a temporal domain $\Omega$ is a finite set of consecutive natural numbers, that is,  
$\Omega =$ $\{ i \in \N \mid a \leq i \leq b \}$ 
for some $a, b \in \N$ such that $a \leq b$.  
\begin{definition}
\label{def:tg}
A temporal property graph (TPG) is a tuple $\G = ( \Omega$, $N$, $E$, $\rho$, $\lambda$, $\xi$, $\sigma)$, where
\begin{itemize}[leftmargin=4mm]
\item $\Omega$ is a temporal domain; $N$ is a finite set of {\em nodes}, $E$ is a finite set of {\em edges}, and \mbox{$V \cap E = \emptyset$;}



\item $\rho: E \to (N \times N)$ is a function that maps an edge to its source and destination nodes;

\item $\lambda: (N \cup E) \to \Lab$ is a function that maps a node or an edge to its label;

\item $\xi: (N \cup E) \times \Omega \to \{\true,\false\}$ is a function that maps a node or an edge, and a time point to a Boolean. Moreover, if $\xi(e,t) = \true$ and $\rho(e) = (v_1,v_2)$, then $\xi(v_1,t) = \true$ and $\xi(v_2,t) = \true$.

\item $\sigma:  (N \cup E) \times \Prop \times \Omega \to \Val$ is a partial function that maps a node or an edge, a property name, and a time point to a value. Moreover, there exists a finite number of triples $(o,p,t) \in (N \cup E) \times \Prop \times \Omega$ such that $\sigma(o,p,t)$ is defined, and if $\sigma(o,p,t)$ is defined, then $\xi(o,t) = \true$. 
\end{itemize}
\end{definition}
Observe that $\Omega$ in Definition~\ref{def:tg} denotes the \emph{temporal domain} of $\G$, a finite set of linearly ordered time points starting from the time associated with the earliest \emph{snapshot} of $\G$, and ending with the time associated with its latest snapshot, where a  snapshot of $\G$ refers to a conventional (non-temporal) property graph that represents the state of  $\G$ at a given time point.  
Function $\rho$ in Definition~\ref{def:tg} is used to provide the starting and ending nodes of an edge,
function $\lambda$ provides the label of a node or an edge, 
and function $\xi$ indicates whether a node or an edge exists at a given time point in $\Omega$ (which 
corresponds to  
$\true$).
Finally, function $\sigma$ indicates the value of a property for a node or an edge at a given time point in $\Omega$.  

Two conditions are imposed on \tpgs to enforce that they conceptually correspond to sequences of valid conventional property graphs. In particular, an edge can only exist at a time when both of the nodes it connects exist, and that a property can only take on a value at a time when the corresponding object exists.
Moreover, observe that by imposing that $\sigma(o,p,t)$ be defined for a \emph{finite} number of triples $(o,p,t)$, we are ensuring that each node or edge can have values for a finite number of properties, so that each $\tpg$ has a finite representation. Finally, Definition~\ref{def:tg} assumes, for simplicity, that property values are drawn from the infinite set $\Val$. That is, we do not distinguish between different data types. If a distinction is necessary, then $\Val$ can be replaced by a domain of values of some $k$ different data types, $\Val_1$, $\dots$, $\Val_k$.  

Recall our running example discussed in Section~\ref{sec-intro-example} and shown in Figure~\ref{fig:covid}.  This example illustrates Definition~\ref{def:tg}; it shows a \tpg used for contact tracing for a communicable disease, with airborne transmission between people (represented by nodes with label \insql{Person}) in enclosed locations (\textit{\eg} nodes with label \insql{Room}). This \tpg has a temporal domain $\Omega = \{1, \dots, 11\}$, although any set of consecutive natural numbers containing $\Omega$ can serve as the temporal domain of this $\tpg$, for example the set $\{0, \dots, 15 \}$. \rev{The TPG is a multi-graph: $n_2$ and $n_3$ are connected by two edges, $e_2$ and $e_5$.}

In the $\tpg$ in Figure~\ref{fig:covid}, \insql{Person} nodes have properties name, risk ({\small \lstinline{'low'}} or {\small \lstinline{'high'}}), and test ({\small \lstinline{'pos'}} or \mbox{{\small \lstinline{'neg'}})}. For example, Eve, represented by node $n_6$, is known to have tested positive for the disease at time 9. Note that each node and edge refers to a specific time-invariant real-life object or event.  A \tpg records observed states of these objects.  In fact, real-life objects correspond to a sequence of temporal objects, each with a set of properties.  For instance, node $n_2$ corresponds to a sequence of 9 temporal objects, one for each time point $1$ through $9$. These are represented in the figure by two boxes inside the outer box for $n_2$, one for each interval during which no change occurred: $[1, 4]$ with name Bob, and low risk, and $[5, 9]$ with name Bob, and high risk.  To simplify the figure, we do not show internal boxes for nodes or edges associated with a single time interval, such as $n_1$ and $e_6$.

\subsection{Interval-timestamped temporal property graphs}
\label{sec-itpg-red}

\rev{An interval of $\mathbb{N}$ is a term of the form $[a, b]$ with $a, b \in \N$ and $a \leq b$, which is used  as a concise representation of the set 
$\{i \in \N \ \vert \ a \leq i \leq b \}$ between its starting point $a$ and its ending point $b$. 
Each TPG $\G = (\Omega, N, E, \rho, \lambda, \xi, \sigma)$ can be transformed 
into an Interval-timestamped Temporal Property Graph (ITPG), by putting the consecutive time points with the same values into the interval. More precisely, an ITPG $I = ( \Omega', N, E, \rho, \lambda, \xi', \sigma')$ encoding $G$ is defined in the following way. The temporal domain $\Omega = \{ i \in \mathbb{N} \mid a \leq i \leq b \}$ of $G$ is replaced by the interval $\Omega' = [a,b]$, and $N$, $E$, $\rho$, $\lambda$ are the same as in $G$. Moreover, $\xi'$ is a function that maps each object $o \in (N \cup E)$ to a set of maximal intervals where $o$ exists according to function $\xi$. For example, for $\Omega = \{1,2,3,4,5\}$ and node $n$ such that $\xi(n,1) = \xi(n,2) = \xi(n,3) = \xi(n,5) = \true$ and $\xi(n,4) = \false$, it holds that $\Omega' = [1,5]$ and $\xi'(n) = \{[1,3], [5,5]\}$. Notice that $\xi'(n)$ could not be defined as $\{[1,2], [3,3], [5,5]\}$ since $[1,2]$ is not a maximal interval where $n$ exists. In other words, the set of intervals in $\xi'(n)$ has to be \emph{coalesced}. Finally, function $\sigma'$ is generated from $\sigma$ in a similar way as $\xi'$. The formal definition of ITPG can be found in the Appendix.} 

\section[Adding Time to a Practical Graph Query Language]{Adding Time to a Practical\\ Graph Query Language}
\label{sec-standard}


The main goal of this paper is to introduce a simple yet general query language for temporal property graphs. 
In this section, we give a guided tour of the query language, using the  \tpg shown in Figure~\ref{fig:covid} as the running example.  All queries, except those presented alongside their equivalent rewritings, are numbered \insql{Q1} through \insql{Q12}, and will be used in the experimental evaluation in Section~\ref{sec-experiments}.

The \insql{MATCH} clause is a fundamental construct in popular graph query languages such as Cypher \cite{Cypher18}, PGQL \cite{PGQL16}, and G-Core \cite{GCore18}. By using graph patterns, the \insql{MATCH} clause allows to bind variables with objects in a property graph, giving rise to {\em binding tables} that are subsequently processed by the other components of the query language. As an important step towards the construction of a temporal graph query language, we show how the \insql{MATCH} clause can be extended to bind variables with temporal objects in a \tpg. In particular, we show how the syntax and semantics of the query language G-Core \cite{GCore18} can be extended to accommodate temporal graph patterns. As the syntax and semantics of G-Core are compatible with those of Cypher \cite{Cypher18} and PGQL \cite{PGQL16}, these languages can accommodate such temporal graph patterns as well. These languages play a fundamental role in the ongoing graph query language standardization effort~\cite{gql}, and our proposal can provide a natural temporal extension for this standard.

\rev{Our proposed syntax for temporal regular path queries can be summarized as the following extension of the \insql{MATCH} clause:}

{\footnotesize \begin{lstlisting}
MATCH (x)-/path/-(y) ON graph
\end{lstlisting}}

\noindent
\rev{Here, \insql{graph} is either a $\tpg$ or an $\ctpg$, and \insql{path} is an expression that can contain temporal and structural navigation operators, together with some other functionalities like testing the label of a node or an edge, and verifying the value of a property of a node or an edge. We will present the formal semantics of the language in Section~\ref{sec-t-rpq}.}

\rev{As a first example,} assume that \insql{contact_tracing} is the \tpg shown in Figure~\ref{fig:covid}. Then, the following G-Core expression extracts the list of people from \mbox{\insql{contact_tracing}:}

{\footnotesize \begin{lstlisting}
Q1 MATCH (x:Person) ON contact_tracing
\end{lstlisting}}

\noindent
The operator \insql{ON} specifies that \insql{contact_tracing} is the input graph, and  \mbox{\insql{(x:Person)}} indicates that \insql{x} is a variable to be assigned nodes with label \insql{Person} from the input graph. The evaluation of a \insql{MATCH} clause in G-Core results in a table consisting of bindings that assign to each variable an object from the input graph: a node, an edge, a label, or a property value. The result of evaluating \insql{Q1} is the binding table:
\begin{center}
{\footnotesize
    \begin{tabular}{|c|}
        \hline
        {\color{blue}{\bfseries\ttfamily x}}\\
        \hline
        \hline
        $n_1$\\
        $n_2$\\
        $n_3$\\
        $n_6$\\
        $n_7$\\
        \hline
    \end{tabular}
}
\end{center}
At this point, two observations should be made: (i) G-Core does not consider \insql{contact_tracing} as a temporal property graph, so no explicit time is associated with the objects in a binding table; (ii) Cypher \cite{Cypher18} and PGQL \cite{PGQL16} produce the same bindings as G-Core when evaluating the previous \insql{MATCH} clause.  How should this clause be evaluated if \insql{contact_tracing} is considered as a temporal property graph? 
The first issue is that variables in the \insql{MATCH} clause are to be assigned temporal objects; for example, \insql{(x:Person)} indicates that \insql{x} is a variable to be assigned a temporal object $(v,t)$, where $v$ is a node with label \insql{Person} that exists at time point $t$. 
This issue is addressed by adding an extra column for each variable to indicate the time point when that variable exists (table entries appear side-by-side to save vertical space):
\begin{center}\footnotesize
    \begin{tabular}{|cc|}
        \hline
        {\color{blue}{\bfseries\ttfamily x}} & {\color{blue}{\bfseries\ttfamily x\_time}}\\
        \hline
        \hline
        $n_1$ & 1\\
        \multicolumn{2}{|c|}{$\cdots$}\\
        $n_1$ & 9\\
        \hline
    \end{tabular}
    \begin{tabular}{|cc|}
        \hline
        {\color{blue}{\bfseries\ttfamily x}} & {\color{blue}{\bfseries\ttfamily x\_time}}\\
        \hline
        \hline
        $n_2$ & 1\\
        \multicolumn{2}{|c|}{$\cdots$}\\
        $n_7$ & 8\\
        \hline
    \end{tabular}
\end{center}
Observe that the time point $t$ for each value $v$ of \insql{x} is stored in the column \insql{x_time}.
Hence, the binding \insql{x}~$\mapsto n_1$, \mbox{\insql{x_time}~$\mapsto 1$} is in the resulting table, since $n_1$ is a node with label \insql{Person} that exists at time point 1 in \insql{contact_tracing}, and similarly for the other bindings. 
\rev{This illustrates that TRPQs without temporal navigation operate under snapshot reducibility, a design principle discussed in Section~\ref{sec-intro-summary}.}

Having explained how bindings to temporal objects are represented, we can now illustrate the main features of our query language. As in other popular graph query languages, we use curly brackets to indicate restrictions on property values. As our first example, consider the following \insql{MATCH} clause:

{\footnotesize \begin{lstlisting}
Q2 MATCH (x:Person {risk = 'low'}) 
   ON contact_tracing
\end{lstlisting}}

\noindent
The expression \insql{\{risk = 'low'\}} is used to indicate that the value of property \insql{risk} must be \insql{'low'}. The following binding table is the result of evaluating the previous \insql{MATCH} clause:
\begin{center}\footnotesize 
    \begin{tabular}{|cc|}
        \hline
        {\color{blue}{\bfseries\ttfamily x}} & {\color{blue}{\bfseries\ttfamily x\_time}}\\
        \hline
        \hline
        $n_1$ & 1\\
        \multicolumn{2}{|c|}{$\cdots$}\\
        $n_1$ & 9\\
        \hline
    \end{tabular}
    \begin{tabular}{|cc|}
        \hline
        {\color{blue}{\bfseries\ttfamily x}} & {\color{blue}{\bfseries\ttfamily x\_time}}\\
        \hline
        \hline
        $n_2$ & 1\\
        \multicolumn{2}{|c|}{$\cdots$}\\
        $n_2$ & 4\\
        \hline
    \end{tabular}
    \begin{tabular}{|cc|}
        \hline
        {\color{blue}{\bfseries\ttfamily x}} & {\color{blue}{\bfseries\ttfamily x\_time}}\\
        \hline
        \hline
        $n_6$ & 2\\
        \multicolumn{2}{|c|}{$\cdots$}\\
        $n_6$ & 11\\
        \hline
    \end{tabular}
\end{center}
Observe that the binding \insql{x}~$\mapsto n_2$, \insql{x_time}~$\mapsto 4$ is in this table, since $n_2$ is a node such that the label of $n_2$ is \insql{Person}, $n_2$ exists at time point 4, and the value of property \insql{risk} is \mbox{\insql{'low'}} for $n_2$ at time point 4, and likewise for the other bindings in this table. As a second example, consider the following query:

{\footnotesize \begin{lstlisting}
Q3 MATCH (x:Person {risk = 'low' AND time = '1'})
   ON contact_tracing
\end{lstlisting}}

\noindent
In this case, we use the reserved word \insql{time} to indicate that we are considering temporal objects at time point 1. The following is the result of evaluating this \insql{MATCH} clause:
\begin{center}\footnotesize 
    \begin{tabular}{|cc|}
        \hline
        {\color{blue}{\bfseries\ttfamily x}} & {\color{blue}{\bfseries\ttfamily x\_time}}\\
        \hline
        \hline
        $n_1$ & 1\\
        $n_2$ & 1\\
        \hline
    \end{tabular}
\end{center}
Other operators can limit the time under consideration, for example, to consider temporal objects at time less than 10:

{\footnotesize \begin{lstlisting}
Q4 MATCH (x:Person {risk = 'low' AND time < '10'}) 
   ON contact_tracing
\end{lstlisting}}

%
\noindent
Now, suppose that we want to retrieve the pairs of low- and high-risk people who have met, along with information about their meeting. For this, we can use the following query:
{\footnotesize \begin{lstlisting}
Q5 MATCH (x:Person {risk = 'low'})-
         [z:meets]->(y:Person {risk = 'high'})
   ON contact_tracing
\end{lstlisting}}

\noindent
The result of evaluating this \insql{MATCH} clause is:
\begin{center}\footnotesize 
    \begin{tabular}{|cccccc|}
        \hline
        {\color{blue}{\bfseries\ttfamily x}} & {\color{blue}{\bfseries\ttfamily x\_time}} & 
        {\color{blue}{\bfseries\ttfamily z}} & {\color{blue}{\bfseries\ttfamily z\_time}} & 
        {\color{blue}{\bfseries\ttfamily y}} & {\color{blue}{\bfseries\ttfamily y\_time}} \\
        \hline
        \hline
        $n_1$ & 5 & $e_1$ & 5 & $n_2$ & 5\\
        $n_1$ & 6 & $e_1$ & 6 & $n_2$ & 6\\
        $n_2$ & 1 & $e_2$ & 1 & $n_3$ & 1\\
        $n_2$ & 2 & $e_2$ & 2 & $n_3$ & 2\\
        \hline
    \end{tabular}
\end{center}
As in other popular graph query languages~\cite{PGQL16,Cypher18,GCore18}, an expression of the form \insql{-[:meets]->} indicates the existence of an edge with label \insql{meets}. We assign the variable \insql{z} to the temporal object that represents that edge.

Importantly, an expression of the form \insql{-[...]->} represents the structural navigation operator that is conceptually evaluated over the snapshots (temporal states) of the graph. 
This is the reason why each binding in the resulting table has the same value in columns  \insql{x_time}, \insql{z_time}, and \insql{y_time} . For example, the binding \insql{x}~$\mapsto n_1$, \insql{x_time}~$\mapsto 5$, \insql{z}~$\mapsto e_1$, \insql{z_time}~$\mapsto 5$, \insql{y}~$\mapsto n_2$, \insql{y_time}~$\mapsto 5$ is in this table, since $n_1$ is a low-risk person at time point 5, $n_2$ is a high-risk person at time point 5, and there exists an edge $e_1$ with label \insql{meets} between $n_1$ and $n_2$  at time point 5.


To ensure that our proposal is practically useful, a minimum requirement is that queries can be evaluated in polynomial time over \tpgs.
Hence, we have to choose very carefully how structural navigation is combined with temporal navigation, and how we refer to time in the query language, as the complexity 
can quickly become intractable when navigation patterns are combined with functionalities for comparing property values~\cite{LMV16}. In fact, 
there is even a fixed query $Q$ for which this negative result holds~\cite{LMV16}. This means that the problem of computing, given a graph $G$ as input, the answer to $Q$ over $G$ is intractable in data complexity~\cite{V82}.

The basic temporal navigation operators in our language are \insql{PREV} and \insql{NEXT} that move by one unit of time into the past and into the future, respectively. 
Consider the following query:

{\footnotesize \begin{lstlisting}
Q6 MATCH (x:Person {test = 'pos'})-
         /PREV/-(y:Person)
   ON contact_tracing
\end{lstlisting}}

\noindent
Here, \insql{x} and \insql{y} are temporal objects that correspond to \emph{the same} real-world object ---a node of type \insql{Person}. In this case, \insql{x} has the value \insql{'pos'} in the property \insql{test}, meaning that \insql{x} tested positive at some time point, and \insql{y} denotes the same node at the time immediately before testing positive.
\rev{Temporal navigation allows single-step temporal movement, and is orthogonal to structural navigation, following navigation orthogonality, discussed in Section~\ref{sec-intro-summary}. Note that \insql{PREV} and \insql{NEXT} reference timestamps, operating under extended snapshot reducibility~\cite{Bohlen2000}, discussed in Section~\ref{sec-related}.}

This example illustrates the use of notation \insql{-/.../-} to specify a pattern that a path connecting objects \insql{x} and \insql{y} must satisfy. In general, such a pattern is a regular expression that can include temporal and structural operators  (see formal definition in Section~\ref{sec-t-rpq}). 
In this example, assuming that  the temporal object $(o_1,t_1)$  corresponds to \mbox{\insql{(x:Person \{test = 'pos'\})}}, and  the temporal object $(o_2,t_2)$  corresponds to \insql{(y:Person)}, then the expression \insql{-/PREV/-} indicates that $(o_1,t_1)$ must be connected with $(o_2,t_2)$ through a path conforming to \mbox{\insql{PREV},} that is, $t_2 = t_1 - 1$.
Importantly,  \insql{-/PREV/-} is evaluated under the restriction that no structural navigation must have occurred, given the separation between temporal and structural navigation that we are arguing for in this work. Hence, we conclude that 
$o_2 = o_1$. The following binding table is the result of evaluating \insql{Q6}:
\begin{center}\footnotesize
    \begin{tabular}{|cccc|}
        \hline
        {\color{blue}{\bfseries\ttfamily x}} & {\color{blue}{\bfseries\ttfamily x\_time}} & 
        {\color{blue}{\bfseries\ttfamily y}} & {\color{blue}{\bfseries\ttfamily y\_time}} \\
        \hline
        \hline
        $n_6$ & 9 & $n_6$ & 8\\
        \hline
    \end{tabular}
\end{center}
Temporal and structural navigation can be combined to retrieve information about which room person \insql{x} was visiting immediately before she received a positive test result:

{\footnotesize \begin{lstlisting}
MATCH (x:Person {test = 'pos'})-
      /PREV/-(y:Person)-[:visits]->(z:Room)
ON contact_tracing
\end{lstlisting}}

\noindent
The result of evaluating this \insql{MATCH} clause is:
\begin{center}\footnotesize
    \begin{tabular}{|cccccc|}
        \hline
        {\color{blue}{\bfseries\ttfamily x}} & {\color{blue}{\bfseries\ttfamily x\_time}} & 
        {\color{blue}{\bfseries\ttfamily y}} & {\color{blue}{\bfseries\ttfamily y\_time}} &
        {\color{blue}{\bfseries\ttfamily z}} & {\color{blue}{\bfseries\ttfamily z\_time}} \\
        \hline
        \hline
        $n_6$ & 9 & $n_6$ & 8 & $n_4$ & 8\\
        \hline
\end{tabular}
\end{center}
Observe that the temporal operator \insql{PREV} moves from \mbox{(\insql{x}, \insql{x_time})} to (\insql{y}, \insql{y_time}), while the structural operator \mbox{\insql{-[:visits]->}} moves from (\insql{y}, \insql{y_time}) to \mbox{(\insql{z}, \insql{z_time})}. Hence, temporal and structural navigation are carried out separately. Besides, observe that the intermediate variable \insql{y} is not needed when retrieving the list of rooms that person \insql{x} was visiting, we just included it to show the paths that are constructed when using different operators. The following simplified \insql{MATCH} clause

{\footnotesize \begin{lstlisting}
MATCH (x:Person {test = 'pos'})-
      /PREV/-()-[:visits]->(z:Room)
ON contact_tracing
\end{lstlisting}}

\noindent
can be used to obtain the desired answer:
\begin{center}\footnotesize 
\begin{tabular}{|cccc|}
    \hline
    {\color{blue}{\bfseries\ttfamily x}} & {\color{blue}{\bfseries\ttfamily x\_time}} & 
    {\color{blue}{\bfseries\ttfamily z}} & {\color{blue}{\bfseries\ttfamily z\_time}} \\
    \hline
    \hline
    $n_6$ & 9 & $n_4$ & 8\\
    \hline
\end{tabular}
\end{center}
%
At this point the reader may be wondering why the language is asymmetric, and it includes different notation for temporal and structural navigation. We have kept the notation \mbox{\insql{-[...]->} to} be compatible with graph query languages used today \cite{Cypher18,PGQL16,GCore18}, but an important feature of our proposal is the use of notation \insql{-/.../-} to include regular expressions combining temporal and structural operators. Hence, we include  two basic structural navigation operators, \insql{BWD} (``backward'') and \insql{FWD} (``forward''), that are analogous to the temporal operators \insql{PREV} and \insql{NEXT}. Assume that an edge is given
\begin{eqnarray}
\label{eq-edge-bf}
(n,t) & \xrightarrow{(e,t)} & (n',t),
\end{eqnarray}
which, in the formal $\tpgs$ notation (see Definition~\ref{def:tg}), represents the fact that $\rho(e) = (n,n')$, $\xi(n,t) = \true$, $\xi(e,t) = \true$, and $\xi(n',t) = \true$. Then, operator \insql{FWD} moves forward from node $n$ to edge $e$, or from edge $e$ to node $n'$, while keeping time $t$ unchanged.  That is,  \insql{FWD} operates in a $\tpg$ snapshot corresponding to time $t$. 
Similarly, operator \insql{BWD} moves backwards from node $n'$ to edge $e$, and from edge $e$ to node $n$ in a $\tpg$ snapshot corresponding to time $t$.
Thus, we can rewrite the previous \insql{MATCH} clause as follows:

{\footnotesize \begin{lstlisting}
Q7 MATCH (x:Person {test = 'pos'})-
         /PREV/FWD/:visits/FWD/-(z:Room)
   ON contact_tracing
\end{lstlisting}}

\noindent
The regular expression \insql{PREV/FWD/:meets/FWD} uses the concatenation operator \insql{/} to indicate that operator \insql{PREV} has to be executed first followed by the expression \insql{FWD/:visits/FWD}, which is executed in the same way. (The precise syntax and semantics of such expressions are presented in Section \ref{sec-t-rpq}.)
Observe that in our query language, the expression \mbox{\insql{-[:visits]->}} is equivalent to \insql{-/FWD/:visits/FWD/-}.  This is because, given an edge of the form of Expression \eqref{eq-edge-bf}, the first operator \insql{FWD} moves from $n$ to $e$, then \insql{:visits} checks that the label of $e$ is \insql{visits}, and finally the last operator \insql{FWD} moves from $e$ to $n'$, thus obtaining the same result as using the operator \insql{-[:visits]->} in an edge of the form of Expression~\eqref{eq-edge-bf}.

So far we only looked at expressions that navigate one step at a time, temporally or structurally.  Our language also supports the Kleene star, indicating zero or more occurrences of an operator.  For example, \insql{Q8} retrieves the list of rooms person \insql{x} visited at any time prior to receiving a positive test (including also at the time when \insql{x} received the test):

{\footnotesize \begin{lstlisting}
Q8 MATCH (x:Person {test = 'pos'})-
         /PREV*/FWD/:visits/FWD/-(z:Room)
   ON contact_tracing
\end{lstlisting}}

\noindent
producing the following temporal bindings:
\begin{center}\footnotesize 
    \begin{tabular}{|cccc|}
        \hline
        {\color{blue}{\bfseries\ttfamily x}} & {\color{blue}{\bfseries\ttfamily x\_time}} & 
        {\color{blue}{\bfseries\ttfamily z}} & {\color{blue}{\bfseries\ttfamily z\_time}} \\
        \hline
        \hline
        $n_6$ & 9 & $n_4$ & 8\\
        $n_6$ & 9 & $n_4$ & 7\\
        $n_6$ & 9 & $n_5$ & 6\\
        $n_6$ & 9 & $n_5$ & 5\\
        \hline
    \end{tabular}
\end{center}
As another example, we can retrieve the high-risk people who met someone who subsequently tested positive for an infectious disease:

{\footnotesize \begin{lstlisting}
Q9 MATCH (x:Person {risk = 'high'})-
         /FWD/:meets/FWD/NEXT*/-({test = 'pos'})
   ON contact_tracing
\end{lstlisting}}

\noindent
Recall that the temporal operator \insql{NEXT} moves in time by one unit into the future.
This query returns the following temporal bindings when evaluated over the graph in Figure~\ref{fig:covid}:
\begin{center}\footnotesize 
\begin{tabular}{|cc|}\hline
{\color{blue}{\bfseries\ttfamily x}} & {\color{blue}{\bfseries\ttfamily x\_time}}  \\
\hline
\hline
$n_3$ & 4 \\
$n_7$ & 5 \\
$n_7$ & 6 \\
\hline
\end{tabular}
\end{center}
Observe 
that the term \insql{(\{test = 'pos'\})} does not include a variable, as we are not storing the contacts who tested positive to avoid stigmatizing them, and only record those who are potentially at risk for complications. 

Moreover, our query language allows to specify the number of times an operator is used. Thus, assuming that the time unit in \insql{contact_tracing} is 5 minutes, we can retrieve the list of high-risk people who met someone who tested positive for an infectious disease 1 hour prior to the meeting:

{\footnotesize \begin{lstlisting}
Q10 MATCH (x:Person {risk = 'high'})-
          /FWD/:meets/FWD/PREV[0,12]-
          ({test = 'pos'})
    ON contact_tracing
\end{lstlisting}}

\noindent
Next, consider the following notion of close contact for an infectious disease: If person $a$ visits the same room as person $b$, and $b$ tests positive for this disease at most two weeks after they visited the same room as $a$, then $a$ is considered to have been in close contact with an infected person. The \mbox{\insql{MATCH} clause} below retrieves high-risk people who have been in close contact with an infected~person:

{\footnotesize \begin{lstlisting}
Q11 MATCH (x:Person {risk = 'high'})-
          /FWD/:visits/FWD/:Room/BWD/:visits/
              BWD/NEXT[0,12]/-({test = 'pos'})
    ON contact_tracing
\end{lstlisting}}

\noindent
Observe that, as was the case for edge labels, node labels can be used inside an expression \insql{-/.../-}, and so  \mbox{\insql{-/:Room/-} in} the expression above is equivalent to \insql{-(:Room)-}.
The query \insql{Q11} produces the following binding table:
\begin{center}\footnotesize
    \begin{tabular}{|cc|}
        \hline
        {\color{blue}{\bfseries\ttfamily x}} & {\color{blue}{\bfseries\ttfamily x\_time}}\\
        \hline
        \hline
        $n_3$ & 7\\
        $n_7$ & 7\\
        $n_7$ & 8\\
        \hline
    \end{tabular}
\end{center}
As the final example, assume that if person $a$ meets with  person $b$, and $b$ tests positive for an infections disease at most two weeks after their meeting, then $a$ should also be considered to have been in close contact with an infected person. \insql{Q11} can be extended to consider this additional case:

{\footnotesize \begin{lstlisting}
 MATCH (x:Person {risk = 'high'})-
       /(FWD/:meets/FWD/NEXT[0,12]) + 
        (FWD/:visits/FWD/:Room/BWD/:visits/
         BWD/NEXT[0,12])/-({test = 'pos'})
ON contact_tracing
\end{lstlisting}}

\noindent
This query produces the following bindings:
\begin{center}\footnotesize
    \begin{tabular}{|cc|}
        \hline
        {\color{blue}{\bfseries\ttfamily x}} & {\color{blue}{\bfseries\ttfamily x\_time}}\\
        \hline
        \hline
        $n_3$ & 4\\
        $n_3$ & 7\\
        $n_7$ & 5\\
        \hline
    \end{tabular}
    \begin{tabular}{|cc|}
        \hline
        {\color{blue}{\bfseries\ttfamily x}} & {\color{blue}{\bfseries\ttfamily x\_time}}\\
        \hline
        \hline
        $n_7$ & 6\\
        $n_7$ & 7\\
        $n_7$ & 8\\
        \hline
    \end{tabular}
\end{center}
As usual in regular expressions, operator \insql{+} represents union. Thus, the regular expression in the previous \insql{MATCH} clause indicates that the results of \insql{FWD/:meets/FWD/NEXT[0,12]} should be put together with the results of \insql{FWD/:visits/FWD/:Room/BWD/:visits/BWD/NEXT[0,12]}. Observe that parentheses are used to have unambiguous expressions that can be parsed in a unique way. For example, the previous expression can be rewritten as follows to avoid using the temporal operator \mbox{\insql{NEXT[0,12]} twice}. (Observe the required use of parentheses to get the desired effect.)

{\footnotesize \begin{lstlisting}
Q12 MATCH (x:Person {risk = 'high'})-
          /(FWD/:meets/FWD + 
            FWD/:visits/FWD/:Room/BWD/:visits/
            BWD)/NEXT[0,12]/-({test = 'pos'})
    ON contact_tracing
\end{lstlisting}}

\noindent
In this section, we illustrated the main features of our proposed language and  showed how popular graph query languages ~\cite{Cypher18,PGQL16,GCore18} can be extended to include these features. We will define the syntax and the semantics of our language next.

\section{Temporal Regular Path Queries}
\label{sec-t-rpq}




\rev{In this section, we provide a formal syntax and semantics for the expression \insql{path} described in the previous section, and study the complexity of evaluating it. In Section~\ref{sec-fss-NavL}, we extend} the widely used notion of regular path query \cite{AV97,calvanese2002rewriting,B13,DBLP:journals/csur/AnglesABHRV17} to deal with temporal objects in \tpgs, which gives rise to the language \Lthree. Moreover, we show in Section \ref{sec-fss-NavL} how \Lthree provides a formalization of the practical query language proposed \rev{in the previous section}. 
Then 
we define the semantics of \Lthree in Section \ref{sec:sem-comp-navl}, by following the definition 
of widely used query languages such as XPath and regular path queries \cite{CD99,M05,GKP05,RDS17,AV97,calvanese2002rewriting,B13,DBLP:journals/csur/AnglesABHRV17}.  Moreover, we study in Section \ref{sec:sem-comp-navl} the complexity of the evaluation problem for \Lthree
for TPGs and ITPGs. \rev{Finally, we provide in Section \ref{sec-comp-other} a comparison of our proposal with other temporal query languages.}
Proofs and additional results can be found in the Appendix.

\subsection{Syntax of NavL[PC, NOI], and its  relationship with  the  practical query language}
\label{sec-fss-NavL}

Recall that labels, property names, and property values are drawn from the sets $\Lab$, $\Prop$, and $\Val$, respectively. Then the expressions in \Lthree, which are called temporal regular path queries (TRPQs), are defined
by the  grammar:
\begin{multline}
        \pt \  ::=  \  \test \,  \mid \,  \axis \,  \mid \,  (\pt/\pt) \,  \mid \\ 
        (\pt+\pt) \,  \mid \,  
        \pt[ n, m ] \,  \mid \,  \pt[n,\_] 
\label{grammar:path-expr}
\end{multline}
where $n$ and $m$ are natural numbers such that $n \leq m$. Intuitively, $\test$ checks a condition on a given node or edge at a given time point, $\axis$ allows structural or temporal navigation, $(\pt/\pt)$ is used for the concatenation of two TRPQs, $(\pt+\pt)$ allows for the disjunction of two TRPQs, $\pt[n, m]$ allows $\pt$ to be repeated a number of times that is between $n$ and $m$, whereas $\pt[n,\_]$ only imposes a lower bound of at least $n$ repetitions of expression $\pt$. 
The Kleene star $\pt^*$ can be expressed as $\pt[0, \_]$, and the expression $\pt[\_,n]$ is equivalent to $\pt[0,n]$.

\textit{Conditions} on temporal objects are defined by the 
grammar:
\begin{multline}
        \test \ ::= \ \vd \  \mid \  \ed \  \mid \  \ell \  \mid \  \propt{p}{v} \  \mid \  < \mt \  \mid \  \ex \  \mid \\  (?\pt) \  \mid \
        (\test \vee \test) \  \mid \  (\test \wedge \test) \ \mid (\neg \test)
    \label{grammar:test}
\end{multline}
where $\ell \in \Lab$, $p \in \Prop$, $v \in \Val$, and $\mt \in \mathbb{N}$. Intuitively, $\test$ is meant to be applied to a temporal object, that is, to a pair $(o, t)$ with  object $o$ and time point $t$. $\vd$ and $\ed$ test whether the object is a node or an edge, respectively; the term $\ell$ checks 
whether the label of the object is $\ell$; the term $\propt{p}{v}$ checks 
whether the value of property $p$ is $v$ for the object at the given time point; $\ex$ checks 
whether the object exists at the given time point; and $< \mt$ checks 
whether the current time point is less than $\mt$. Further, $\test$ can be $(?\pt)$, where $\pt$ is an expression satisfying grammar \eqref{grammar:path-expr}, meaning that there is a path starting on the tested temporal object that satisfies $\pt$. Finally, $\test$ can be a disjunction or a conjunction of a pair of $\test$ expressions, or a negation of a $\test$ expression.

Furthermore, the following grammar defines \emph{navigation}: 
\begin{align}
    \axis \ ::= \ \fw \ \mid \ \bw \ \mid \ \nxt \ \mid \ \prv \label{grammar:axis}
\end{align}
Operators $\fw$, $\bw$ move structurally in a \tpg: $\fw$ moves forward in the direction of an edge, and $\bw$ moves backward in the reverse direction of an edge. Operators $\nxt$, $\prv$ move temporally in a \tpg: $\nxt$  moves to the next time point, and $\prv$ moves to the previous time point.

Having a formal definition of the syntax of  \Lthree, we show that this 
language provides a formalization 
of the practical query language of Section \ref{sec-standard}. 
More precisely, temporal navigation operators \insql{PREV} and \insql{NEXT} in the practical query language correspond to the analogous operators $\prv$ and $\nxt$ in \Lthree, respectively, while structural navigation operators \insql{BWD} and \insql{FWD} in the practical query language correspond to the operators $\bw$ and $\fw$ in \Lthree, respectively. 
Then consider the following \insql{MATCH} clause over an arbitrary TPG:

{\footnotesize \begin{lstlisting}
MATCH (x:Person {test = 'pos'})-/PREV/-(y) 
ON graph
\end{lstlisting}}

\noindent
Our task is to construct a query $\pt$ in \Lthree such that the evaluation of this \insql{MATCH} clause over \mbox{\insql{graph} is} equivalent to the evaluation of $\pt$ over this \tpg. 
The following expression satisfies this condition:
\begin{align*}
(\vd \wedge \text{Person} \wedge \propt{\text{test}}{\text{pos}})
/\prv/(\vd \wedge \exists)
\end{align*}
Observe that $(\vd \wedge \text{Person} \wedge \propt{\text{test}}{\text{pos}})$ is used to check whether the following conditions are satisfied for a temporal object $(o,t)$: $o$ is a node with label $\text{Person}$ and with value $\text{pos}$ in the property $\text{test}$ at time point $t$. Notice that, by definition of $\tpgs$, the fact that $\propt{\text{test}}{\text{pos}}$ holds 
at time $t$ implies that node $o$ exists at this time point.
Hence, $(\vd \wedge \text{Person} \wedge \propt{\text{test}}{\text{pos}})$ is used to represent the expression \mbox{\insql{(x:Person \{test = 'pos'\})}}. Moreover, temporal navigation operator $\prv$ is used to move from the temporal object $(o,t)$ to a temporal object $(o,t')$ such that $t' = t-1$, so that it is used to represent the expression \insql{-/PREV/-}. 
Finally, the condition $(\vd \wedge \exists)$ is used to test that $o$ is a node that exists at time $t'$. Observe that we explicitly need to mention the condition $\exists$, as expressions in \Lthree do not enforce the existence of temporal objects by default. The main reason to choose such a semantics is that there are many scenarios where moving through temporal objects that do not exists is useful, in particular when these temporal objects only exist at certain time points. For example, 
if a room is unavailable for some time, then the temporal path expression \[
    (\text{Room} \wedge \neg \exists)/(\nxt/ \neg \exists)[0,\_]/(\text{Room} \wedge \exists)
\]
can be used to look for the next time the room is available. Here, $(\nxt/\neg \exists)[0,\_]$ moves through an arbitrary number of time points during which the room is unavailable, until the condition $\exists$ holds, and the room becomes available.

As a second example, consider query \insql{Q8} from Section \ref{sec-standard}.
%
%
Based on the previous discussion, 
such a query can be represented as the following TRPQ:
\begin{multline*}
    (\vd \wedge \text{Person} \wedge \propt{\text{test}}{\text{pos}})/\\
    (\prv/\exists)[0,\_]/\fw/(\text{visits} \wedge \exists)/\fw/(\vd \wedge \text{Room}),
\end{multline*}
where all temporal objects must exist, as required in 
Section~\ref{sec-standard}. Note that we have not explicitly included the existence condition on the last room node, as the existence of an edge at time point $t$ implies, according to the definition of $\tpgs$, the existence of its starting and ending nodes. 

As an additional example, consider query \insql{Q12} from Section~\ref{sec-standard}, which uses many of the features of \Lthree.
%
%
This 
query corresponds to the temporal path expression:
\begin{align*}
    & (\vd \wedge \text{Person} \wedge \propt{\text{risk}}{\text{high}})/(\fw/(\text{meets} \wedge \exists)/\fw \ +\\
    & \hspace{45pt}\fw/(\text{visits} \wedge \exists)/\fw/\text{Room}/\bw/(\text{visits} \wedge \exists)/\bw)/\\
    & \hspace{90pt}(\nxt/\exists)[0,12]/(\vd \wedge \propt{\text{test}}{\text{pos}})
\end{align*}
As our final example, consider query \insql{Q4} 
from Section \ref{sec-standard}.
%
%
The use of a condition over the reserved word \insql{time} is represented in \Lthree by the condition $< \mt$. For example, \insql{time < '10'} is represented by the condition $<\!10$, as a temporal object $(o,t)$ satisfies $<\!10$ if, and only if, $t < 10$.
Hence, \insql{Q4} is equivalent to the following query in \Lthree:
\begin{align*}
    (\vd \wedge \text{Person} \wedge \propt{\text{risk}}{\text{low}} \ \wedge <\!10)
\end{align*}
Notice that abbreviations can be introduced for some of the operators described in this section, and some other common operators, to make notation of the formal language easier to use. For example, we could use condition $= \!\mt$, which is written in \Lthree as $(<\!\mt+1 \wedge \neg(<\!\mt))$, and operator $\mathbf{NE}$ that moves by one unit into the future if the object that is reached exists. However, as such operators are expressible in \Lthree, we prefer to use a minimal notation in this formal language to simplify its definition and analysis.

\subsection{Semantics and complexity of NavL[PC,NOI]}
\label{sec:sem-comp-navl}
Let $\G = (\Omega,N,E,\rho,\lambda,\xi,\sigma)$ be a $\tpg$. 
Given an expression $\pt$ in \Lthree,
the evaluation 
of $\pt$ 
over 
$\G$, denoted by $\sem{ \pt }$, is defined by the set of tuples $(o,t,o',t')$ such that there exists a sequence of temporal objects starting in $(o,t)$, ending in $(o',t')$, and conforming to $\pt$.  
More precisely, assume that $\src(e) = v_1$ and $\tgt(e) = v_2$ whenever $\rho(e) = (v_1,v_2)$, and assume that  $\Pt(\G) = (N \cup E) \times \Omega \times (N \cup E) \times \Omega$. Then the evaluation of the axes in grammar~\eqref{grammar:path-expr} is defined as:
\begin{eqnarray*}
    \sem{\fw} & = & \{ (v, t, e, t) \in \Pt(\G) \mid  \src(e) = v \} \ \cup \\
    && \{ (e, t, v, t) \in \Pt(\G) \mid \tgt(e) = v \} \\
    \sem{\bw} & = & \{ (v, t, e, t) \in \Pt(\G) \mid \tgt(e) = v \} \ \cup \\
    && \{ (e,t, v,t) \in \Pt(\G) \mid  \src(e) = v \} \\
    \sem{\nxt} & = & \{ (o, t_1, o, t_2) \in \Pt(\G)  \mid  t_2 = t_1 + 1 \} \\
    \sem{\past} & = & \{ (o, t_1, o, t_2) \in \Pt(\G)  \mid  t_2 = t_1 - 1 \}
\end{eqnarray*}
Moreover, assuming that, $\pt$, $\pt_1$ and $\pt_2$ are expressions in \Lthree, we have that:
\begin{align*}
    \sem{(\pt_1 / \pt_2)} \ &= \ \{(o_1, t_1, o_2, t_2) \in \Pt(\G) \mid \\
    &\hspace{25pt}\exists (o,t) \,:\, (o_1, t_1, o, t) \in \sem{\pt_1}\\ 
    &\hspace{45pt}\text{and } (o, t, o_2, t_2) \in \sem{\pt_2} \},\\
    \sem{(\pt_1 + \pt_2)} \ &= \ \sem{\pt_1} \cup \sem{\pt_2},\\
    \sem{\pt[n, m]} \ &= \  \bigcup_{k=n}^m \sem{\pt^k}, \\
    \sem{\pt[n,\_]} \ &= \ \bigcup_{k \geq n} \sem{\pt^k},
\end{align*}
where $\pt^k$ is defined as the concatenation of $\pt$ with itself $k$ times. Finally, the evaluation of an expression $\test$, defined according to grammar~\eqref{grammar:test}, is a navigation expression  that stays in the same temporal object if $\test$ is satisfied:
    $\sem{\test}  =  \{ (o, t, o , t) \in \Pt(\G) \mid  (o,t) \models \test\}$. 
Hence, to conclude the definition of the semantic of \Lthree, we need to indicate when a temporal object $(o,t)$ satisfies a condition $\test$, which is denoted by $(o,t) \models \test$. Formally, this is recursively defined as follows (omitting the usual semantics for Boolean~connectives):
\begin{itemize}[leftmargin=4mm]
    \item If $\test = \vd$, then $(o,t) \models \test$ if $o \in N$;
    
    \item If $\test = \ed$, then $(o,t) \models \test$ if $o \in E$;
    
    \item If $\test = \ell$, with $\ell \in \Lab$, then $(o,t) \models \test$ if $\lambda(o) = \ell$;
    
    \item If $\test = \propt{p}{v}$, with $p \in \Prop$ and $v \in \Val$, then $(o,t) \models \test$ if $\sigma(o,p,t)$ is defined and  $\sigma(o,p,t) = v$;
    
    \item If $\test = \ex$, then $(o, t) \models \test$ if $\xi(o, t) = \true$;
    
    \item If $\test = \ < \mt$, then $(o, t) \models \test$ if $t < \mt$;
    
    
    
       
    \item If $\test = (?\pt)$ for an expression $\pt$ conforming to grammar \eqref{grammar:path-expr}, then $(o, t) \models \test$ if there exists a temporal object $(o',t')$ in $\G$ such that 
    $(o,t,o',t') \in \sem{\pt}$.
\end{itemize}


\noindent
To define the evaluation of an expression $\pt$ over a interval-timestamped temporal property graph $I$, we just need to translate $I$ into an equivalent \tpg and consider the previous definition.
Formally, assuming that $\can(\cdot)$ is a canonical translation from an \ctpg into an equivalent \tpg, we have that:
$\semp{\pt}{I}  =  \semp{\pt}{\can(I)}$.

Having a formal definition of TRPQs allows not only to provide an unambiguous definition of the practical query language of Section \ref{sec-standard}, but also to formally study the complexity of evaluating this language. Assuming that $\mathcal{G}$ is a class of graphs and $\mathcal{L}$ is a query language, define $\tupleeval(\mathcal{G}$, $\mathcal{L})$ as the problem of verifying whether $(o, t, o', t') \in \semp{\pt}{G}$, for an input consisting of a graph $G \in \mathcal{G}$, an expression $\pt$ in $\mathcal{L}$ and a pair $(o,t)$, $(o',t')$ of temporal objects in $G$.
By studying the complexity  of $\tupleeval(\mathcal{G},\mathcal{L})$ for different fragments $\mathcal{L}$ of \Lthree, we can understand how the use of the operators in \Lthree affects the complexity of the evaluation problem, and which operators are mode difficult to~implement. 

Assume that \Ltwo is the fragment of \Lthree obtained by disallowing numerical occurrence indicators, while
\Lonep~is the fragment of \Lthree obtained by disallowing path conditions.
\begin{theorem}\label{th:main-summary}
The following results hold.
\begin{enumerate}
    \item {\em \tupleeval({\rm TPG}, \Lthree)} and 
    {\em \tupleeval({\rm ITPG}, \Ltwo)} can be solved in polynomial time.
   
   \item {\em \tupleeval({\rm ITPG}, \Lonep)} is \sigmatwop-hard, and {\em \tupleeval({\rm ITPG}, \Lthree)} is \pspace-complete.
\end{enumerate}
\end{theorem}
The results of this section can guide future implementations of \Lthree over interval-timestamped \tpgs. The main insight is that, while \tupleeval({\rm ITPG}, \Lonep) and \tupleeval({\rm ITPG}, \Lthree) are intractable, the language including only path conditions can be efficiently evaluated over such graphs. 

\subsection{\rev{A comparison with T-GQL and Cypher}}
\label{sec-comp-other}

\rev{T-GQL is a recently proposed temporal query language~\cite{Debrouvier21} developed on top of Cypher~\cite{Cypher18}, a popular graph query language. 
We now compare our TRPQs with T-GQL, and with the alternative of implementing  a temporal graph query language that encodes time intervals as lists  directly in Cypher.}

\rev{First, consider the five design principles of our language, described in Section \ref{sec-intro-summary}. Since Cypher's data model does not explicitly consider time, it is not surprising that it does not satisfy navigability,  navigation orthogonality,  static testability, or snapshot reducibility, and only node-edge symmetry is satisfied. 
T-GQL satisfies navigability, navigation orthogonality and snapshot reducibility, but it treats nodes and edges differently, violating node-edge symmetry. Moreover, T-GQL test conditions do not satisfy static testability.}

\rev{Second, consider the complexity of the query evaluation problem. As shown in Theorem \ref{th:main-summary}, our query language can be evaluated in polynomial time over temporal property graphs. In contrast, the evaluation problem for Cypher is intractable, even if we focus on non-temporal property graphs (i.e., a temporal property graph consisting of a single timestamp). In fact, a fixed query that checks for the existence of two disjoint paths from the same source node to the same destination node can be expressed in Cypher and is known to be NP-hard~\cite{Cypher18}. Whether these intractability results carry over \mbox{T-GQL} is not clear, as an exact characterization of T-GQL as a fragment of Cypher has not yet been provided.}


\rev{Finally, we compare the expressive power of our proposal with Cypher and T-GQL. As Cypher is a general purpose graph query language, it is not surprising that every query in our proposal can be expressed in it, but at the cost of using unnatural and expensive time interval encodings. However, we can show that some natural TRPQs cannot be expressed in \mbox{T-GQL}. First, consider a graph for travel scheduling that includes different transportation services, such as flights, trains, and buses. By the definition of  consecutive path in \cite{Debrouvier21}, it is not possible to express a query in T-GQL that indicates how to go from one city to another combining different transportation services, which can be easily expressed in our proposal. As a more fundamental example, 
consider a query that  retrieves paths that combine an arbitrary number of temporal journeys, some of them moving to the future and some to the past. Such a combination of temporal journeys cannot be specified in T-GQL, while it can be handled by our proposal.}


\section{Implementation}
\label{sec-system}
We implement a fragment of 
\Lthree that includes all queries of Section~\ref{sec-standard} over interval-timestamped \tpgs.  We use Rust and the Itertools library~\cite{rust-itertools}, which efficiently implements dataflow operators, supports lazy evaluation of expressions, and collects data only when necessary. \rev{For multithreaded implementation, we use Rayon-Rs~\cite{rayon-rs}, an interface over dataflow operators.}
Our algorithms can be implemented using any system that supports the dataflow model, such as Apache Spark~\cite{DBLP:journals/cacm/ZahariaXWDADMRV16}, Apache Flink~\cite{carbone2015apache}, Timely~\cite{murray2013naiad} and Differential dataflow~\cite{mcsherry2013differential}. 

We represent a \tpg as a pair of interval-timestamped temporal relations  
{\small $\vtab(\underline{\text{\bfseries\ttfamily id}}, \text{\bfseries\ttfamily label}, \text{\bfseries\ttfamily properties}, \text{\bfseries\ttfamily time})$} and {\small $\etab(\underline{\text{\bfseries\ttfamily id}}, \text{\bfseries\ttfamily src}, \text{\bfseries\ttfamily tgt}, \text{\bfseries\ttfamily label}, \text{\bfseries\ttfamily properties}, \text{\bfseries\ttfamily time})$}, 
where {\small \text{\bfseries\ttfamily properties}} are a set of key-value pairs. For example, for node $n_2$ and edge $e_1$ from Figure~\ref{fig:covid}, we have:


\begin{center}
\footnotesize  
    $\vtab$\\\vspace{2pt}
    \setlength{\tabcolsep}{4pt}
    \begin{tabular}{|cccc|}
        \hline
        \underline{\bfseries\ttfamily id} &  {\bfseries\ttfamily label} & 
         {\bfseries\ttfamily properties} & {\bfseries\ttfamily time} \\
        \hline
        \hline
        $n_2$ & \insql{Person} & \insql{\{name = 'Bob', risk = 'low'\}} & [1, 4]\\
        $n_2$ & \insql{Person} & \insql{\{name = 'Bob', risk = 'high'\}} & [5, 9]\\
        \hline
    \end{tabular}
\end{center}

\setlength{\tabcolsep}{7.5pt}
\begin{center}
\footnotesize 
    $\etab$\\\vspace{2pt}
    \begin{tabular}{|cccccc|}
        \hline
        \underline{\bfseries\ttfamily id} & {\bfseries\ttfamily src} &
        {\bfseries\ttfamily tgt} & {\bfseries\ttfamily label} & 
         {\bfseries\ttfamily properties} & {\bfseries\ttfamily time} \\
        \hline
        \hline
        $e_1$ & $n_1$ & $n_2$ & \insql{meets} & \insql{\{loc = 'cafe'\}} & [3, 3]\\
        $e_1$ & $n_1$ & $n_2$ & \insql{meets} & \insql{\{loc = 'park'\}} & [5, 6]\\
        \hline
    \end{tabular}
\end{center}


            
            
%

By the formal definition of TRPQs in Section \ref{sec-t-rpq}, we know that temporal and structural navigation operators are orthogonal, in the sense that the language allows non-simultaneous single-step time and structural movements. Hence, we break down the evaluation of a TRPQ into \rev{{\bf Step 1:} evaluating the \textit{structural navigation} 
portion of the path expression over the interval-based \tpg; 
{\bf Step 2:} evaluating the \textit{temporal navigation} portion of the path expression over the interval-based  intermediate result; and {\bf Step 3:} if needed, transforming the intermediate result into a point-wise representation for the final portion of evaluation and materialization.} 

Evaluation of conventional path queries in \rev{\bf Step 1} is a well-studied problem~\cite{B13,DBLP:journals/csur/AnglesABHRV17}. In this work,  we select an optimized select-project-join execution plan for each query in Section~\ref{sec-standard}, and then implement these plans using Itertools operators in Rust.\rev{We implement in-memory hash-join that uses interval-based reasoning to identify temporally-aligned~\cite{Dignos2012} matches. For example, 
for \insql{Q5}, 
we  compute the intersection of the validity intervals for \insql{x}, \insql{y} and \insql{z}.}
%
\rev{For TRPQs without temporal navigation (\insql{Q1}-\insql{Q5}), the final bindings table can be returned after this step, and it can remain temporally coalesced.  For example, the coalesced binding table for \insql{Q5} will contain:}

\begin{center}\footnotesize 
    \begin{tabular}{|cccccc|}
        \hline
        {\color{blue}{\bfseries\ttfamily x}} & {\color{blue}{\bfseries\ttfamily x\_time}} & 
        {\color{blue}{\bfseries\ttfamily z}} & {\color{blue}{\bfseries\ttfamily z\_time}} & 
        {\color{blue}{\bfseries\ttfamily y}} & {\color{blue}{\bfseries\ttfamily y\_time}} \\
        \hline
        \hline
        $n_1$ & [5,6] & $e_1$ & [5,6] & $n_2$ & [5,6]\\
        $n_2$ & [1,2] & $e_2$ & [1,2] & $n_3$ & [1,2]\\
        \hline
    \end{tabular}
\end{center}
\rev{The interpretation of this temporally coalesced result is snapshot-based: we bind \insql{x} $=n_1$, \insql{z} $=e_1$, \insql{y}$=n_2$, with \insql{x_time} = \insql{y_time} = \insql{z_time} = 5, and similarly for time 6.}

\begin{table}[!t]
\centering
\footnotesize
\caption{\rev{Temporal property graphs used in experiments.}} 
\label{tbg:ds-desc}
\begin{tabular}{l|rrrr}
 & \# nodes & \# edges & \# temp. nodes & \# temp. edges      \\
 \hline 
G1 & 1,000  &  12,000      & 3,500             & 14,000        \\
G2 & 2,000  &  30,000      & 7,000             & 35,000        \\
G3 & 4,000  &  84,000      & 14,000            & 94,000         \\
G4 & 6,000  &  158,000     & 20,000            & 180,000       \\
G5 & 8,000  &  253,000     & 28,000            & 282,000       \\
G6 & 10,000 &  371,000     & 34,000            & 413,000      \\
G7 & 25,000 &  2,046,000   & 85,000            & 2,215,000    \\ 
G8 & 50,000 &  7,370,000   & 170,000           & 8,048,000     \\
G9 & 75,000 &  15,717,000  & 256,000           & 17,554,000      \\
G10 & 100,000 & 28,996,000 & 340,000           & 32,255,000      \\
\end{tabular}
\vspace{-0.5cm}
\end{table}

\rev{{\bf Step 2:} To evaluate the \textit{temporal navigation} portion of the path expression, we use interval-based reasoning to join and prune out potential  matches that do not satisfy the temporal constraint.  For example, for \insql{Q7}, 
we can limit the validity interval of \insql{z} to the time immediately before \insql{x} was tested positive. Note that interval intersection and union can be computed in constant time based on interval boundaries. }

\rev{{\bf Step 3:} For the final portion of query evaluation, we may need to use point-wise reasoning for \textit{temporal navigation}.  For example, \insql{Q8} retrieves the list of rooms \insql{z} that person \insql{x} visited at or prior to the time of testing positive.  The \insql{PREV} operator is defined over time points, and we need to compare pairs of time points of \insql{x} and \insql{z} to correctly identify person-room pairs.
Furthermore, \textit{result generation} for TRPQs that use temporal navigation must compute point-based bindings. Returning to our example, in the result of \insql{Q8}, \insql{x_time} may or may not be the same as \insql{z_time}, and so we cannot use an interval representation for the output bindings such as ($n_6$, [5,6], $n_5$, [3,5]), because such a representation is inherently snapshot-based and it does not uniquely map to a set of point-wise temporal bindings over $n_6$ and $n_5$.}

\rev{An exception are TRPQs that return a single variable, such as \insql{Q9}-\insql{Q12}. Results of such queries can be returned temporally coalesced for compactness, although this rarely translates to savings in the running time of query execution, because temporal constraints must be check over a point-based representation for these queries in Step 3, as discussed above.}

\section{Experimental Evaluation}
\label{sec-experiments}
\begin{table}[!t]
\footnotesize
\centering
\caption{\rev{Execution time of queries Q1 through Q12 for graph G10.}}
\label{fig:exp-q}
\begin{tabular}{l|cc|r}
   &  interval-based time (s)  &  total time (s)  &  output size \\ \hline
\multicolumn{1}{l|}{Q1} & 0.004 & \multicolumn{1}{r|}{0.004} & 341,278 \\
\multicolumn{1}{l|}{Q2} & 0.017 & \multicolumn{1}{r|}{0.017} & 278,931\\
\multicolumn{1}{l|}{Q3} & 0.016 & \multicolumn{1}{r|}{0.016} & 26,494\\
\multicolumn{1}{l|}{Q4} & 0.038 & \multicolumn{1}{r|}{0.038} & 116,021\\
\multicolumn{1}{l|}{Q5} & 4.546 & \multicolumn{1}{r|}{4.546} & 743,714\\
\multicolumn{1}{l|}{Q6} & 0.096 & \multicolumn{1}{r|}{0.173} & 86,553\\
\multicolumn{1}{l|}{Q7} & 0.036 & \multicolumn{1}{r|}{0.079} & 47,287 \\
\multicolumn{1}{l|}{Q8} & 0.025 & \multicolumn{1}{r|}{0.379} & 1,277,729 \\
\multicolumn{1}{l|}{Q9} & 0.828 & \multicolumn{1}{r|}{0.983} & 1,234,922 \\
\multicolumn{1}{l|}{Q10} & 0.899 & \multicolumn{1}{r|}{1.509} & 3,927,763 \\
\multicolumn{1}{l|}{Q11} & 1.375 & \multicolumn{1}{r|}{4.986} & 22,961,108 \\
\multicolumn{1}{l|}{Q12} & 2.434 & \multicolumn{1}{r|}{6.455} & 26,888,871
\end{tabular}
\vspace{-0.5cm}
\end{table}

\begin{figure}
    \begin{minipage}{.235\textwidth}
       \includegraphics[width=0.95\linewidth]{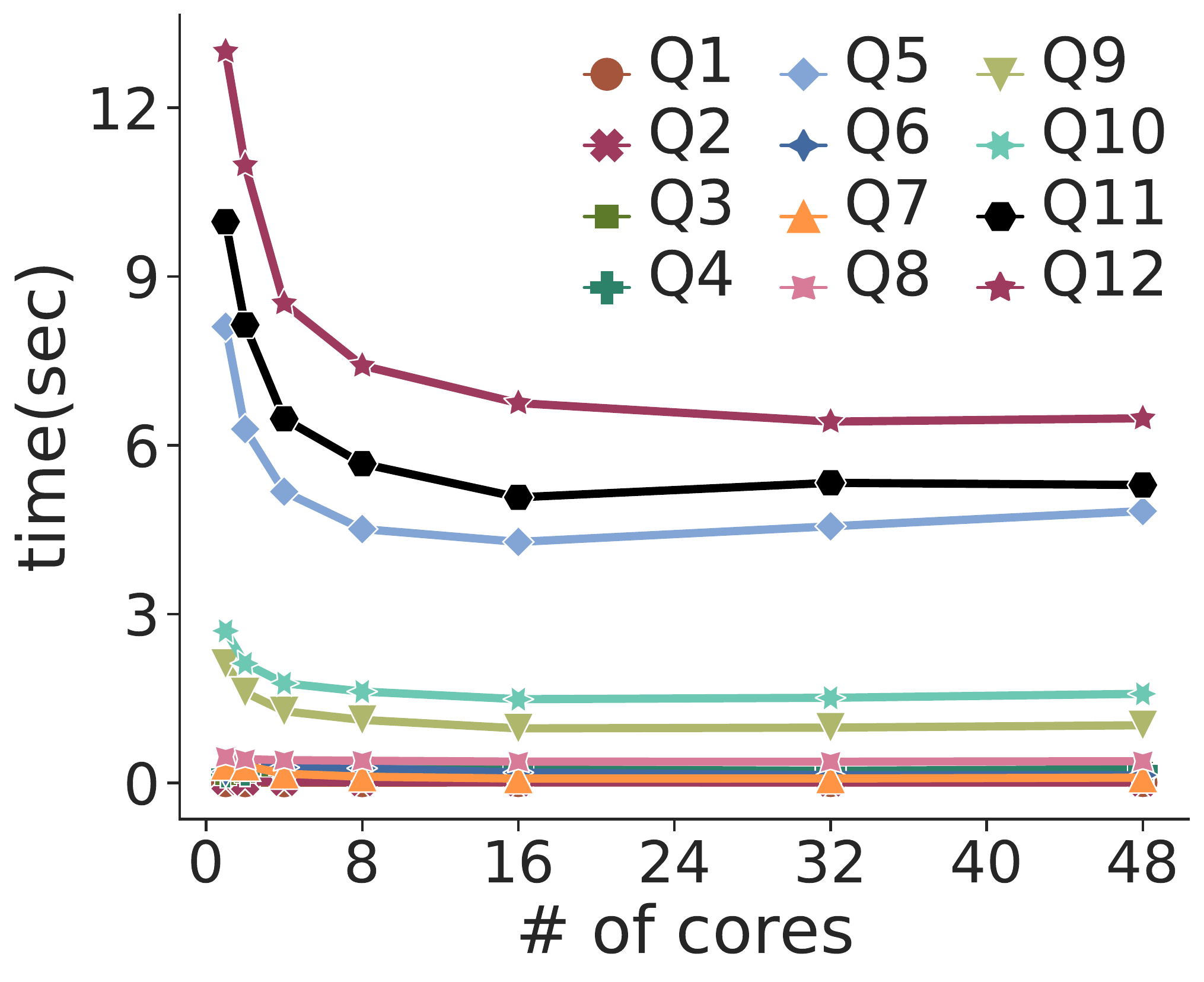}
\vspace{-0.3cm}       
\caption{\rev{Effect of parallelism on G10. }}
\label{fig:exp-cpu}
  \vspace{-0.3cm}
\end{minipage}
\begin{minipage}{.235\textwidth}
 \center
  \vspace{-0.35cm}
  \includegraphics[width=0.95\linewidth]{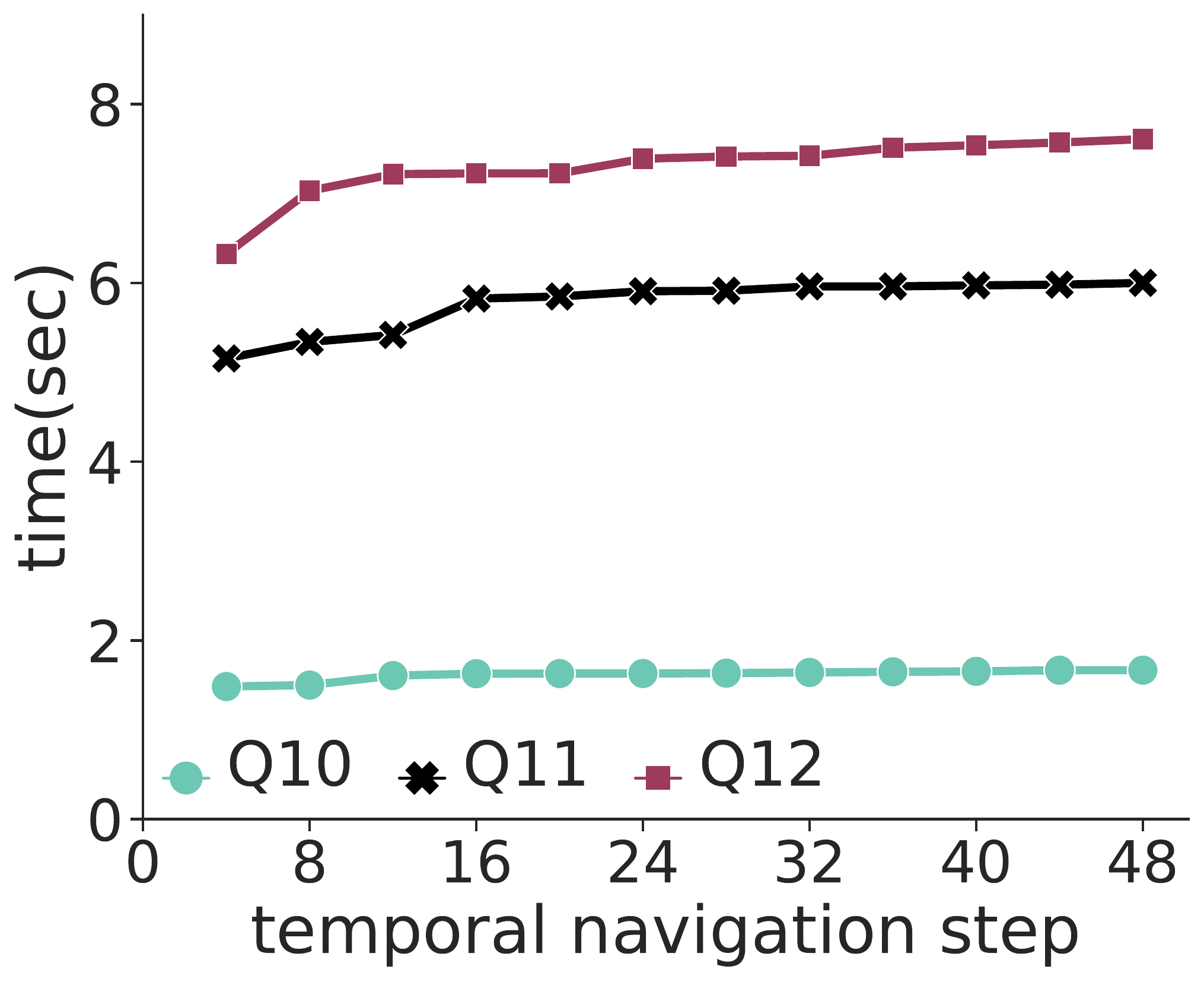}
  \vspace{-0.25cm}
\caption{\rev{Effect of temp. nav. on G10.}} 
\label{fig:exp-dur}
  \vspace{-0.3cm}
\end{minipage}
\end{figure}

All experiments were run as a \rev{multi-threaded} Rust application on a single cluster node with \rev{64 GB} of RAM and an Intel Xeon Platinum 8268 CPU, using the Slurm scheduler~\cite{yoo2003slurm}. \rev{According to our results (Figure~\ref{fig:exp-cpu}), performance for demanding queries was best at 16 CPU cores, and we use this setting in all experiments, unless noted otherwise. }
  Reported execution times are averages of 5 runs. In most cases, the coefficient of variation of the running time was less than 6\% (max 10\%).

\subsection{Experimental datasets}
\label{sec-experiments-data}

We built interval-timestamped \tpgs (per 
Sec.~\ref{sec-itpg-red})
similar to Figure~\ref{fig:covid} using a trajectory dataset generated by Ojagh~\etal~\cite{ojagh2021person} to study COVID-19 contact tracing. The authors tracked 20 individuals on the University of Calgary campus, and used that data to simulate trajectories of individuals visiting campus locations, recording the times when individuals entered and exited those locations.  The synthetic dataset of Ojagh~\etal records time up to a second.  To make this data more realistic, we (i) made temporal resolution coarser, mapping timestamps to 5-min windows, and (ii) associated individuals with locations where they spent at least 2.5 min.

Our goal was to have an  interval-timestamped graph with two types of nodes, \insql{Person} and \insql{Room} (representing classrooms), and two types of edges, \insql{visits} and \mbox{\insql{meets}.}
To achieve this, we represented \rev{100,000} individuals as \mbox{\insql{Person} nodes}, with their periods of validity corresponding to visits of classrooms. 
Next, from among \rev{410} unique locations in the dataset, we selected \rev{100} most frequently visited as nodes of type \insql{Room}, with periods of validity defined by the times of first entrance and last exit. Then, we added a \mbox{\insql{visits}} edge between each person and each room they visit, with an appropriate time interval.  We used information about the remaining \rev{310} locations to add bi-directional \insql{meets} edges between a pair of individuals who were at the same location at the same time.
Finally, we randomly selected 18\% of the \insql{Person} nodes (proportion of the Canadian population aged 65+) as high risk for disease complications, and fixed this property over the lifespan of those nodes.

To study the impact of graph size on performance, we created graphs at different scale factors by randomly selecting a subset of the \mbox{\insql{Person}} nodes of a given size, and keeping only the valid edges. 
To study the impact of query selectivity on performance, we selected between 2\% and 10\% of the \mbox{\insql{Person}} nodes as positive for COVID-19, assigning the time of a positive test uniformly at random from the temporal domain of the graph, and keeping the selected nodes as positive for the remainder of their lifespan.

Table~\ref{tbg:ds-desc} summarized the temporal graphs used in our experiments. The largest graph has \rev{100,000} unique \insql{Person} nodes, 
\rev{100} unique \insql{Room} nodes, and a temporal domain of 48 time points, each representing a 5-minute window.  This corresponds to \rev{340,000} temporal nodes and over \rev{32} million temporal edges.  

\begin{figure}[!t]
    \centering
    \begin{minipage}{.5\textwidth}
       \includegraphics[width=0.9\linewidth]{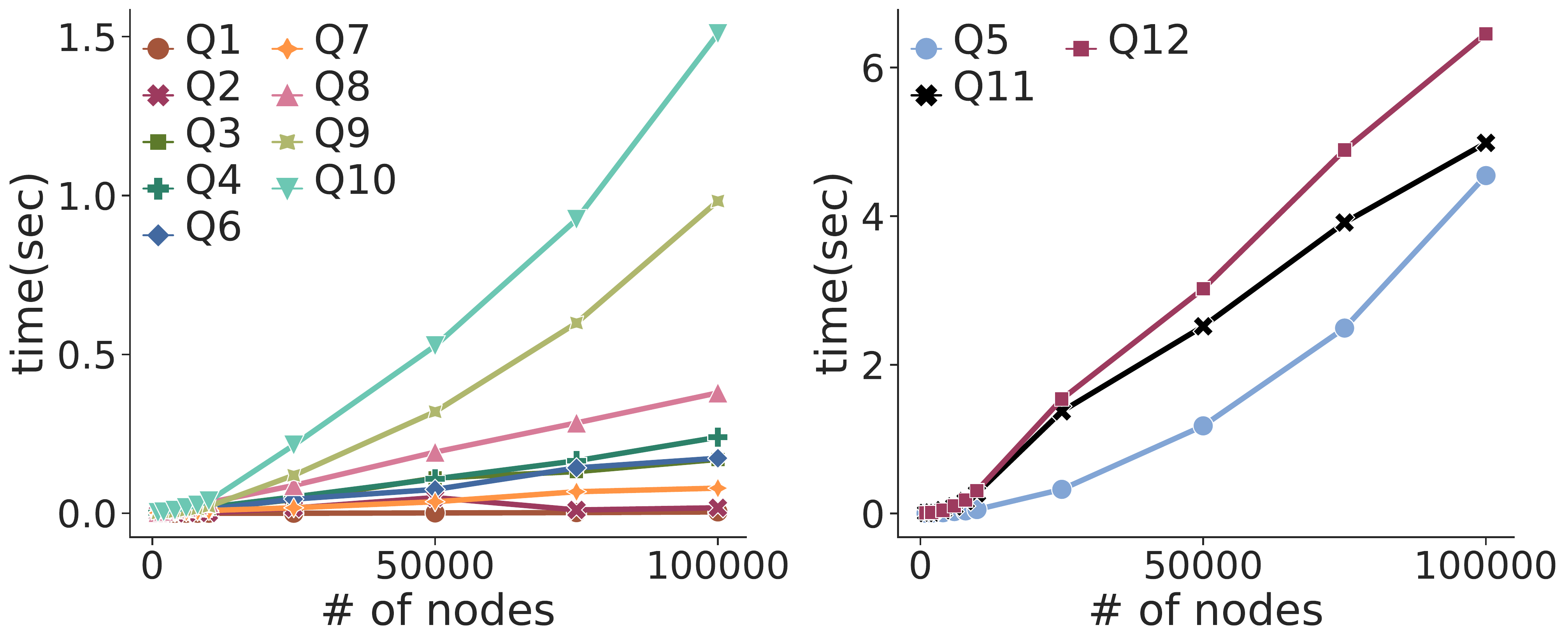}
\vspace{-0.3cm}
\caption{\rev{Effect of graph size on query execution time, on G1-G10.}}
\label{fig:exp-nodes}
    \end{minipage}%
\end{figure}

\subsection{Results}
\label{sec-experiments-res}
For the first experiment, we executed queries \insql{Q1}-\insql{Q12}, discussed in Section~\ref{sec-standard}, over graph \rev{G10} (Table~\ref{tbg:ds-desc}). Table~\ref{fig:exp-q} shows the execution time of each query in seconds, and its output size in the number of tuples in the bindings table.  
\rev{Recall from Section~\ref{sec-system} that Steps 1 and 2 of query evaluation act on the interval representation or TRPG, while Step 3 expands the output of Step 2 into a point-based representation to check any remaining temporal constraints.  Our implementation uses lazy evaluation. Decoupling the execution times of Steps 1 and 2 for the purpose of measurement would degrade performance, and we report these times jointly as ``interval-based time'' in Table~\ref{fig:exp-q}.}
\rev{Queries \insql{Q1}-\insql{Q5} do not use temporal navigation, and so interval-based time and total time coincide and the output can remain temporally coalesced.  In contrast, \insql{Q6}-\insql{Q12} use temporal navigation; they require both interval-based and point-based processing, and the output for these queries is point-based.} 

\begin{figure}[!t]
    \centering
    \begin{minipage}{.5\textwidth}
       \includegraphics[width=0.9\linewidth]{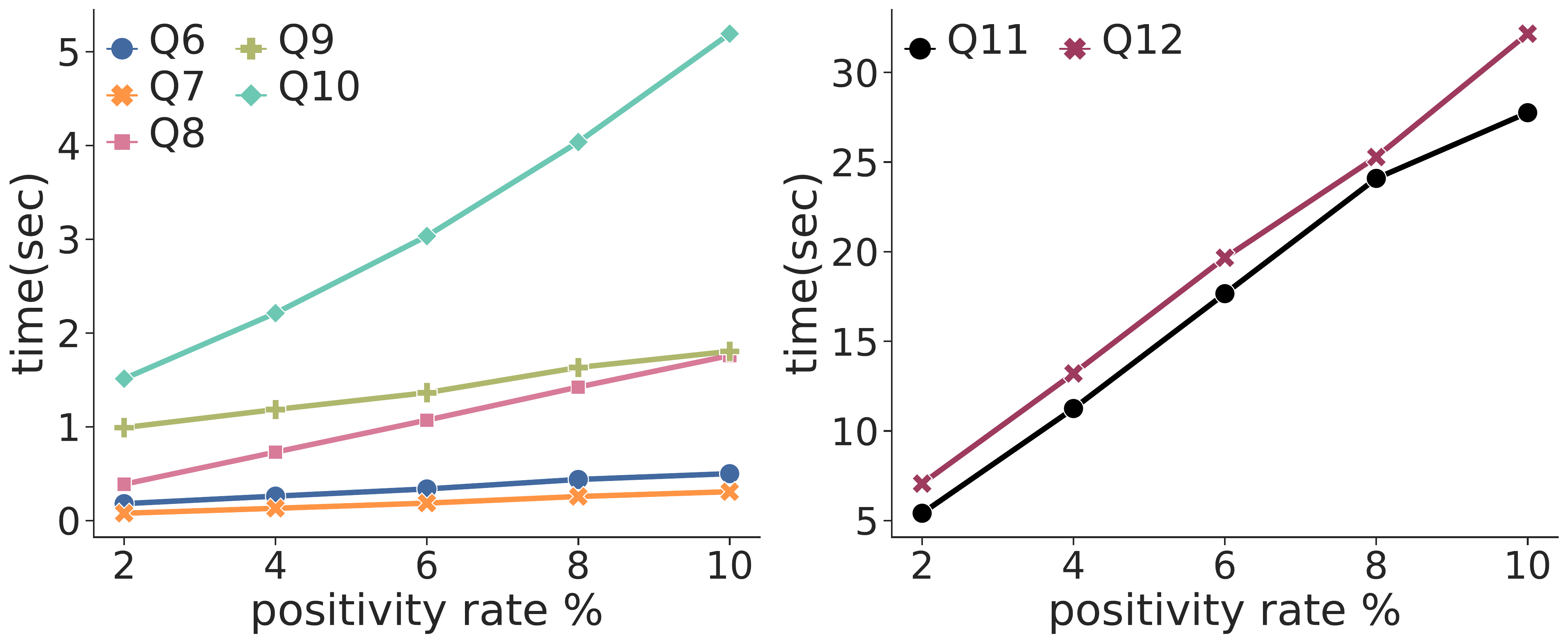}
\vspace{-0.3cm}       
\caption{\rev{Effect of positivity rate on query execution time, on G10. }}
\label{fig:exp-pos}
\vspace{-0.3cm}     
    \end{minipage}%
\end{figure}

We observe that most queries execute in less than \rev{1} sec.  The most challenging queries, \insql{Q11} and \insql{Q12}, both produce over \rev{22} million tuples in the output and take at most \rev{6.5} sec.

In the second experiment, we execute all queries over graphs G1-G10 to study the impact of graph size on query performance. Figure~\ref{fig:exp-nodes} shows this result, with the number of unique \insql{Person} nodes on the $x$-axis, and execution time in 
seconds on the $y$-axis. Observe that the running time increases linearly for all queries except \insql{Q5}, \insql{Q9}, and \mbox{\insql{Q10}} where the time increases approximately quadratically with increasing graph size.  Increase in the running time is nearly perfectly explained by the increase in the size of the output.  For example, increasing  input size by a factor of x10 nodes and x100 edges (G6 to G10) increases output size of \insql{Q11} (resp. \insql{Q12}) by a factor of \rev{18.39} (resp. \rev{19.29}), and it increases the execution time by a factor of \rev{18.89} (resp. \rev{19.29}). 

\rev{In our third experiment, we studied the impact of parallelism on performance. Figure~\ref{fig:exp-cpu} shows the result of this experiment over the largest graph, G10, with the number of CPU cores on the $x$-axis and execution time in seconds on the $y$-axis. (The number of threads is the number of CPUs + 1.)  Observe that the most demanding queries \insql{Q5}, \insql{Q10}, \insql{Q11}, and \mbox{\insql{Q12}} substantially benefit from increased parallelism, with best performance at 16 cores.  For example, \insql{Q12} executes in 6.45 sec on 16 cores, down from 13 sec on 1 core.} 



Queries \insql{Q6}-\insql{Q11} all select \insql{Person} nodes that at some point had a positive COVID-19 test. In our next  experiment, we vary the positivity rate from 2\% to 10\%, thus impacting query selectivity, and study its effect on execution time. Figure~\ref{fig:exp-pos} shows the result of this experiment over the largest graph, \rev{G10}, with positivity rate on the $x$-axis and execution time on the $y$-axis.  We observe a linear relationship between positivity rate and execution time for all queries.

In our final experiment, we consider the effect of temporal navigation on query performance.  We select queries \insql{Q10}, \mbox{\insql{Q11}} and \insql{Q12} because they all contain a temporal navigation operator with a numerical occurrence indicator (\mbox{\insql{PREV[n,m]}} in \insql{Q10} and \insql{NEXT[n,m]} in \insql{Q11} and \insql{Q12}). We set $n=0$, and vary the maximum number of temporal navigation steps $m$ between 4 and 48 in increments of 4. Figure~\ref{fig:exp-dur} shows the result over \rev{G10} with $m$ on the $x$-axis and query execution time  on the $y$-axis. We observe that increasing the number of temporal navigation steps increases the execution time.  This increase is initially linear, but plateaus when $m$ reaches 16, because of the cumulative effect of increasing $m$. 


\section{Conclusions and \rev{Future Work}}
\label{sec-conc}
We considered temporal property graphs (\tpgs) 
and proposed temporal regular path queries (TRPQs) that incorporate time into \tpg navigation.
%
%
Starting with design principles,
we proposed a natural syntactic extension of the MATCH clause of popular query languages, 
formally presented the semantics of TRPQs, and studied the complexity of their evaluation. We also demonstrated that a fragment of the TRPQ language can be implemented efficiently.  We hope that our work on the syntax and semantics,  the positive complexity results, and our implementation and evaluation will pave the way to usable and practical production-level implementations of TRPQs.

\rev{An interesting future direction is to add support for aggregation and grouping. Another natural direction is to incorporate our methods into existing graph processing systems like GraphX~\cite{Gonzalez2012}, Portal~\cite{DBLP:conf/edbt/AghasadeghiMSS20} or Neo4j~\cite{Neo4j_dm}, and to investigate a range of systems questions, including the impact of different object timestamping strategies, temporal coalescing strategies, and indexing methods on performance.}




\balance 

\bibliographystyle{IEEEtran}
\bibliography{references}

\begin{thebibliography}{}

\end{thebibliography}


\begin{thebibliography}{10}
\providecommand{\url}[1]{#1}
\csname url@samestyle\endcsname
\providecommand{\newblock}{\relax}
\providecommand{\bibinfo}[2]{#2}
\providecommand{\BIBentrySTDinterwordspacing}{\spaceskip=0pt\relax}
\providecommand{\BIBentryALTinterwordstretchfactor}{4}
\providecommand{\BIBentryALTinterwordspacing}{\spaceskip=\fontdimen2\font plus
\BIBentryALTinterwordstretchfactor\fontdimen3\font minus
  \fontdimen4\font\relax}
\providecommand{\BIBforeignlanguage}[2]{{%
\expandafter\ifx\csname l@#1\endcsname\relax
\typeout{** WARNING: IEEEtran.bst: No hyphenation pattern has been}%
\typeout{** loaded for the language `#1'. Using the pattern for}%
\typeout{** the default language instead.}%
\else
\language=\csname l@#1\endcsname
\fi
#2}}
\providecommand{\BIBdecl}{\relax}
\BIBdecl

\bibitem{DBLP:journals/csur/AnglesABHRV17}
\BIBentryALTinterwordspacing
R.~Angles, M.~Arenas, P.~Barcel{\'{o}}, A.~Hogan, J.~L. Reutter, and D.~Vrgoc,
  ``Foundations of modern query languages for graph databases,'' \emph{{ACM}
  Comput. Surv.}, vol.~50, no.~5, pp. 68:1--68:40, 2017. [Online]. Available:
  \url{http://doi.acm.org/10.1145/3104031}
\BIBentrySTDinterwordspacing

\bibitem{GCore18}
\BIBentryALTinterwordspacing
R.~Angles, M.~Arenas, P.~Barcelo, P.~Boncz, G.~Fletcher, C.~Gutierrez,
  T.~Lindaaker, M.~Paradies, S.~Plantikow, J.~Sequeda, O.~van Rest, and
  H.~Voigt, ``G-core: A core for future graph query languages,'' in
  \emph{Proceedings of the 2018 International Conference on Management of
  Data}, ser. SIGMOD '18.\hskip 1em plus 0.5em minus 0.4em\relax New York, NY,
  USA: Association for Computing Machinery, 2018, p. 1421–1432. [Online].
  Available: \url{https://doi.org/10.1145/3183713.3190654}
\BIBentrySTDinterwordspacing

\bibitem{DBLP:conf/icwsm/GoetzLMF09}
\BIBentryALTinterwordspacing
M.~Goetz, J.~Leskovec, M.~McGlohon, and C.~Faloutsos, ``Modeling blog
  dynamics,'' in \emph{Proceedings of the Third International Conference on
  Weblogs and Social Media, {ICWSM} 2009, San Jose, California, USA, May 17-20,
  2009}, E.~Adar, M.~Hurst, T.~Finin, N.~S. Glance, N.~Nicolov, and B.~L.
  Tseng, Eds.\hskip 1em plus 0.5em minus 0.4em\relax San Jose, CA: The {AAAI}
  Press, 2009, pp. 26--33. [Online]. Available:
  \url{http://aaai.org/ocs/index.php/ICWSM/09/paper/view/152}
\BIBentrySTDinterwordspacing

\bibitem{DBLP:journals/tweb/LeskovecAH07}
\BIBentryALTinterwordspacing
J.~Leskovec, L.~A. Adamic, and B.~A. Huberman, ``The dynamics of viral
  marketing,'' \emph{ACM Trans. Web}, vol.~1, no.~1, p. 5–es, May 2007.
  [Online]. Available: \url{https://doi.org/10.1145/1232722.1232727}
\BIBentrySTDinterwordspacing

\bibitem{DBLP:conf/kdd/LeskovecBKT08}
\BIBentryALTinterwordspacing
J.~Leskovec, L.~Backstrom, R.~Kumar, and A.~Tomkins, ``Microscopic evolution of
  social networks,'' in \emph{Proceedings of the 14th ACM SIGKDD International
  Conference on Knowledge Discovery and Data Mining}, ser. KDD '08.\hskip 1em
  plus 0.5em minus 0.4em\relax New York, NY, USA: Association for Computing
  Machinery, 2008, p. 462–470. [Online]. Available:
  \url{https://doi.org/10.1145/1401890.1401948}
\BIBentrySTDinterwordspacing

\bibitem{DBLP:conf/icml/SarkarCJ12}
P.~Sarkar, D.~Chakrabarti, and M.~I. Jordan, ``Nonparametric link prediction in
  dynamic networks,'' in \emph{Proceedings of the 29th International Coference
  on International Conference on Machine Learning}, ser. ICML'12.\hskip 1em
  plus 0.5em minus 0.4em\relax Madison, WI, USA: Omnipress, 2012, p.
  1897–1904.

\bibitem{DBLP:journals/tkdd/AsurPU09}
\BIBentryALTinterwordspacing
S.~Asur, S.~Parthasarathy, and D.~Ucar, ``An event-based framework for
  characterizing the evolutionary behavior of interaction graphs,'' \emph{ACM
  Trans. Knowl. Discov. Data}, vol.~3, no.~4, Dec. 2009. [Online]. Available:
  \url{https://doi.org/10.1145/1631162.1631164}
\BIBentrySTDinterwordspacing

\bibitem{DBLP:journals/tcsb/BeyerTLSF10}
\BIBentryALTinterwordspacing
A.~Beyer, P.~Thomason, X.~Li, J.~Scott, and J.~Fisher, ``Mechanistic insights
  into metabolic disturbance during type-2 diabetes and obesity using
  qualitative networks,'' \emph{Transactions on Computational Systems Biology
  XII, Special Issue on Modeling Methodologies}, vol.~12, pp. 146--162, 2010.
  [Online]. Available: \url{http://dx.doi.org/10.1007/978-3-642-11712-1_4}
\BIBentrySTDinterwordspacing

\bibitem{Stuart2003}
J.~M. Stuart, E.~Segal, D.~Koller, and S.~K. Kim, ``A gene-coexpression network
  for global discovery of conserved genetic modules,'' \emph{Science}, vol.
  5643, no. 302, pp. 249--–255, 2003.

\bibitem{Chan2008}
J.~Chan, J.~Bailey, and C.~Leckie, ``{Discovering correlated spatio-temporal
  changes in evolving graphs},'' \emph{Knowledge and Information Systems},
  vol.~16, no.~1, pp. 53--96, 2008.

\bibitem{DBLP:journals/jisa/PapadimitriouDG10}
\BIBentryALTinterwordspacing
P.~Papadimitriou, A.~Dasdan, and H.~Garcia{-}Molina, ``Web graph similarity for
  anomaly detection,'' \emph{J. Internet Services and Applications}, vol.~1,
  no.~1, pp. 19--30, 2010. [Online]. Available:
  \url{http://dx.doi.org/10.1007/s13174-010-0003-x}
\BIBentrySTDinterwordspacing

\bibitem{DBLP:journals/tkde/ByunWK20}
\BIBentryALTinterwordspacing
J.~Byun, S.~Woo, and D.~Kim, ``Chronograph: Enabling temporal graph traversals
  for efficient information diffusion analysis over time,'' \emph{{IEEE} Trans.
  Knowl. Data Eng.}, vol.~32, no.~3, pp. 424--437, 2020. [Online]. Available:
  \url{https://doi.org/10.1109/TKDE.2019.2891565}
\BIBentrySTDinterwordspacing

\bibitem{Debrouvier21}
A.~Debrouvier, E.~Parodi, M.~Perazzo, V.~Soliani, and A.~Vaisman, ``A model and
  query language for temporal graph databases,'' \emph{VLDB Journal}, 2021.

\bibitem{DBLP:conf/sigmod/JohnsonKLS16}
\BIBentryALTinterwordspacing
T.~Johnson, Y.~Kanza, L.~V.~S. Lakshmanan, and V.~Shkapenyuk, ``Nepal: a path
  query language for communication networks,'' in \emph{Proceedings of the 1st
  {ACM} {SIGMOD} Workshop on Network Data Analytics, NDA@SIGMOD 2016, San
  Francisco, California, USA, July 1, 2016}, A.~Arora, S.~Roy, and S.~Mehta,
  Eds.\hskip 1em plus 0.5em minus 0.4em\relax {ACM}, 2016, pp. 6:1--6:8.
  [Online]. Available: \url{https://doi.org/10.1145/2980523.2980530}
\BIBentrySTDinterwordspacing

\bibitem{Labouseur2015}
\BIBentryALTinterwordspacing
A.~G. Labouseur, J.~Birnbaum, P.~W. Olsen, S.~R. Spillane, J.~Vijayan, J.~H.
  Hwang, and W.~S. Han, ``{The G* graph database: efficiently managing large
  distributed dynamic graphs},'' \emph{Distributed and Parallel Databases},
  vol.~33, no.~4, pp. 479--514, 2014. [Online]. Available:
  \url{http://dx.doi.org/10.1007/s10619-014-7140-3}
\BIBentrySTDinterwordspacing

\bibitem{DBLP:conf/dbpl/MoffittS17}
\BIBentryALTinterwordspacing
V.~Z. Moffitt and J.~Stoyanovich, ``Temporal graph algebra,'' in
  \emph{Proceedings of The 16th International Symposium on Database Programming
  Languages}, ser. DBPL '17.\hskip 1em plus 0.5em minus 0.4em\relax New York,
  NY, USA: Association for Computing Machinery, 2017. [Online]. Available:
  \url{https://doi.org/10.1145/3122831.3122838}
\BIBentrySTDinterwordspacing

\bibitem{Bohlen2000}
M.~H. B{\"{o}}hlen, C.~S. Jensen, and R.~T. Snodgrass, ``{Temporal Statement
  Modifiers},'' \emph{ACM Transactions on Database Systems}, vol.~25, no.~4,
  pp. 407--456, 2000.

\bibitem{Montanari2009}
\BIBentryALTinterwordspacing
A.~Montanari and J.~Chomicki, \emph{Time Domain}.\hskip 1em plus 0.5em minus
  0.4em\relax Boston, MA: Springer US, 2009, pp. 3103--3107. [Online].
  Available: \url{http://dx.doi.org/10.1007/978-0-387-39940-9_427}
\BIBentrySTDinterwordspacing

\bibitem{Liu:2009:EDS:1804422}
L.~Liu and M.~T. Zsu, \emph{Encyclopedia of Database Systems}, 1st~ed.\hskip
  1em plus 0.5em minus 0.4em\relax Boston, MA: Springer Publishing Company,
  Incorporated, 2009.

\bibitem{Clifford1985}
\BIBentryALTinterwordspacing
J.~Clifford and A.~U. Tansel, ``On an algebra for historical relational
  databases: Two views,'' in \emph{Proceedings of the 1985 ACM SIGMOD
  International Conference on Management of Data}, ser. SIGMOD '85.\hskip 1em
  plus 0.5em minus 0.4em\relax New York, NY, USA: Association for Computing
  Machinery, 1985, p. 247–265. [Online]. Available:
  \url{https://doi.org/10.1145/318898.318922}
\BIBentrySTDinterwordspacing

\bibitem{JENSEN1994513}
\BIBentryALTinterwordspacing
C.~S. Jensen, M.~D. Soo, and R.~T. Snodgrass, ``Unifying temporal data models
  via a conceptual model,'' \emph{Information Systems}, vol.~19, no.~7, pp. 513
  -- 547, 1994. [Online]. Available:
  \url{http://www.sciencedirect.com/science/article/pii/0306437994900132}
\BIBentrySTDinterwordspacing

\bibitem{Snodgrass:1985:TTD:318898.318921}
\BIBentryALTinterwordspacing
R.~Snodgrass and I.~Ahn, ``A taxonomy of time databases,'' in \emph{Proceedings
  of the 1985 ACM SIGMOD International Conference on Management of Data}, ser.
  SIGMOD '85.\hskip 1em plus 0.5em minus 0.4em\relax New York, NY, USA: ACM,
  1985, pp. 236--246. [Online]. Available:
  \url{http://doi.acm.org/10.1145/318898.318921}
\BIBentrySTDinterwordspacing

\bibitem{Bohlen1998}
\BIBentryALTinterwordspacing
M.~H. B{\"{o}}hlen, R.~Busatto, and C.~S. Jensen, ``{Point Versus
  Interval-based Temporal Data Models},'' in \emph{Proceedings of the 14th IEEE
  ICDE}.\hskip 1em plus 0.5em minus 0.4em\relax Orlando, FL: IEEE, 1998, pp.
  192--200. [Online]. Available:
  \url{http://people.cs.aau.dk/{~}csj/Thesis/pdf/chapter7.pdf}
\BIBentrySTDinterwordspacing

\bibitem{Dignos2012}
\BIBentryALTinterwordspacing
A.~Dign\"{o}s, M.~H. B\"{o}hlen, and J.~Gamper, ``Temporal alignment,'' in
  \emph{Proceedings of the 2012 ACM SIGMOD International Conference on
  Management of Data}, ser. SIGMOD '12.\hskip 1em plus 0.5em minus 0.4em\relax
  New York, NY, USA: Association for Computing Machinery, 2012, p. 433–444.
  [Online]. Available: \url{https://doi.org/10.1145/2213836.2213886}
\BIBentrySTDinterwordspacing

\bibitem{Salzberg1999}
\BIBentryALTinterwordspacing
B.~Salzberg and V.~J. Tsotras, ``{Comparison of access methods for
  time-evolving data},'' \emph{ACM Computing Surveys}, vol.~31, no.~2, pp.
  158--221, jun 1999. [Online]. Available:
  \url{http://portal.acm.org/citation.cfm?doid=319806.319816}
\BIBentrySTDinterwordspacing

\bibitem{DBLP:journals/sigmod/KulkarniM12}
\BIBentryALTinterwordspacing
K.~G. Kulkarni and J.~Michels, ``Temporal features in {SQL:} 2011,''
  \emph{{SIGMOD} Record}, vol.~41, no.~3, pp. 34--43, 2012. [Online].
  Available: \url{http://doi.acm.org/10.1145/2380776.2380786}
\BIBentrySTDinterwordspacing

\bibitem{Borgwardt2006}
\BIBentryALTinterwordspacing
K.~M. Borgwardt, H.-P. Kriegel, and P.~Wackersreuther, ``Pattern mining in
  frequent dynamic subgraphs,'' in \emph{Proceedings of the Sixth International
  Conference on Data Mining}, ser. ICDM '06.\hskip 1em plus 0.5em minus
  0.4em\relax USA: IEEE Computer Society, 2006, p. 818–822. [Online].
  Available: \url{https://doi.org/10.1109/ICDM.2006.124}
\BIBentrySTDinterwordspacing

\bibitem{Fard2012}
\BIBentryALTinterwordspacing
A.~Fard, A.~Abdolrashidi, L.~Ramaswamy, and J.~Miller, ``{Towards Efficient
  Query Processing on Massive Time-Evolving Graphs},'' in \emph{Proceedings of
  the 8th IEEE International Conference on Collaborative Computing: Networking,
  Applications and Worksharing}, 2012, pp. 567--574. [Online]. Available:
  \url{http://eudl.eu/doi/10.4108/icst.collaboratecom.2012.250532}
\BIBentrySTDinterwordspacing

\bibitem{Ferreira2004}
A.~Ferreira, ``{Building a reference combinatorial model for MANETs},''
  \emph{IEEE Network}, vol.~18, no.~5, pp. 24--29, 2004.

\bibitem{Kan2009}
A.~Kan, J.~Chan, J.~Bailey, and C.~Leckie, ``A query based approach for mining
  evolving graphs,'' in \emph{Proceedings of the Eighth Australasian Data
  Mining Conference - Volume 101}, ser. AusDM '09.\hskip 1em plus 0.5em minus
  0.4em\relax AUS: Australian Computer Society, Inc., 2009, p. 139–150.

\bibitem{Khurana2013}
\BIBentryALTinterwordspacing
U.~Khurana and A.~Deshpande, ``Efficient snapshot retrieval over historical
  graph data,'' in \emph{Proceedings of the 2013 IEEE International Conference
  on Data Engineering (ICDE 2013)}, ser. ICDE '13.\hskip 1em plus 0.5em minus
  0.4em\relax USA: IEEE Computer Society, 2013, p. 997–1008. [Online].
  Available: \url{https://doi.org/10.1109/ICDE.2013.6544892}
\BIBentrySTDinterwordspacing

\bibitem{Khurana2016}
\BIBentryALTinterwordspacing
------, ``{Storing and Analyzing Historical Graph Data at Scale},'' in
  \emph{Proceedings of the 19th International Conference on Extending Database
  Technology, EDBT'16}, Bordeaux, France, 2016, pp. 65--76. [Online].
  Available: \url{http://arxiv.org/abs/1509.08960}
\BIBentrySTDinterwordspacing

\bibitem{Lahiri2008}
M.~Lahiri and T.~Berger-Wolf, ``{Mining Periodic Behavior in Dynamic Social
  Networks},'' in \emph{2008 Eighth IEEE International Conference on Data
  Mining}, 2008, pp. 373--382.

\bibitem{Ren2011}
C.~Ren, E.~Lo, B.~Kao, X.~Zhu, and R.~Cheng, ``{On Querying Historical Evolving
  Graph Sequences},'' \emph{Proceedings of the VLDB Endowment}, vol.~4, no.~11,
  pp. 726--737, 2011.

\bibitem{Semertzidis2015}
\BIBentryALTinterwordspacing
K.~Semertzidis, E.~Pitoura, and K.~Lillis, ``Timereach: Historical reachability
  queries on evolving graphs,'' in \emph{Proceedings of the 18th International
  Conference on Extending Database Technology, {EDBT} 2015, Brussels, Belgium,
  March 23-27, 2015}, G.~Alonso, F.~Geerts, L.~Popa, P.~Barcel{\'{o}},
  J.~Teubner, M.~Ugarte, J.~V. den Bussche, and J.~Paredaens, Eds.\hskip 1em
  plus 0.5em minus 0.4em\relax Brussels, Belgium: OpenProceedings.org, 2015,
  pp. 121--132. [Online]. Available:
  \url{https://doi.org/10.5441/002/edbt.2015.12}
\BIBentrySTDinterwordspacing

\bibitem{Sricharan2014}
\BIBentryALTinterwordspacing
K.~Sricharan and K.~Das, ``{Localizing anomalous changes in time-evolving
  graphs},'' in \emph{Proceedings of the 2014 ACM SIGMOD international
  conference on Management of data}, Snowbird, Utah USA, 2014, pp. 1347--1358.
  [Online]. Available:
  \url{http://dl.acm.org/citation.cfm?doid=2588555.2612184}
\BIBentrySTDinterwordspacing

\bibitem{DBLP:conf/cikm/YangQZGL07}
L.~Yang, L.~Qi, Y.~Zhao, B.~Gao, and T.~Liu, ``Link analysis using time series
  of web graphs,'' in \emph{Proceedings of the Sixteenth {ACM} Conference on
  Information and Knowledge Management, {CIKM} 2007, Lisbon, Portugal, November
  6-10, 2007}, M.~J. Silva, A.~H.~F. Laender, R.~A. Baeza{-}Yates, D.~L.
  McGuinness, B.~Olstad, {\O}.~H. Olsen, and A.~O. Falc{\~{a}}o, Eds.\hskip 1em
  plus 0.5em minus 0.4em\relax {ACM}, 2007, pp. 1011--1014.

\bibitem{DBLP:journals/pvldb/WuCHKLX14}
\BIBentryALTinterwordspacing
H.~Wu, J.~Cheng, S.~Huang, Y.~Ke, Y.~Lu, and Y.~Xu, ``Path problems in temporal
  graphs,'' \emph{Proc. {VLDB} Endow.}, vol.~7, no.~9, pp. 721--732, 2014.
  [Online]. Available: \url{http://www.vldb.org/pvldb/vol7/p721-wu.pdf}
\BIBentrySTDinterwordspacing

\bibitem{DBLP:journals/tkde/WuCKHHW16}
\BIBentryALTinterwordspacing
H.~Wu, J.~Cheng, Y.~Ke, S.~Huang, Y.~Huang, and H.~Wu, ``Efficient algorithms
  for temporal path computation,'' \emph{{IEEE} Trans. Knowl. Data Eng.},
  vol.~28, no.~11, pp. 2927--2942, 2016. [Online]. Available:
  \url{https://doi.org/10.1109/TKDE.2016.2594065}
\BIBentrySTDinterwordspacing

\bibitem{DBLP:conf/icde/WuHCLK16}
\BIBentryALTinterwordspacing
H.~Wu, Y.~Huang, J.~Cheng, J.~Li, and Y.~Ke, ``Reachability and time-based path
  queries in temporal graphs,'' in \emph{32nd {IEEE} International Conference
  on Data Engineering, {ICDE} 2016, Helsinki, Finland, May 16-20, 2016}.\hskip
  1em plus 0.5em minus 0.4em\relax {IEEE} Computer Society, 2016, pp. 145--156.
  [Online]. Available: \url{https://doi.org/10.1109/ICDE.2016.7498236}
\BIBentrySTDinterwordspacing

\bibitem{Cypher18}
\BIBentryALTinterwordspacing
N.~Francis, A.~Green, P.~Guagliardo, L.~Libkin, T.~Lindaaker, V.~Marsault,
  S.~Plantikow, M.~Rydberg, P.~Selmer, and A.~Taylor, ``Cypher: An evolving
  query language for property graphs,'' in \emph{Proceedings of the 2018
  International Conference on Management of Data}, ser. SIGMOD '18.\hskip 1em
  plus 0.5em minus 0.4em\relax New York, NY, USA: Association for Computing
  Machinery, 2018, p. 1433–1445. [Online]. Available:
  \url{https://doi.org/10.1145/3183713.3190657}
\BIBentrySTDinterwordspacing

\bibitem{DBLP:conf/sigmod/DignosBG12}
\BIBentryALTinterwordspacing
A.~Dign{\"{o}}s, M.~H. B{\"{o}}hlen, and J.~Gamper, ``Temporal alignment,'' in
  \emph{Proceedings of the {ACM} {SIGMOD} International Conference on
  Management of Data, {SIGMOD} 2012, Scottsdale, AZ, USA, May 20-24, 2012},
  K.~S. Candan, Y.~Chen, R.~T. Snodgrass, L.~Gravano, and A.~Fuxman, Eds.\hskip
  1em plus 0.5em minus 0.4em\relax {ACM}, 2012, pp. 433--444. [Online].
  Available: \url{https://doi.org/10.1145/2213836.2213886}
\BIBentrySTDinterwordspacing

\bibitem{DBLP:conf/vldb/BohlenSS96}
M.~H. B{\"{o}}hlen, R.~T. Snodgrass, and M.~D. Soo, ``Coalescing in temporal
  databases,'' in \emph{VLDB'96, Proceedings of 22th International Conference
  on Very Large Data Bases, September 3-6, 1996, Mumbai (Bombay), India}, 1996,
  pp. 180--191.

\bibitem{PGQL16}
\BIBentryALTinterwordspacing
O.~van Rest, S.~Hong, J.~Kim, X.~Meng, and H.~Chafi, ``Pgql: A property graph
  query language,'' in \emph{Proceedings of the Fourth International Workshop
  on Graph Data Management Experiences and Systems}, ser. GRADES '16.\hskip 1em
  plus 0.5em minus 0.4em\relax New York, NY, USA: Association for Computing
  Machinery, 2016. [Online]. Available:
  \url{https://doi.org/10.1145/2960414.2960421}
\BIBentrySTDinterwordspacing

\bibitem{gql}
{Association of ISO Graph Query Language Proponents}, ``{GQL} standard,'' 2020,
  \url{https://www.gqlstandards.org}.

\bibitem{LMV16}
L.~Libkin, W.~Martens, and D.~Vrgoc, ``Querying graphs with data,'' \emph{J.
  {ACM}}, vol.~63, no.~2, pp. 14:1--14:53, 2016.

\bibitem{V82}
\BIBentryALTinterwordspacing
M.~Y. Vardi, ``The complexity of relational query languages (extended
  abstract),'' in \emph{Proceedings of the Fourteenth Annual ACM Symposium on
  Theory of Computing}, ser. STOC '82.\hskip 1em plus 0.5em minus 0.4em\relax
  New York, NY, USA: Association for Computing Machinery, 1982, p. 137–146.
  [Online]. Available: \url{https://doi.org/10.1145/800070.802186}
\BIBentrySTDinterwordspacing

\bibitem{AV97}
\BIBentryALTinterwordspacing
S.~Abiteboul and V.~Vianu, ``Regular path queries with constraints,'' in
  \emph{Proceedings of the Sixteenth ACM SIGACT-SIGMOD-SIGART Symposium on
  Principles of Database Systems}, ser. PODS '97.\hskip 1em plus 0.5em minus
  0.4em\relax New York, NY, USA: Association for Computing Machinery, 1997, p.
  122–133. [Online]. Available: \url{https://doi.org/10.1145/263661.263676}
\BIBentrySTDinterwordspacing

\bibitem{calvanese2002rewriting}
D.~Calvanese, G.~De~Giacomo, M.~Lenzerini, and M.~Y. Vardi, ``Rewriting of
  regular expressions and regular path queries,'' \emph{Journal of Computer and
  System Sciences}, vol.~64, no.~3, pp. 443--465, 2002.

\bibitem{B13}
\BIBentryALTinterwordspacing
P.~Barcel\'{o}~Baeza, ``Querying graph databases,'' in \emph{Proceedings of the
  32nd ACM SIGMOD-SIGACT-SIGAI Symposium on Principles of Database Systems},
  ser. PODS '13.\hskip 1em plus 0.5em minus 0.4em\relax New York, NY, USA:
  Association for Computing Machinery, 2013, p. 175–188. [Online]. Available:
  \url{https://doi.org/10.1145/2463664.2465216}
\BIBentrySTDinterwordspacing

\bibitem{CD99}
J.~Clark and S.~DeRose, ``{XML} path language {(XPath)} version 1.0,'' W3C
  Recommendation 16 November 1999.

\bibitem{M05}
M.~Marx, ``Conditional {XPath},'' \emph{{ACM} Trans. Database Syst.}, vol.~30,
  no.~4, pp. 929--959, 2005.

\bibitem{GKP05}
G.~Gottlob, C.~Koch, and R.~Pichler, ``Efficient algorithms for processing
  {XPath} queries,'' \emph{{ACM} Trans. Database Syst.}, vol.~30, no.~2, pp.
  444--491, 2005.

\bibitem{RDS17}
J.~Robie, M.~Dyck, and J.~Spiegel, ``{XML} path language {(XPath)} 3.1,'' W3C
  Recommendation 21 March 2017.

\bibitem{rust-itertools}
\BIBentryALTinterwordspacing
Rust-Itertools, ``rust-itertools/itertools.'' [Online]. Available:
  \url{https://github.com/rust-itertools/itertools}
\BIBentrySTDinterwordspacing

\bibitem{rayon-rs}
\BIBentryALTinterwordspacing
Rayon-Rs, ``Rayon-rs/rayon: Rayon: A data parallelism library for rust.''
  [Online]. Available: \url{https://github.com/rayon-rs/rayon/}
\BIBentrySTDinterwordspacing

\bibitem{DBLP:journals/cacm/ZahariaXWDADMRV16}
\BIBentryALTinterwordspacing
M.~Zaharia, R.~S. Xin, P.~Wendell, T.~Das, M.~Armbrust, A.~Dave, X.~Meng,
  J.~Rosen, S.~Venkataraman, M.~J. Franklin, A.~Ghodsi, J.~Gonzalez,
  S.~Shenker, and I.~Stoica, ``Apache spark: a unified engine for big data
  processing,'' \emph{Commun. {ACM}}, vol.~59, no.~11, pp. 56--65, 2016.
  [Online]. Available: \url{http://doi.acm.org/10.1145/2934664}
\BIBentrySTDinterwordspacing

\bibitem{carbone2015apache}
P.~Carbone, A.~Katsifodimos, S.~Ewen, V.~Markl, S.~Haridi, and K.~Tzoumas,
  ``Apache flink: Stream and batch processing in a single engine,''
  \emph{Bulletin of the IEEE Computer Society Technical Committee on Data
  Engineering}, vol.~36, no.~4, 2015.

\bibitem{murray2013naiad}
D.~G. Murray, F.~McSherry, R.~Isaacs, M.~Isard, P.~Barham, and M.~Abadi,
  ``Naiad: a timely dataflow system,'' in \emph{Proceedings of the
  Twenty-Fourth ACM Symposium on Operating Systems Principles}, 2013, pp.
  439--455.

\bibitem{mcsherry2013differential}
F.~McSherry, D.~G. Murray, R.~Isaacs, and M.~Isard, ``Differential dataflow,''
  in \emph{CIDR}, 2013.

\bibitem{yoo2003slurm}
A.~B. Yoo, M.~A. Jette, and M.~Grondona, ``Slurm: Simple linux utility for
  resource management,'' in \emph{Workshop on job scheduling strategies for
  parallel processing}.\hskip 1em plus 0.5em minus 0.4em\relax Springer, 2003,
  pp. 44--60.

\bibitem{ojagh2021person}
S.~Ojagh, S.~Saeedi, and S.~H. Liang, ``A person-to-person and person-to-place
  covid-19 contact tracing system based on ogc indoorgml,'' \emph{ISPRS
  International Journal of Geo-Information}, vol.~10, no.~1, p.~2, 2021.

\bibitem{Gonzalez2012}
\BIBentryALTinterwordspacing
J.~Gonzalez, Y.~Low, and H.~Gu, ``{Powergraph: Distributed graph-parallel
  computation on natural graphs},'' in \emph{OSDI'12 Proceedings of the 10th
  USENIX conference on Operating Systems Design and Implementation}, 2012, pp.
  17--30. [Online]. Available:
  \url{https://www.usenix.org/system/files/conference/osdi12/osdi12-final-167.pdf}
\BIBentrySTDinterwordspacing

\bibitem{DBLP:conf/edbt/AghasadeghiMSS20}
\BIBentryALTinterwordspacing
A.~Aghasadeghi, V.~Z. Moffitt, S.~Schelter, and J.~Stoyanovich, ``Zooming out
  on an evolving graph,'' in \emph{Proceedings of the 23rd International
  Conference on Extending Database Technology, {EDBT} 2020, Copenhagen,
  Denmark, March 30 - April 02, 2020}, A.~Bonifati, Y.~Zhou, M.~A.~V. Salles,
  A.~B{\"{o}}hm, D.~Olteanu, G.~H.~L. Fletcher, A.~Khan, and B.~Yang,
  Eds.\hskip 1em plus 0.5em minus 0.4em\relax OpenProceedings.org, 2020, pp.
  25--36. [Online]. Available: \url{https://doi.org/10.5441/002/edbt.2020.04}
\BIBentrySTDinterwordspacing

\bibitem{Neo4j_dm}
``Neo4j: What is a graph database?''
  \url{https://neo4j.com/developer/graph-database/#property-graph}, [Online;
  accessed 18-July-2017].

\bibitem{Allen1983}
J.~F. Allen, ``{Maintaining Knowledge about Temporal Intervals},''
  \emph{Communications of the ACM}, vol.~26, no.~11, pp. 832--843, 1983.

\bibitem{DBLP:reference/db/Bohlen09}
\BIBentryALTinterwordspacing
M.~B{\"o}hlen, \emph{Temporal Coalescing}.\hskip 1em plus 0.5em minus
  0.4em\relax Boston, MA: Springer US, 2009, pp. 2932--2936. [Online].
  Available: \url{http://dx.doi.org/10.1007/978-0-387-39940-9_388}
\BIBentrySTDinterwordspacing

\bibitem{DBLP:books/daglib/0023376}
T.~H. Cormen, C.~E. Leiserson, R.~L. Rivest, and C.~Stein, \emph{Introduction
  to Algorithms, 3rd Edition}.\hskip 1em plus 0.5em minus 0.4em\relax {MIT}
  Press, 2009.

\bibitem{Berman2001}
P.~Berman, M.~Karpinski, L.~L. Larmore, W.~Plandowski, and W.~Rytter, ``On the
  complexity of pattern matching for highly compressed two-dimensional texts,''
  in \emph{Proceedings of the 8th Annual Symposium on Combinatorial Pattern
  Matching}, ser. CPM '97.\hskip 1em plus 0.5em minus 0.4em\relax Berlin,
  Heidelberg: Springer-Verlag, 1997, p. 40–51.

\bibitem{StockmeyerMeyer1973}
\BIBentryALTinterwordspacing
L.~J. Stockmeyer and A.~R. Meyer, ``Word problems requiring exponential
  time(preliminary report),'' in \emph{Proceedings of the Fifth Annual ACM
  Symposium on Theory of Computing}, ser. STOC '73.\hskip 1em plus 0.5em minus
  0.4em\relax New York, NY, USA: Association for Computing Machinery, 1973, p.
  1–9. [Online]. Available: \url{https://doi.org/10.1145/800125.804029}
\BIBentrySTDinterwordspacing

\bibitem{SubsetSum1990}
M.~R. Garey and D.~S. Johnson, \emph{Computers and Intractability; A Guide to
  the Theory of NP-Completeness}.\hskip 1em plus 0.5em minus 0.4em\relax USA:
  W. H. Freeman \& Co., 1990.

\end{thebibliography}

\newpage

\onecolumn

\appendices

\section{Formal definition of Interval-timestamped temporal property graphs}
\label{sec-itpg-comp-def}

An interval of $\mathbb{N}$ is a term of the form $[a, b]$ with $a, b \in \N$ and $a \leq b$, which is used  as a concise representation of the set of natural numbers $\{i \in \N \ \vert \ a \leq i \leq b \}$ (that is, to specify this interval, we just need to mention its starting point $a$ and its ending point $b$). Using Allen's interval algebra~\cite{Allen1983}, given two intervals $[a_1,b_1]$ and $[a_2, b_2]$, we say that $[a_1, b_1]$ occurs during $[a_2, b_2]$ if $a_2 \leq a_1$ and $b_1 \leq b_2$, $[a_1, b_1]$ meets $[a_2, b_2]$ if $b_1 + 1 = b_2$, and $[a_1, b_1]$ is before $[a_2, b_2]$ if $b_1 + 1 < a_2$.   

A finite family $\F$ of intervals is said to be 
\emph{coalesced}~\cite{DBLP:reference/db/Bohlen09}
if $\F = \{ [a_1, b_1], \ldots, [a_n, b_n] \}$ and $[a_j, b_j]$ is before $[a_{j+1}, b_{j+1}]$ for every $j \in \{1, \ldots, n-1\}$.   For example, $\F_1 = \{[1,4], [6,8]\}$ is \coal, while  $\F_2 = \{[1,2], [3,4], [6,8] \}$ is not, because $[1,2]$ meets $[3,4]$.
The set of all finite \coal families of intervals is denoted by $\fnc$.  Observe that $\emptyset \in \fnc$.
Moreover, given $\F_1, \F_2 \in \fnc$, family $\F_1$ is said to be contained in family $\F_2$, denoted by $\F_1 \sqsubseteq \F_2$, if for every $[a_1, b_1] \in \F_1$, there exists $[a_2, b_2] \in \F_2$ such that $[a_1, b_1]$ occurs during $[a_2, b_2]$. 
Finally, given an interval $\Omega$, we use $\fnc(\Omega)$ to denote the set of all families $\F \in \fnc$ such that for every $[a, b] \in \F$, it holds that $[a, b]$ occurs during $\Omega$.

Given an interval $[a, b]$ and $v \in \Val$, the pair $(v,[a, b])$ is a {\em valued} interval.  
A finite family $\F$ of valued intervals is said to be \coal if $\F = \{(v_1,[a_1, b_1])$, $\ldots$, $(v_n,[a_n, b_n])\}$ and for every $j \in \{1, \ldots, n-1\}$, either $[a_j, b_j]$ is before $[a_{j+1}, b_{j+1}]$, or $[a_j, b_j]$ meets $[a_{j+1}, b_{j+1}]$ and \mbox{$v_j \neq v_{j+1}$}.  For example, $\F_1 = \{(v, [1,2])$, $(v, [5,8]) \}$ and $\F_2 = \{(v, [1,2])$, $(w, [3,4]) \}$ are both \coal (assuming that $v \neq w$). On the other hand, $\F_3 = \{(v, [1,2])$, $(v, [3,4]) \}$ is not \coal because $[1,2]$ meets $[3,4]$ and these intervals have the same value in $\F_3$.
Moreover, the set of all finite \coal families of valued intervals is denoted by $\vfnc$.
Finally, given an interval $\Omega$, we use $\vfnc(\Omega)$ to denote the set of all families $\F \in \vfnc$ such that for every $(v,[a, b]) \in \F$, it holds that $[a, b]$ occurs during~$\Omega$. 

With these 
ingredients, we can introduce the notion of interval-timestamped temporal property graph.
\begin{definition}
\label{def:tg1}
An interval-timestamped 
temporal property graph (ITPG) is a tuple $I = ( \Omega, N, E, \rho, \lambda, \xi, \sigma)$, where $N$, $E$, $\rho$ and $\lambda$ are defined exactly as for the case of TPGs (see Definition \ref{def:tg}). Moreover,
\begin{itemize}
\item $\Omega$ is an interval of $\mathbb{N}$;
\item $\xi: (N \cup E) \to \fnc(\Omega)$ is a function that maps a node or an edge to a finite \coal family of intervals occurring during~$\Omega$;

\item $\sigma:  (N \cup E) \times \Prop \to \vfnc(\Omega)$ is a function that maps a node or an edge, and a property name to a finite \coal family of valued intervals occurring during $\Omega$.
\end{itemize}
In addition, $I$ satisfies the following conditions:
\begin{itemize}
\item If $\rho(e) = (n_1,n_2)$, then  $\xi(e) \sqsubseteq \xi(n_1)$
and
$\xi(e) \sqsubseteq \xi(n_2)$.

\item 
There exists a finite set of pairs $(o,p) \in (N \cup E) \times \Prop$ such that $\sigma(o,p) \neq \emptyset$. Moreover, if $\sigma(o,p) = \{(v_1, [a_1, b_1])$, $\ldots$, $(v_n, [a_n, b_n])\}$, then $\{[a_1, b_1], \ldots, [a_n, b_n] \} \sqsubseteq \xi(o)$.
\end{itemize}
\end{definition}
In the definition of an \ctpg, given a node or edge $o$, function $\xi$  indicates 
the time intervals where $o$ exists,
and function $\sigma$ indicates the values of a property $p$ for 
$o$. More precisely, if $\sigma(o,p) = \{(v_1, [a_1, b_1]), \ldots, (v_n, [a_n, b_n])\}$, then the value of property $p$ for $o$ is $v_j$ in every time point in the interval $[a_j, b_j]$ ($1 \leq j \leq n$). 
Moreover,
observe that two additional conditions are imposed on $I$, which enforce that an \ctpg conceptually corresponds to a finite sequence of valid conventional property graphs. In particular, as was the case for \tpgs, an edge can only exist at a time when both of the nodes it connects exist, and a property can only take on a value at a time when the corresponding node or edge exists. 
%
%
For instance, assume that $I = ( \Omega, N, E, \rho, \lambda, \xi, \sigma)$ is an $\ctpg$ corresponding to our running example in Figure~\ref{fig:covid}. Then, we have that $\Omega = [1,11]$, $\xi(n_2) = \{[1,9]\}$, $\xi(n_3) = \{[1,7]\}$ and $\xi(e_2) = \{[1,2]\}$, so that $\xi(e_2) \sqsubseteq \xi(n_2)$ and $\xi(e_2) \sqsubseteq \xi(n_3)$. Moreover, for the property risk, we have that $\sigma(n_2,\text{risk}) = \{(\text{low},[1,4]), (\text{high}, [5,9])\}$.

We conclude this section by observing that there is a one-to-one correspondence between TPGs and ITPGs. On the one hand, each TPG can be transformed in polynomial-time into a ITPG, by putting in the same interval consecutive time points with the same values. On the other hand, each ITPG can be transformed in exponential-time into a TPG, by replacing each interval by the set of time points represented by it.

\section{Formal definition of some fragments of \Lthree}

{\bf Removing numerical occurrence indicators.}
We start by considering a restriction of our query language in which numerical occurrence indicators are not allowed. Formally, this means that grammar \eqref{grammar:path-expr} is replaced by:
\begin{eqnarray}
        \pt & ::= & \test \, \mid \, \axis \, \mid \, (\pt/\pt) \, \mid \, (\pt+\pt)
    \label{grammar:ax-path-expr-no-noi}
\end{eqnarray}
The resulting language is called \Ltwo.

\bigskip

{\bf Removing path conditions.}
Consider a second restriction of our language in which there are no path conditions. Formally, this means that instead of grammar \eqref{grammar:test}, we use: 
\begin{eqnarray}
        \test & ::= &  \vd \  \mid \  \ed \  \mid \  \ell \  \mid \  \propt{p}{v} \  \mid \  < \mt \  \mid \  \ex \  \mid \
        (\test \vee \test) \  \mid \  (\test \wedge \test) \ \mid (\neg \test)
    \label{grammar:test-no-pc}
\end{eqnarray}
The resulting language is called \Lonep. 

\bigskip

{\bf Allowing numerical occurrence indicators only in the axes.}
Consider a grammar for tests as in  \eqref{grammar:test-no-pc}, where path conditions are not allowed, and a grammar for path expressions where numerical occurrence indicators are only used in the axes:
\begin{eqnarray*}
    \pt & ::= &  \test \ \mid \ \axis \ \mid \ \axis[ n, m ] \ \mid \ \axis[n,\_] \ \mid\
    (\pt/\pt) \ \mid \ (\pt+\pt)
    \label{grammar:ax-path-expr-anoi}
\end{eqnarray*}
The resulting language is called \Lonepp, where ANOI refers to numerical occurrence indicators used only in the axes.

\section{Proof of Theorem \ref{th:main-summary}}

\subsection{{\em \tupleeval({\rm TPG}, \Lthree)} can be solved in polynomial time}
\label{sec-app-tpg-pol}
 
In this section, we show that $\Lthree$ can be evaluated in polynomial time,
considering a computational model where accessing the distinct elements of a $\tpg$
takes time $O(1)$. More precisely, for a $\tpg$ $\G = (\Omega,N,E,\rho,\lambda,\xi,\sigma)$, it is assumed that the following operations can be performed in time $O(1)$: given $e \in E$ and $n_1,n_2 \in N$, check whether $\rho(e) = (n_1,n_2)$; given $e \in E$, compute $\src(e)$ and $\tgt(G)$; given $o \in (N \cup E)$ and $\ell \in \Lab$, check whether $\lambda(o) = \ell$; given $o \in (N \cup E)$ and $t \in \Omega$, check whether $\xi(o,t) = \true$; and given $o \in (N \cup E)$, $p \in \Prop$ and $v \in \Val$, check whether $\sigma(o,p,t) = v$.
Moreover, we use notation $\|\pt\|$ for the length of $\Lthree$-expression $\pt$ as an input string over an appropriate alphabet. 
Then, it is possible to prove the following.

\begin{theorem}\label{theo-pol-time-tpg}
There exists an algorithm that, given a temporal property graph $\G$
and a $\Lthree$-expression $\pt$, computes 
$\sem{\pt}$ in time $\tilde{O}(\|\pt\|^2 \cdot |\Omega|^2 \cdot (|N| + |E|)^2)$.
\end{theorem}
%
%

\begin{figure}
    \begin{center}
        \begin{tikzpicture} 
\node[] (n) {$/$};
\node[below left = 4mm and 18mm of n] (n0) {$\wedge$};
\node[below right = 4mm and 18mm of n] (n1) {$/$};
\node[below left = 4mm and 4mm of n0] (n00) {$\vd$};
\node[below right = 4mm and 4mm of n0] (n01) {$\wedge$};
\node[below left = 4mm and 4mm of n1] (n10) {$[5,9]$};
\node[below right = 4mm and 6mm of n1] (n11) {$\wedge$};
\node[below left = 5mm and 1.5mm of n01] (n010) {$\text{Person}$};
\node[below right = 5mm and 1.5mm of n01] (n011) {$\propt{\text{\rm test}}{\text{\rm pos}}$};
\node[below left = 4mm and 2mm of n11] (n110) {$\vd$};
\node[below right = 4mm and 3mm of n11] (n111) {$\exists$};
\node[below = 3.2mm of n10] (n000) {$\prv$};

\path[arrout] (n) edge node[above] {} (n0);
\path[arrout] (n) edge node[above] {} (n1);
\path[arrout] (n0) edge node[above] {} (n00);
\path[arrout] (n0) edge node[above] {} (n01);
\path[arrout] (n1) edge node[above] {} (n10);
\path[arrout] (n1) edge node[above] {} (n11);
\path[arrout] (n01) edge node[above] {} (n010);
\path[arrout] (n01) edge node[above] {} (n011);
\path[arrout] (n11) edge node[above] {} (n110);
\path[arrout] (n11) edge node[above] {} (n111);
\path[arrout] (n10) edge node[above] {} (n000);
        \end{tikzpicture}
    \end{center}
    \caption{A parsing tree of \Lthree-expression $\pt = (\vd \wedge \text{\rm Person} \wedge \propt{\text{\rm test}}{\text{\rm pos}})
/\prv[5,9]/(\vd \wedge \exists)$. \label{fig-par-tree}}
\end{figure}
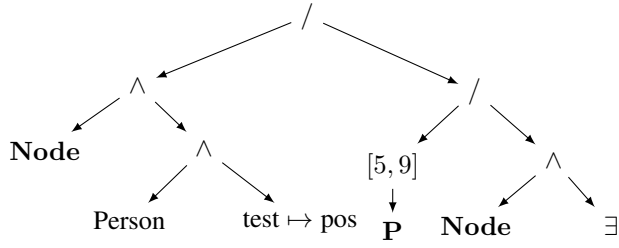

In the rest of this section, we describe the polynomial-time algorithm in the statement of Theorem \ref{theo-pol-time-tpg}. Let 
$G = (\Omega,N,E,\rho,\lambda,\xi,\sigma)$ be a $\tpg$ and $\pt$ be an expression in \Lthree. 
Moreover, assume that $M = | \Omega | \cdot (| N | + | E |)$ is the number of distinct (existing or non-existing) temporal objects in $\G$.
The algorithm constructs a parsing tree of $\pt$, where each node is associated with an operator in \Lthree, and then, by using  a bottom-up approach, computes, for each node $u$, the set of tuples $(o,t,o',t')$ that satisfy the operator labeling $u$. For example, a parsing tree of \Lthree-expression $(\vd \wedge \text{\rm Person} \wedge \propt{\text{\rm test}}{\text{\rm pos}})
/\prv[5,9]/(\vd \wedge \exists)$ is shown in Figure \ref{fig-par-tree}. The algorithm starts by evaluating the leaves: $\sem{\vd}$, $\sem{\text{Person}}$, 
$\sem{\propt{\text{\rm test}}{\text{\rm pos}}}$, $\sem{\prv}$ and $\sem{\exists}$, according to the semantics defined in Section \ref{sec-fss-NavL}, and then it combines the resulting tables by using the operators $\wedge$, $[5,9]$ and~$/$ in the order specified by the parsing tree. For instance, in the right-hand side of the tree, once $\sem{\prv}$, $\sem{\vd}$ and $\sem{\exists}$ have been computed, the algorithm continues by constructing $\sem{\vd \wedge \exists}$, followed by
$\sem{\prv[5,9]}$ and then by $\sem{\prv[5,9]/(\vd \wedge \exists)}$. Notice that at any given moment, the result of at most $\| \pt \|$ nodes is stored in the form of a table, each one with as many pairs of temporal objects as there are available, \textit{i.e.}, with at most $M^2$ tuples.




We now explain in detail the different components of the algorithm, paying particular attention to the expressions of the form $\pt_1[n,m]$ and $\pt_2[n,\_]$, as they are the most expensive to evaluate in \Lthree.
Initially, for each leaf of the parsing tree of $\pt$, we have to process either a test $\vd$, $\ed$, $\ell$, $\propt{p}{v}$, $\exists$ or $< \mt$, or a navigation operator $\nxt$, $\prv$, $\fw$, or $\bw$. Basic tests can be evaluated in time $O(M)$ 
just by considering each tuple of the form $(o,t,o,t) \in \Pt(\G)$ and checking in $O(1)$  whether $(o,t)$ satisfies the test.
As for navigation operators, each one of them can also be evaluated in time $O(M)$ (recall that the existence of nodes or edges at a given time point is not required in the language). For example, $\sem{\prv}$ can be constructed just by considering all objects $o \in V \cup E$ and then generating tuples $(o,t,o,t-1)$ such that $t \in \Omega$ and $t-1 \in \Omega$, while $\sem{\fw}$ can be constructed by considering all edges $e \in E$, and then generating tuples $(\src(e),t,e,t)$ and $(e,t,\tgt(e),t)$ for each $t \in \Omega$.

For each internal node $u$ of the parsing tree of $\pt$, we must consider one of the following two cases. Assume first that the label of $u$ is either $\wedge$, $\vee$, $\neg$ or $?$, so that $u$ represents a more complex test expression $(\test_1 \wedge \test_2)$, $(\test_1 \vee \test_2)$, $(\neg \test_1)$ or $(?\pt_1)$. If $u$ represents test $(\test_1 \wedge \test_2)$, then the algorithm has already computed $T_1 = \sem{\test_1}$ and $T_2 = \sem{\test_2}$. Hence, to construct $\sem{\test_1 \wedge \test_2}$, the algorithm  needs to compute the intersection of $T_1$ and $T_2$, which can be done in time $\tilde{O}(M^2)$ by sorting both tables (each of size at most $M^2$) and iterating with two pointers, one on each table, to see which elements occur in both. Recall that the notation $\tilde{O}(M^2)$ ignores the logarithmic factors, which in this case appear when sorting tables $T_1$ and $T_2$. The case where $u$ represents either $(\test_1 \vee \test_2)$ or $(\neg \test_1)$ can be treated in a similar way.
Finally, if $u$ represents test $(?\pt_1)$, for each tuple $(o,t,o',t') \in \sem{\pt_1}$, we need to include the tuple $(o,t,o,t)$ in the table for $u$, as $(o,t) \models (?\pt_1)$ if and only if $(o,t,o',t') \in \sem{\pt_1}$ for some temporal object $(o',t')$ in $\G$. This can be done in time $O(M^2)$ as $\sem{\pt_1}$ contains at most $M^2$ tuples.

\begin{sloppypar}
Assume now that the label of $u$ is either $/$, $+$, $[n,m]$ or $[n,\_]$, so that $u$ represents a more complex path expression $(\pt_1 / \pt_2)$, $(\pt_1 + \pt_2)$, $\pt_1[m, n]$ or $\pt_1[m, \_]$. If $u$ represents expression $(\pt_1 / \pt_2)$, then the algorithm has already computed $T_1 = \sem{\pt_1}$ and $T_2 = \sem{\pt_2}$. Hence, to construct $\sem{\pt_1 / \pt_2}$, the algorithm just need to sort $T_1$ by the third and fourth columns (the second pair of temporal objects), sort $T_2$ by the first and second column (the first pair of temporal objects), and then join $T_1$ with $T_2$ by looking at matching temporal objects on those columns. The overall time for this construction is $\tilde{O}(M^2)$, as it corresponds to a sort-merge join on two tables with at most $M^2$ tuples. 
If $u$ represents the expression $(\pt_1 + \pt_2)$, the algorithm  computes the union of $T_1$ and $T_2$.
If $u$ represents the expression $\pt_1[n, m]$, then the algorithm proceeds as follows, assuming that $T_1 = \sem{\pt_1}$ has already been computed. Given that $\pt_1[n,m]$ is equivalent to the expression $\pt_1[n, n] / \pt_1[0, m -n]$, the procedure first runs Algorithm~\ref{algo:NOInn} $\textsc{ComputeRepetition}(G, T_1,n)$ to compute $\sem{\pt_1[n,n]}$ in a similar way to the exponentiation by squaring algorithm~\cite{DBLP:books/daglib/0023376}.
If $n = m$, then we are ready in time $\tilde{O}(\| \pt_1 \| \cdot M^2)$, since the most expensive operation is the sort-merge join, which is carried out in time $\tilde{O}(M^2)$ and at most $O(\log(n))$ times, that is, $O(\| \pt_1 \|)$ times. Otherwise, Algorithm~\ref{algo:NOI0n} $\textsc{ComputeIntervalRepetition}(G, T_1, T_2, m - n)$ is called to compute $\sem{\pt_1[n,n] / \pt_1[0, m -n]}$, where $T_2 = \sem{\pt_1[n,n]}$ is the result of invoking $\textsc{ComputeRepetition}(G, T_1,n)$.
Here, $O(\log(m-n))$ sort-merge joins have to be carried out, which is again $O(\| \pt_1 \|)$, so this takes a total time of $\tilde{O}(\| \pt_1 \| \cdot M^2)$.
\end{sloppypar}
\setcounter{algocf}{0}
\renewcommand{\thealgocf}{\arabic{algocf}}

\begin{algorithm}[t!]
    \SetAlgoLined
    \SetKwInOut{Input}{Input}
    \SetKwInOut{Output}{Output}
    \Input{A $\tpg$ $\G$, a table $T_1$ such that $T_1 = \sem{\pt_1}$ for some \Lthree-expression $\pt_1$, and $n \geq 0$}
    \Output{$\sem{\pt_1[n,n]}$}
    \uIf{$n = 0$}{
        \Return $\{(o,t,o,t) \in \Pt(G)\}$
    }
    \uElseIf{$n = 1$}{
        \Return $T_1$
    }
    $n' \gets \lfloor n/2 \rfloor$ \\
    $T_2 \gets \textsc{ComputeRepetition}(G, T_1, n')$ \\
    Compute $T_3 = \{(o_1,t_1,o_2,t_2) \mid \exists (o,t): (o_1,t_1,o,t) \in T_2$ and $(o,t,o_2,t_2) \in T_2\}$ by doing a sort-merge join\\
    \uIf{$n$ is even}{
        \Return $T_3$
    }
    Compute $T_4 = \{(o_1,t_1,o_2,t_2) \mid \exists (o,t): (o_1,t_1,o,t) \in T_3$ and $(o,t,o_2,t_2) \in T_1\}$ by doing a sort-merge join\\
    \Return $T_4$
    \caption{$\textsc{ComputeRepetition}(G, T_1, n)$}
    \label{algo:NOInn}
\end{algorithm}

\begin{algorithm}[t!]
    \SetAlgoLined
    \SetKwInOut{Input}{Input}
    \SetKwInOut{Output}{Output}
    \Input{
        A $\tpg$ $\G$, a table $T_i$ such that $T_i = \sem{\pt_i}$ for some \Lthree-expression $\pt_i$ for $i = 1,2$,
        and $n >0$
    }
    \Output{$\sem{\pt_1/\pt_2[0, n]}$}
    \uIf{$n = 1$}{
        \Return $T_1 \cup T_2$
    }
    $n' \gets \lfloor n/2 \rfloor$ \\
    $T_3 \gets  \textsc{ComputeIntervalRepetition}(G, 
    T_1, T_2, n')$ \\
    Compute $T_4 = \{(o_1,t_1,o_2,t_2) \mid \exists (o,t): (o_1,t_1,o,t) \in T_3$ and $(o,t,o_2,t_2) \in T_3\}$ by doing a sort-merge join\\
    $T_5 \gets T_3 \cup T_4$ \\
    \uIf{$n$ is even}{
         \Return $T_5$
    }
    Compute $T_6 = \{(o_1,t_1,o_2,t_2) \mid \exists (o,t): (o_1,t_1,o,t) \in T_5$ and $(o,t,o_2,t_2) \in T_2\}$ by doing a sort-merge join\\
    \Return $T_5 \cup T_6$
    \caption{$\textsc{ComputeIntervalRepetition}(G, T_1, T_2, n)$}
    \label{algo:NOI0n}
\end{algorithm}
\begin{sloppypar}
    Finally, if $u$ represents the expression $\pt_1[n, \_  ]$, then the computation process is similar to the previous one, assuming that $T_1 = \sem{\pt_1}$ has already been computed. As before, we act as if we have to compute the table for another, but equivalent, expression, $\pt_1[n, n] / \pt_1[0, M^2]$, which is done by first computing $T_2 = \textsc{ComputeRepetition}(G, T_1,n)$, and then invoking 
    $\textsc{ComputeIntervalRepetition}(G, T_1, T_2, M^2)$. This takes time $\tilde{O}(\| \pt_1 \| \cdot M^2)$ as the sort-merge  join is  carried  out  in  time $\tilde{O}(M^2)$, and $O(\log(n) + \log(M^2))$ such joins need to be computed, that is $\tilde{O}(\|\pt_1\|)$ such joins. Notice that $\sem{\pt_1[0,\_]} = \sem{\pt_1[0, M^2]}$ because: (a) $\sem{\pt_1[0, k]} \subseteq \sem{\pt_1[0, k+1]}$ for every $k \geq 0$; (b) $\big|\sem{\pt_1[0,k]}\big| \leq M^2$ for every $k \geq 0$; and (c) if $\sem{\pt_1[0, k]} = \sem{\pt_1[0, k+1]}$, then $\sem{\pt_1[0,k]} = \sem{\pt_1[0,k']}$ for every $k' > k$.
\end{sloppypar}
In summary, the table associated to each node of the parsing tree of $\pt$ can be computed in time $\tilde{O}(\|\pt\| \cdot M^2)$. Given that there are at most $O(\|\pt\|)$ such nodes, the total computation time is $\tilde{O}(\|\pt\|^2 \cdot M^2)$, that is, $\tilde{O}(\|\pt\|^2 \cdot |\Omega|^2 \cdot (|N| + |E|)^2)$. This concludes the proof of Theorem \ref{theo-pol-time-tpg}.





\subsection{{\em \tupleeval({\rm ITPG}, \Ltwo)} can be solved in polynomial time}

First of all notice that basic tests such as $\vd$, $\ed$, $\ell$, $\propt{p}{v}$, $<\mt$ and $\ex$ can be checked in $O(1)$, so in absence of numerical occurrence indicators, checking a test is equivalent to checking the satisfaction of a Boolean formula on a given valuation, which can be done efficiently in time $O(n)$, where $n$ is the size of the formula. We will assume in the following then that we have a linear-time function $\textsc{CheckTestNoPC}(\C, (o, t), \test)$ that takes as input an \ctpg $\C$, a temporal object $(o, t)$ in $\C$ and a test expression $\test$ in $\Ltwo$, and returns $\true$ if $(o, t) \models \test$ in $\C$, and $\false$ otherwise.

To show that \tupleeval(\ctpg, \Ltwo) is in \ptime, we present a polynomial-time procedure in Algorithm \ref{algo:TupleEvalSolveOnlyPC}, called $\textsc{TupleEvalSolveOnlyPC}$, that, given an \ctpg $\C$, a tuple representing a pair of temporal objects $(o_1, t_1, o_2, t_2)$ and an expression $r$ in $\Ltwo$, checks whether $(o_1, t_1, o_2, t_2) \in \semp{r}{\C}$. In what follows, we show that
Algorithm \ref{algo:TupleEvalSolveOnlyPC} works in polynomial time.

\LinesNumbered

\begin{algorithm}
    \SetAlgoLined
    \SetKwInOut{Input}{Input}
    \SetKwInOut{Output}{Output}
    \SetKwInOut{Initialization}{Initialization}
    \Input{\ An \ctpg $\C = \left( \Omega, N, E, \rho, \lambda, \xi, \sigma \right)$, an expression $r$ in $\Ltwo$ and a pair of temporal objects $\left( o_1, t_1, o_2, t_2 \right) \in \Pt(\C)$}
    \Output{\ $\true$ if $\left( o_1, t_1, o_2, t_2 \right) \in \semp{r}{\C}$, and $\false$ otherwise}
    \Initialization{
        \ A global variable $H$ initially empty, storing a hash table with the currently computed values for $\textsc{TupleEvalSolveOnlyPC}$. \\
    }
    \uIf{$\left( o_1, t_1, o_2, t_2, r \right)$ is a key in hash table $H$}{
        \Return $H[\left( o_1, t_1, o_2, t_2, r \right)]$
    }
    \uIf{$r = \nxt$}{
        $A \gets \left(
            o_1 = o_2 \ \textbf{and} \ t_2 = t_1 + 1
        \right)$
    }
    \uElseIf{$r = \prv$}{
        $A \gets \left(
            o_1 = o_2 \ \textbf{and} \ t_2 = t_1 - 1
        \right)$
    }
    \uElseIf{$r = \fw$}{
        $A \gets \left(
            t_1 = t_2 \ \textbf{and} \ \left(
                \left( o_1 \in E \ \textbf{and} \ o_2 = \tgt\left( o_1 \right) \right)
                \ \textbf{or} \
                \left( o_2 \in E \ \textbf{and} \ o_1 = \src\left( o_2 \right) \right)
            \right)
        \right)$
    }
    \uElseIf{$r = \bw$}{
        $A \gets \left(
            t_1 = t_2 \ \textbf{and} \ \left(
                (o_1 \in E \ \textbf{and} \ o_2 = \src(o_1))
                \ \textbf{or} \
                (o_2 \in E \ \textbf{and} \ o_1 = \tgt(o_2))
            \right)
        \right)$
    }
    \uElseIf{$r$ is a test}{
        $A \gets \left( 
            ( o_1, t_1) = ( o_2, t_2)
            \ \textbf{and} \
            \textsc{CheckTestNoPC}( \C, ( o_1, t_1), r)
        \right)$
    }
    \uElseIf{$r = (r_1 + r_2)$}{
        $A \gets \textsc{TupleEvalSolveOnlyPC}(
            \C, ( o_1, t_1, o_2, t_2), r_1
        )$ \\
        \uIf {$\textbf{not} A$}{
            $A \gets \textsc{TupleEvalSolveOnlyPC}(
                \C, ( o_1, t_1, o_2, t_2), r_2
            )$
        }
    }
    \uElseIf{$r = (r_1 \ / \ r_2)$}{
        $A \gets \false$ \\
        $l_1 \gets \| r_1 \|$ \\
        $l_2 \gets \| r_2 \|$ \\
        \ForEach{$(o, t) \in (N \cup E) \times \{ t \in \Omega \ \mid \ \left( |t - t_1| \leq l_1 \ \wedge \ |t - t_2| \leq l_2 \right) \}$}{
            \uIf{\textsc{TupleEvalSolveOnlyPC}$(\C, (o_1, t_1, o, t), r_1)$}{
                \uIf{\textsc{TupleEvalSolveOnlyPC}$(\C, (o, t, o_2, t_2), r_2)$} {
                    $A \gets \true$ \\
                    \textbf{break}
                }
            }
        }
    }
    Store the value $A$ for $\textsc{TupleEvalSolveOnlyPC}(\C, (o_1, t_1, o_2, t_2), r)$ in the hash table $H$ with key $(o_1, t_1, o_2, t_2, r)$ \\
    \Return $A$
    \caption{$\textsc{TupleEvalSolveOnlyPC}(\C, (o_1, t_1, o_2, t_2), r)$}
    \label{algo:TupleEvalSolveOnlyPC}
\end{algorithm}

Notice first that we do not directly return the result. Instead, we first look at a hashing table that stores previously stored results, and only if this was not previously computed, we compute the result for the input, and store it in the table before returning the value. We employ this to avoid an exponential number of calls when recursively calling the algorithm.
This is possible since, in absence of numerical occurrence indicators, navigation is done at most one step at a time. Thus, if $(o_2,t_2)$ is a temporal object reached from $(o_1, t_1)$ using an expression $r$, then $|t_1 - t_2|$ is at most 
the number of 
symbols $\nxt$ and $\prv$ occurring in $r$. Hence, if $\|r\|$ is the length of expression $r$, then there are at most $O(\| r \| \cdot | N \cup E |)$ temporal objects that we will need to consider for this call, which means, at most $O(\| r \|^2 \cdot | N \cup E |^2)$ tuples representing pairs of temporal objects. Hence, given that there are at most $\| r \|$ sub-expressions of $r$ that can be reached
in the tree decomposition of $r$, we need to store at most $O(\| r \|^3 \cdot | N \cup E |^2)$ different results for $\textsc{TupleEvalSolveOnlyPC}$.

The rest of the algorithm is quite straightforward, and it considers the case when the 
result has not been precomputed. If $r$ is a temporal navigation operator, then by the definitions of $\nxt$ and $\prv$, a single temporal object can be reached, $(o_1, t_1+1)$ or $(o_1, t_1-1)$, respectively, so we check that $(o_2, t_2)$ is equal to that respective temporal object. This can be easily done in $O(1)$.
If $r$ is a spatial navigation operator, then we have to look at the mapping from edges to source and destination nodes, $\rho$, to determine the set of objects that can be reached. If $o$ is an edge, and we move forward, we look for its destination node, if we move backward, for its source node. If $o$ is a node, and we move forward, we are looking for the edges that have $o$ as their source, and if we move backward, then we look for those who have $o$ as their destination. The whole process can be done in time $O(\| \rho \|)$, which is $O(\| \C \|)$.
When $r$ is testing a condition, we just call the previously mentioned algorithm $\textsc{CheckTestNoPC}$ to check whether $(o, t) \models r$ in $\C$. This last base case can be done in time $O(\|r\|)$.

When $r$ is of the form $(r_1 + r_2)$, where $r_1$ and $r_2$ are expressions in $\Ltwo$, we have that $(o_1, t_1, o_2, t_2) \in \semp{(r_1 + r_2)}{\C}$ if and only if $(o_1, t_1, o_2, t_2) \in \semp{r_1}{\C}$ or $(o_1, t_1, o_2, t_2)  \in \semp{r_2}{\C}$, so if suffices that the call to $\textsc{TupleEvalSolveOnlyPC}$ with any of inputs $r_1$ or $r_2$ returns $\true$, and this can be easily checked by the algorithm.
The last case is when $r$ is of the form $(r_1 \ / \ r_2)$ where $r_1$ and $r_2$ are expressions in $\Ltwo$, where we have that $(o_1, t_1, o_2, t_2) \in \semp{r}{\C}$ if there exists a temporal object $(o, t)$ such that $(o_1, t_1, o, t) \in \semp{r_1}{\C}$ and $(o, t, o_2, t_2) \in \semp{r_2}{\C}$. We know that in absence of numerical occurrence indicators, $t$ will be at distance at most $\|r_1\|$ from $t_1$ and at most $\|r_2\|$ from $t_2$, since one can move only as many times as there are $\nxt$ and $\prv$ symbols in the respective formulas, so this gives us a polynomial-size set from which we can extract candidates to satisfy this condition.

Finally, notice that at every call to $\textsc{TupleEvalSolveOnlyPC}(\C, (o_1, t_1, o_2, t_2), r)$, we either already have computed the value for the key $(o_1, t_1, o_2, t_2, r)$, in which case we can give an answer immediately, or we are computing a new value to store in the hash table $H$, which has size bounded by $O(\| r \|^3 \cdot | N \cup E |^2)$. In any case, since the most expensive step performs at most $O(| N \cup E | \cdot \| r \|)$ recursive calls, we will be getting an answer in time $O(\| r \|^4 \cdot | N \cup E |^3)$, which is polynomial in the size of the input.

\subsection{{\em \tupleeval({\rm ITPG}, \Lonep)} is \sigmatwop-hard} 

Consider the following decision problem called Generalized Subset Sum (\gsss) which is known to be \sigmatwop-complete \cite{Berman2001}:

\begin{center}
	\framebox{
	    \begin{tabular}{p{1.6cm} p{11cm}}
			\textbf{Problem:} & $\gsss$ \\
		    \textbf{Input:} & 
		    Natural numbers vectors $u$ and $w$ of dimensions $\dim(u)$ and $\dim(w)$, respectively, 
		    and a positive integer $S \in \mathbb{N}$ \\
		    \textbf{Output:} & $\true$ if there exists $x \in \{0, 1\}^{\dim(u)}$ such that, for all $y \in \{0, 1\}^{\dim(w)}$, it holds that $x\cdot u + y \cdot w \neq S$, and $\false$ otherwise.
		\end{tabular}
	}
\end{center}
In this problem, $a \cdot b$ represents the inner product between vectors $a$ and $b$.
Given vectors $u = (u_1, \dots, u_n) \in \mathbb{N}^n$ and $w = (w_1, \dots, w_m) \in \mathbb{N}^m$, and the integer $S \in \mathbb{N}$, the goal is to provide a polynomial-time algorithm that returns 
a \ctpg $\C$, a tuple $(o_1, t_1, o_2, t_2)$, and an expression $r$ in $\Lonep$ such that $(o_1, t_1, o_2, t_2) \in \semp{r}{\C}$ if and only if $\exists x \in \{0, 1\}^n \ \forall y \in \{ 0, 1 \}^m \ x \cdot u + y \cdot w \neq S$.

Let $M = 2 \cdot \left(\sum_{i=1}^n u_i + \sum_{j=1}^m w_j \right)$, which can be easily computed in polynomial time from $u$ and $w$. Then $\C$ will be the \ctpg $\C = \left( \Omega, N, E, \rho, \lambda, \xi, \sigma \right)$ where $\Omega = [0, 2 \cdot M]$, $N = \lbrace v \rbrace$, $E = \varnothing$, $\rho$ is an empty function, $\lambda(v) = l$, $\xi(v)=\left\lbrace [0, 2 \cdot M] \right\rbrace$ and $\sigma$ is an empty function. In other words, $\C$ is a \ctpg consisting of only one node existing from time $0$ to time $2 \cdot M$, with no edges or properties. The tuple $(o_1, t_1, o_2, t_2)$ in our reduction will be given by $(v, M, v, 2 \cdot M)$.
As for the expression $r$, it will be defined recursively as follows. First define an expression for each component $u_i$ of $u$ that will represent whether $u_i \cdot x_i$ will be chosen to be $u_i$ or $0$:
\begin{eqnarray*} 
r_{u_i} &=& \nxt [u_i, u_i] [0, 1].
\end{eqnarray*}
The idea is that the time $t$ of the temporal object that is being reached will store the sum given by $x \cdot u + y \cdot w$, plus $M$ to avoid having negative numbers on the time dimension when testing that the result 
is different from $S$ (this will be explained in more detail later). 
Define then an expression for $u$, representing the sum accumulated by the $\exists x \in \{0, 1\}^n$ part of the problem:
\begin{eqnarray*}
r_u &=& r_{u_1} \ / \ \dots \ / \ r_{u_n}.
\end{eqnarray*}
We will now use a recursive construction to represent the sum accumulated by the $\forall y \in \{ 0, 1 \}^m$ part of the problem. First define condition $r_{t \neq S + M}$, that represents that the accumulated sum is not $S$:
\begin{eqnarray*}
    r_{t\neq S + M}  & = & (< S + M \vee \neg < S + M + 1)
\end{eqnarray*}
By taking $r_0 := r_{t \neq S + M}$, now recursively define $r_{j + 1}$ from $r_j$, for $j \in \{ 0, \dots, m - 1 \}$, as follows:
\begin{eqnarray*}
r_{j + 1} &=& \left( \nxt [w_{j + 1}, w_{j + 1}] \ / \ r_j \ / \ \prv [2 \cdot w_{j + 1}, 2 \cdot w_{j + 1}] \right) \left[ 2, 2 \right] \ / \ \nxt [ 2 \cdot w_{j + 1}, 2 \cdot w_{j + 1} ]
\end{eqnarray*}
The formula $r_w = r_m$ will allow to iterate over all the accumulated sums implied by the $\forall y \in \{ 0, 1 \}^m$ part of the problem. Finally, $r$ is defined as follows:
\begin{eqnarray*}
r &=& r_u \ / \ r_w \ / \ \nxt[0, \_] \ / \ (\neg < 2 \cdot M )
\end{eqnarray*}
We now prove that $(v, M, v, 2 \cdot M) \in \semp{r}{\C}$ if and only if $\exists x \in \{0, 1\}^n \ \forall y \in \{ 0, 1 \}^m \ x \cdot u + y \cdot w \neq S$.

Assume that both $t$ and $t'$ are in $\Omega$. By induction on the definition of numerical occurrence indicators, it is easy to see that $(v, t, v, t') \in \semp{\nxt[k, k]}{\C}$ if and only if $t' = t + k$. Hence, by definition of $r_{u_i}$, we have that $(v, t, v, t') \in \semp{r_{u_i}}{\C}$ if and only if $t' = t$ (0 occurrences) or $t' = t + u_i$ (1 occurrence), or what is equivalent, if there exists $x_i \in \{ 0, 1 \}$ such that $t' = t + x_i \cdot u_i$. 
In fact, it can be proved by induction that $(v, t, v, t') \in \semp{r_u}{\C}$ if and only if $\exists x \in \{ 0, 1 \}^n$ such that $t' = t + x \cdot u$. We will demonstrate something stronger, which is that $(v, t, v, t') \in \semp{r_{u_1} \ / \ \dots \ / r_{u_i}}{\C}$ if and only if there exists $(x_1, \dots, x_i) \in \{ 0, 1 \}^i$ such that $t' = t + \sum_{k=1}^i x_i \cdot u_i$.

Our base case will be checking the property for $r_{u_1}$, which, by the same exact reasoning as above, satisfies that $(v, t, v, t') \in \semp{r_{u_1}}{\C}$ if and only if there exists $x_1 \in \{ 0, 1 \}$ such that $t' = t + x_1 \cdot u_1$.
Suppose now that $(v, t, v, t') \in \semp{r_{u_1} \ / \ \dots \ / r_{u_i}}{\C}$ if and only if there exists $(x_1, \dots, x_i) \in \{ 0, 1 \}^i$ such that $t' = t + \sum_{k=1}^i x_i \cdot u_i$. We also know by the previous reasoning that $(v, t', v, t'') \in \semp{r_{u_{i+1}}}{\C}$ if and only if there exists $x_{i+1} \in \{ 0, 1 \}$ such that $t'' = t' + x_{i+1} \cdot u_{i+1}$. Hence, by definition of $(r_1 / r_2)$, we get that $(v, t, v, t'') \in \semp{r_{u_1} \ / \ \dots \ / r_{u_{i+1}}}{\C}$ if and only if there exists $(v, t')$ such that there exists $(x_1, \dots, x_i) \in \{ 0, 1 \}^i$ such that $t' = t + \sum_{k=1}^i x_i \cdot u_i$ and there exists $x_{i+1} \in \{ 0, 1 \}$ such that $t'' = t' + x_{i+1} \cdot u_{i+1}$, \textit{i.e.}, if and only if there exists $(x_1, \dots, x_{i+1}) \in \{ 0, 1 \}^{i+1}$ such that $t'' = t + \sum_{k=1}^{i+1} x_i \cdot u_i$. This yields the result.


Moreover, by definition of $\semp{(r_1 / r_2)}{\C}$, we get that $(v, M, v, 2 \cdot M) \in \semp{r}{\C}$ if and only if there exists $x \in \{ 0, 1 \}^n$ such that:
\[(v, M + x \cdot u, v, 2 \cdot M) \in \semp{r_v \ / \ \nxt[0, \_] \ / \ (\neg < 2 \cdot M )}{\C}\]
Notice also that the right part of this formula is built in the following way:  $(\neg < 2 \cdot M)$ is a test that is only satisfied by $(v, 2 \cdot M, v, 2 \cdot M)$, whereas $\nxt[0, \_]$ is satisfied by any tuple $(v, t, v, t')$ such that $t \leq t'$. Hence, $(v, t, v', t') \in \semp{ \nxt[0, \_] \ / \ (\neg < 2 \cdot M )}{\C}$ if and only if $t$ is any time point in $\Omega$ and $t' = 2 \cdot M$.
Thus, all we need to prove now is that there exists some time point $t$ such that $(v, M + x \cdot u, v, t) \in \semp{r_v}{\C}$ if and only if $\forall y \in \{0, 1\}^m \ x \cdot u + y \cdot w \neq S $.

First, we show  
by induction that if $(v, t, v, t') \in \semp{r_j}{\C}$, then $t = t'$. The base case is trivial, since $r_0$ is a test.
For the inductive case, 
assume that if $(v, t, v, t') \in \semp{r_j}{\C}$, then $t' = t$.
    For conciseness, define $a := w_{j+1}$. In the case of $j + 1$, we know that if $(v, t_1, v, t_2) \in \semp{\nxt[a,a] \ / \ r_j \ / \ \prv[2a, 2a]}{\C}$, then there exist time points $t_3$ and $t_4$ such that $(v, t_1, v, t_3) \in \semp{\nxt[a, a]}{\C}$, $(v, t_3, v, t_4) \in \semp{r_j}{\C}$ and $(v, t_4, v, t_2) \in \semp{\prv[2a, 2a]}{\C}$. Notice then that these conditions only hold respectively if $t_3 = t_1 + a$, by definition of operator $\nxt$ and numerical occurrence indicators, $t_3 = t_4$ by induction hypothesis and $t_2 = t_4 - 2a$. Thus, $t_2 = t4 - 2a = t_3 - 2a = t_1 - a$.
    It is clear that if $(v, t_1, v, t_5) \in \semp{(\nxt[a,a] \ / \ r_k \ / \ \prv[2a, 2a])[2, 2]}{\C}$, then $t_5 = t_1 - 2a$.
    Finally, if $(v, t_5, v, t') \in \semp{\nxt[2a, 2a]}{\C}$, then $t' = t_5 + 2a$, so by definition of $\semp{(r_1/ r_2)}{\C}$, we conclude  that if $(v, t, v, t') \in \semp{r_{j+1}}{\C}$, then $t' = t$.

Given the conclusion in the previous paragraph,
all we need to prove now is that $(v, M + x \cdot u, v, M + x \cdot u) \in \semp{r_v}{\C}$ if and only if $\forall y \in \{0, 1\}^m \ x \cdot u + y \cdot w \neq S $.
In fact, we will prove by induction a stronger condition:
\[
\text{for every } k \in \{0, \dots, m\}, \text{it holds that }
\forall y \in \{0, 1\}^k \ t + \sum_{j=1}^k y_j \cdot w_j \neq S \text{ if and only if }
(v, M +  t, v, M + t) \in \semp{r_k}{\C} \]
The case when $t = x \cdot u$ and $k = m$ yields
\sigmatwop-hardness of 
\tupleeval({\rm ITPG}, \Lonep) 
as a result.
In the base case $k = 0$, we need to prove that 
$t \neq S$ if and only if $(v, M + t, v, M + t) \in \semp{r_0}{\C}$. Recall that $r_0 = r_{t \neq S + M}$. It can be easily checked that $(v, t) \models r_0$ if and only if $t \neq S + M$. Also, $r_0$ is a test, so $(v, t, v, t') \in \semp{r_0}{\C}$ if and only if $t = t'$ and $t \neq S + M$. Hence, we have that $(v, M + t, v, M + t) \in \semp{r_0}{\C}$ if and only if $ M +  t =  M + t$ (which is trivially satisfied) and $M +  t \neq M + S$,
\textit{i.e.}, if and only if $t \neq S$.
For the inductive case, assume that for $k \in \{0, \dots, m\}$, it holds that:
\[\forall y \in \{0, 1\}^k \ t + \sum_{j=1}^k y_j \cdot w_j \neq S \text{ if and only if }
(v, M + t, v, M + t) \in \semp{r_k}{\C}.\]
Then we have to prove that:
\[\forall y \in \{0, 1\}^{k + 1} \ t + \sum_{j=1}^{k+1} y_j \cdot w_j \neq S 
\text{ if and only if }
(v, M + t, v, M + t) \in \semp{r_{k+1}}{\C} \]
To prove this, define again $a := w_{k+1}$ for conciseness. Recall that $r_{k + 1} = ((\nxt[a, a] \ / \ r_k \ / \ \prv[2a, 2a])[2,2] / \nxt[2a, 2a])$. Notice then that $(v, M + t, v, t') \in \semp{\nxt[a, a] \ / \ r_k \ / \ \prv[2a, 2a]}{\C}$ if and only if there exist time points $t_1$ and $t_2$ such that $(v, M + t, v, t_1) \in \semp{\nxt[a, a]}{\C}$, $(v, t_1, v, t_2) \in \semp{r_k}{\C}$ and $(v, t_2, v, t') \in \semp{\prv[2a, 2a]}{\C}$.
The first condition is equivalent to having that $t_1 = M + t + a$. The second condition implies that $t_1 = t_2$, given what we proved in the previous paragraphs. Hence, given that $(v, M + t + a, v, M + t + a) \in \semp{r_k}{\C}$, we conclude by induction hypothesis that 
$\forall y \in \{ 0, 1 \}^k \ t + a + \sum_{j=1}^k y_j \cdot w_j \neq S$. 
The third condition is equivalent to having that $t' = t_2 - 2a$. Altogether, this means that $(v, M + t, v, t') \in \semp{\nxt[a, a] \ / \ r_k \ / \ \prv[2a, 2a]}{\C}$ if and only if $t' = M + t - a$ and $\forall y \in \{ 0, 1 \}^k$, it holds that $t + a + \sum_{j=1}^k y_j \cdot w_j \neq S$.
Now, this means that $(v, M + t, v, t') \in \semp{(\nxt[a, a] \ / \ r_k \ / \ \prv[2a, 2a])[2, 2]}{\C}$ if there exists a time point $t''$ such that $t'' = M + t - a$, $t' = M + (t-a) - a = M + t - 2a$, $\forall y \in \{ 0, 1 \}^k$, it holds that $t + a + \sum_{j=1}^k y_j \cdot w_j \neq S$, and $\forall y \in \{ 0, 1 \}^k$, it holds that $(t - a) + a + \sum_{j=1}^k y_j \cdot w_j \neq S$.
Therefore,
$(v, M + t, v, t') \in \semp{(\nxt[a, a] \ / \ r_k \ / \ \prv[2a, 2a])[2, 2]}{\C}$ if and only if $t' = M + t - 2a$ and for every $y \in \{ 0, 1 \}^{k+1}$, it holds that $t + \sum_{j=1}^{k+1} y_j \cdot w_j \neq S$.
Given that
for $t_1, t_2 \in \Omega$, it holds that $(v, t_1, v, t_2) \in \semp{\nxt[2a, 2a]}{\C}$ if and only if $t_2 = t_1 + 2a$, we conclude that $(v, M + t, v, t') \in \semp{(\nxt[a, a] \ / \ r_k \ / \ \prv[2a, 2a])[2, 2] \ / \ \nxt [2a, 2a]}{\C}$ if and only if $t' = M + t$ and $\forall y \in \{ 0, 1 \}^{k+1} \ t + \sum_{j=1}^{k+1} y_j \cdot w_j \neq S$.
Finally, this gives us the result we are trying to prove, that is, $(v, M + t, v, M + t) \in \semp{(\nxt[a, a] \ / \ r_k \ / \ \prv[2a, 2a])[2, 2] \ / \ \nxt [2a, 2a]}{\C}$ if and only if $\forall y \in \{ 0, 1 \}^{k+1} \ t + \sum_{j=1}^{k+1} y_j \cdot w_j \neq S$.

To conclude the proof of the theorem, notice that $r$ can be constructed in polynomial time with respect to the sizes of $u$, $v$ and $S$, so the entire reduction can be computed in  polynomial time.

\subsection{{\em \tupleeval({\rm ITPG}, \Lthree)} is \pspace-complete}
\label{sec-proof-itpg-pspace}

Consider the following well-known decision problem called True Quantified Boolean Formula ($\tqbf$), which is well known to be \pspace-complete \cite{StockmeyerMeyer1973}:

\begin{center}
	\framebox{
	    \begin{tabular}{p{1.6cm} p{13cm}}
			\textbf{Problem:} & $\tqbf$ \\
		    \textbf{Input:} & A quantified Boolean formula
		    $ \psi = Q_1 x_1 \dots Q_n x_n \, \varphi (x_1, \dots, x_n)$
		    in prenex normal form where $\varphi (x_1, \dots, x_n)$ is a Boolean formula on variables $x_1, \dots, x_n$, and $Q_1, \dots, Q_n$ are quantifiers ($\forall$ or $\exists$). \\
		    \textbf{Output:} & $\true$ if $\psi$ is valid, and $\false$ otherwise.
		\end{tabular}
	}
\end{center}

Without loss of generality, $\varphi$ can be assumed to be in conjunctive normal form. We will show that $\tqbf$ can be reduced to our problem \tupleeval(\ctpg, \Lthree) by proceeding in three steps. Let $\psi = Q_1 x_1 \dots Q_n x_n \, \varphi (x_1, \dots, x_n)$. First, we will show that a predicate $\bit(i,t)$ (defined below) can be written in our language. Then, by using that predicate, we will show that $\varphi$
can be encoded in our language. Finally, we will show that an expression representing  the quantifiers of $\psi$ can be added to the expression encoding $\varphi$, which yields the result.

We start with a $\qbf$ formula $\psi$ as described above to build an input for the problem
\tupleeval(\ctpg, \Lthree). More precisely, the input \ctpg will be $\C = \left( \Omega, N, E, \rho, \lambda, \xi, \sigma \right)$ where $\Omega = [0, 2^n-1]$, $N = \lbrace v \rbrace$, $E = \varnothing$, $\rho$ is an empty function, $\lambda(v) = l$, $\xi(v)=\left\lbrace [0, 2^n-1] \right\rbrace$ and $\sigma$ is an empty function. In other words, $\C$ is an \ctpg consisting of only one node existing from time $0$ to time $2^n-1$, with no edges or properties. Moreover, we will build an expression $r$ such that $(v, 0, v, 0) \in \semp{r}{\C}$ if and only if $\psi$ is valid, which concludes the reduction.
The steps to construct $r$ are shown next.


\bigskip

{\bf Step 1: Expressing the predicate $\bit$ with an expression in $\Lthree$.}
%
%
Consider predicate $\bit(i, t)$ that tests whether the $i$-th bit of time $t$ (from right to left when written in its binary representation) is 1. For instance, $\bit(1, 0)$ is false, and $\bit(5, 30)$ is true, since the first bit of $0$ is $0$, whereas $30$ is $11110$ in binary, and its fifth bit is $1$.
Now, consider the following expression:
\begin{eqnarray*}
    r_i &  = & ?\left( \past \left[ 2^i, 2^i \right] \left[ 0, \_ \right] \ / \ \left( < 2^i \wedge \neg < 2^{i-1} \right) \right)
\end{eqnarray*}

Notice that $r_i$ is a test. Thus, for a pair of temporal objects $(o_1, t_1, o_2, t_2)$ 
to satisfy $r_i$, $(o_1, t_1)$ must be equal to  $(o_2, t_2)$. The expression to satisfy is a path test, so there must be some temporal object $(o_3, t_3)$, such that $(o_1, t_1, o_3, t_3) \in \semp{ \past \left[ 2^i, 2^i \right] \left[ 0, \_ \right] \ /  \left( < 2^i \wedge  \neg < 2^{i-1} \right) }{\C}$.
Since the right part is a test as well, we can split the expression into two parts. Firstly, we must have that $(o_1, t_1, o_3, t_3) \in \semp{ \past \left[ 2^i, 2^i \right] \left[ 0, \_ \right] }{\C}$, which implies that $t_3 = t_1 - k*2^i$ for some integer $k \geq 0$. 
Secondly, we must have that $(o_3, t_3) \models \left( < 2^i \wedge \neg < 2^{i-1} \right)$, which implies that $2^{i-1} \leq t_3 < 2^i$. This means that by writing $t_3$ in its binary form, we get $1$ as its $i$-th bit. Together, these two conditions imply that $t_1$ also has $1$ as its $i$-th bit, when written in its binary form. In consequence, we get that
\begin{align*}
    (o,t, o, t) \in \semp{r_i}{\C} \text{ if and only if } \bit(i, t) \text{ is } \true
\end{align*}
Besides, we trivially get that:
\begin{align*}
    (o,t, o, t) \in \semp{ \neg r_i }{\C} \text{ if and only if } \bit(i, t) \text{ is } \false
\end{align*}
Finally, notice that both $r_i$ and $\neg r_i$ have linear length with respect to $i$, which will be important later.

\bigskip

\textbf{\bf Step 2: Expressing any CNF formula in $\Lthree$.}
Assume that
\begin{eqnarray*}
    \varphi(x_1, \dots, x_n)  &=& \bigwedge_{j=1}^m \bigvee_{k=1}^{m_j} l_{j,k}
\end{eqnarray*}
where for every $j$ and $k$, $l_{j,k}$ is a literal, \textit{i.e.}, either a variable in $\lbrace x_1, \dots, x_n \rbrace$ or its negation. Then, for every $j \in \lbrace 1, \dots, m\rbrace$ and $k \in \lbrace 1, \dots, m_j \rbrace$, we define:
\begin{eqnarray*}
    L_{j,k} &=& \begin{cases}
        r_i & \text{if }  l_{j,k} = x_i \\
        \neg r_i & \text{if  } l_{j,k} = \neg x_i
    \end{cases}
\end{eqnarray*}
    We can use these expressions to build a regular expression that tests the satisfiability of our formula $\varphi(x_1, \dots, x_n)$ by any valuation $\sigma: \left\lbrace x_1, \dots, x_n \right\rbrace \rightarrow \left\lbrace \false, \true \right\rbrace$. To do this, we will use a time value $t \in \left[ 0, 2^n - 1 \right]$ to represent a valuation $\sigma_t$, where $\sigma_t (x_i) = \true$ if and only if the $i$-th bit of $t$ is $1$. We can do so by employing the expressions $L_{i,k}$ on tests along with conjunctions ($\wedge$) and disjunctions ($\vee$), which are also present in $\Lthree$:
\begin{eqnarray*}
    r_{\varphi(x_1, \dots, x_n)} & = & \bigwedge_{j=1}^m \bigvee_{k=1}^{m_j} L_{j, k}
\end{eqnarray*}
Here again, $r_{\varphi(x_1, \dots, x_n)}$ is a test, so if it is satisfied by $(o_1, t_1, o_2, t_2)$, then $(o_1, t_1) = (o_2,t_2)$
Furthermore, we show that, because of how the expressions $L_{j,k}$ are defined, we have:
\begin{align*}
    (o,t, o, t) \in \semp{ r_{\varphi(x_1, \dots, x_n)} }{\C} \text{ if and only if } \sigma_t \left( \varphi(x_1, \dots, x_n) \right) = \true
\end{align*}
%
To show the previous assertion, first assume that $(o, t, o, t) \in \semp{ r_{\varphi (x_1, \dots, x_n) } }{\C}$. Because of how the expression is defined, for each $j \in \lbrace 1, \dots, m \rbrace$ we must have that $(o, t, o, t) \in \semp{ \bigvee_{k=1}^{m_j} L_{j, k} }{\C}$. Hence, for any arbitrary $j \in \lbrace 1, \dots, m \rbrace$, we immediately get that there must exist $k \in \lbrace 1, \dots, m_j \rbrace$ such that $(o, t, o, t) \in \semp{ L_{j, k} }{\C}$. If $l_{j,k} = x_i$, then we must also have that $(o, t, o, t) \in \semp{r_i}{\C}$, which, as we already showed, is equivalent to having that $\bit(i, t)$ is $\true$, which in turn is equivalent to $\sigma_t(x_i) = \true$. Otherwise, if $l_{j,k} = \neg x_i$, then we must have that $(o, t, o, t) \in \semp{ \neg r_i }{\C}$, which, as we also showed, is equivalent to having that $\bit(i, t)$ is $\false$, which in turn is equivalent to $\sigma_t(x_i) = \false$. In both cases, we get that ${\sigma}_t(l_{j, k}) = \true$, which means that ${\sigma}_t \left( \bigvee_{k=1}^{m_j} l_{j,k} \right) = \true$. Since this holds for an arbitrary value $j$, it holds for the entire conjunction. Hence, ${\sigma}_t \left( \varphi (x_1, \dots, x_n) \right) = \true$.

Now, to show that the inverse is also true, assume that there is some $t \in \Omega$ such that ${\sigma}_t \left( \varphi( x_1, \dots, x_n ) \right) = \true$. By definition, this means that for every $j \in \lbrace 1, \dots, m \rbrace$, ${\sigma}_t \left( \bigvee_{k=1}^{m_j} l_{j, k} \right) = \true$. In turn, this means that for every $j$ there is some $k \in \lbrace 1, \dots, m_j \rbrace$ such that ${\sigma}_t( l_{j, k}) = \true$. As we saw earlier, this condition is equivalent to having that $(o, t, o, t) \in \semp{L_{j,k}}{\C}$, hence, for every $j \in \lbrace 1, \dots, m \rbrace$, we also have that $(o, t, o, t) \in \semp{ \bigvee_{k=1}^{m_j} L_{j,k} }{\C}$. Given that this is true for every $j$, by definition, it is also true for the conjunction of them. Hence, we get that $(o, t, o, t) \in \semp{ r_{\varphi(x_1, \dots, x_n)} }{\C}$. This concludes the proof in Step 2.

\bigskip

\textbf{Observation}: Notice that the previous results already implies $\np$-hardness and $\conp$-hardness for the problem. Since every valuation $\sigma: \lbrace x_1, \dots, x_n \rbrace \rightarrow \lbrace \false, \true \rbrace$ has a corresponding time point $t$ such that $\sigma = \sigma_t$, and a Boolean CNF formula $\varphi (x_1, \dots, x_n)$ on $n$ variables $x_1, \dots, x_n$ is satisfiable if and only if there exists a valuation $\sigma$ such that ${\sigma} \left( \varphi (x_1, \dots, x_n) \right) = \true$, it is also true that $\varphi (x_1, \dots, x_n)$ is satisfiable if and only if $(v, 0, v, 0) \in \semp{ ?(\nxt[0, \_] \ / \ r_{ \varphi (x_1, \dots, x_n) }) }{\C}$ (that is, advance in time to an arbitrary time point $t$ and check the condition that implies that ${\sigma}_t \left( \varphi( x_1, \dots, x_n ) \right) = \true$ for the temporal object $(v, t)$). Similarly, $\varphi(x_1, \dots, x_n) $ is a tautology if and only if $(v, 0, v, 0) \in \semp{\neg ? (\nxt [0, \_] \ / \ \neg r_{\varphi (x_1, \dots, x_n)}) }{\C}$ (there is no path to a time point $t$ such that ${\sigma}_t \left( \varphi( x_1, \dots, x_n ) \right) = \false$, hence there is no valuation that makes $\varphi$ to be $\false$). In what follows, we show $\pspace$-hardness of the problem.

\bigskip

\textbf{Step 3: Expressing satisfiability of quantified Boolean formulae 
with an expression in \Lthree.}
Consider a quantified Boolean formula in prenex normal form
\[ Q_1 x_1 \dots Q_n x_n \ \varphi(x_1, \dots, x_n) \]
Since we already have a way to express $\varphi(x_1, \dots, x_n)$, we only need to express the possible valuations (\textit{i.e.}, time points) generated by the sequence of quantifiers $Q_1, \dots, Q_n$.

First, assume $Q_i$ is the existential quantifier ($\exists$). This means that we can either make $x_i$ take valuation $\true$ or $\false$ to satisfy our formula. Considering time points, this is equivalent to have either $1$ or $0$ 
at the $i$-th bit of the time $t$ at which we are standing. The intuition is that this can easily be expressed by starting at time $0$, and then deciding whether to move into a future time point with the expression $(\nxt [2^{i-1}, 2^{i-1}] + \nxt [0, 0])$ to set the $i$-th bit of the time point. Notice that if there are only expressions of the form $\nxt [2^{i-1}, 2^{i-1}]$, and they are only mentioned once (at least, as prefixes of our test expressions) for each $i$, they will not affect other bits of $t$. Hence, if $s_{i+1}$ represents the part of the subformula $Q_{i+1} x_{i+1} \dots Q_n x_n \ \varphi(x_1, \dots, x_n)$, then the subformula $\exists x_i \ Q_{i+1} x_{i+1} \dots Q_n x_n \ \varphi(x_1, \dots, x_n)$ can be represented by first navigating through time, only affecting the first $i-1$ bits, and then testing that the reached temporal object satisfies the following test:
\begin{eqnarray*} 
s_i & = & ?\left( \left(\nxt [2^{i-1}, 2^{i-1}] + \nxt [0, 0] \right) \ / \ s_{i+1} \right).
\end{eqnarray*}
In turn, if $Q_i$ is the universal quantifier ($\forall$), we will employ the fact that $\forall x \, \psi(x)$ is equivalent to $\neg \exists x \, \neg \psi(x)$ for every formula $\psi(x)$ with free variable $x$. Hence, if $s_{i+1}$ represents the part of the subformula $Q_{i+1} x_{i+1} \dots Q_n x_n \ \varphi(x_1, \dots, x_n)$, then the subformula $\forall x_i \ Q_{i+1} x_{i+1}  \dots Q_n x_n$ $\varphi(x_1, \dots, x_n)$ can be represented by first only navigating through time, only affecting the first $i-1$ bits, and then testing whether the reached temporal object satisfies the following test:
\begin{eqnarray*}
s_i &=& \neg \left( ? \left( (\nxt [2^{i-1}, 2^{i-1}] + \nxt [0, 0]) \ / \ \left(\neg s_{i+1} \right) \right) \right).
\end{eqnarray*}
Finally, define $s_{n+1} = r_{\varphi(x_1, \dots, x_n)}$. We claim that, for $i \in \{ n + 1, n, \dots, 1 \}$ (\textit{i.e.}, $0$ to $n$ quantifiers), if $t < 2^{i - 1}$, then:
\[ (v, t, v, t) \in \semp{s_i}{\C} \ \text{if and only if} \ Q_i x_i \dots Q_n x_n \ \varphi(\sigma_t(x_1), \dots, \sigma_t(x_{i-1}), x_i, \dots, x_n) \ \text{is valid}. \]
In particular, for $n$ quantifiers, \textit{i.e.}, when $i=1$, this result gives us \pspace-hardness for \tupleeval(\ctpg, \Lthree), since we will have that $\psi$ is $\true$ if and only if $(v, 0, v, 0) \in \semp{s_1}{\C}$, where $\C$ and $s_1$ can be constructed in polynomial time in the size of $\psi$. We will prove this claim by induction over the number of quantifiers preceding $\varphi(x_1, \dots, x_n)$.

The base case consists of the formula with no quantified variables, \textit{i.e.}, when $i = n + 1$. We must show that if $t < 2^n$, then $(v, t, v, t) \in \semp{r_{\varphi(x_1, \dots, x_n)}}{\C}$ if and only if $\varphi(\sigma(x_1), \dots, \sigma(x_n))$ is $\true$. Notice that $\varphi(\sigma(x_1), \dots, \sigma(x_n))$ is $\true$
is equivalent to having that ${\sigma}_t(\varphi(x_1, \dots, x_n))$ is $\true$. 
Hence, by step 2, this is equivalent to having that $(v, t, v, t) \in \semp{r_{\varphi(x_1, \dots, x_n)}}{\C}$.

For the inductive case, assume that the claim holds for $k$ quantifiers, for some $k$ such that $0 \leq k \leq n$. This means that for $i = n - k + 1$, if $t < 2^{i-1}$, then the following condition holds:
\[ (v, t, v, t) \in \semp{s_i}{\C} \ \text{if and only if} \ Q_i x_i \dots Q_n x_n \ \varphi(\sigma_t(x_1), \dots, \sigma_t(x_{i-1}), x_i, \dots, x_n) \ \text{is valid} \]
We then have to prove that the condition holds for $k+1$ quantifiers, \textit{i.e.}, for $i - 1 = n - k$. That is, if $t' < 2^{i-2}$, then we have to show that:
\begin{align}
    \label{eq-equiv-i-1}
(v, t', v, t') \in \semp{s_{i-1}}{\C} \ \text{if and only if} \ Q_{i-1} x_{i-1} \dots Q_n x_n \ \varphi(\sigma_{t'}(x_1), \dots, \sigma_{t'}(x_{i-2}), x_{i-1}, \dots, x_n) \ \text{is valid}
\end{align}
Let $t' < 2^{i-2}$, and consider the following cases.
\begin{itemize}
    \item If $Q_{i-1} = \exists$, recall that
    \begin{eqnarray*}
        s_{i-1} &=& ? \left( \left(\nxt [2^{i-2}, 2^{i-2}] + \nxt [0, 0] \right) \ / \ s_{i} \right).
    \end{eqnarray*}
    \begin{sloppypar}
        Notice then that by definition of $s_{i-1}$,
        we have that $(v, t', v, t') \in \semp{s_{i-1}}{\C}$ if and only if there exists $t_1 \in \Omega$ such that $(v, t', v, t_1) \in \semp{ \left(\nxt [2^{i-2}, 2^{i-2}] + \nxt [0, 0] \right) \ / \ s_i}{\C}$. By definition of $s_i$, 
        the previous condition holds if and only if $(v, t', v, t_1) \in \semp{\left(\nxt [2^{i-2}, 2^{i-2}] + \nxt [0, 0] \right)}{\C}$ and $(v, t_1, v, t_1) \in \semp{s_i}{\C}$.
        Now, this means that $(v, t', v, t') \in \semp{s_{i-1}}{\C}$ if and only if there exists $t_1 \in \{ t', t' + 2^{i-2} \}$ satisfying that $(v, t_1, v, t_1) \in \semp{s_i}{\C}$.
        
        To prove the direction $(\Rightarrow)$ of \eqref{eq-equiv-i-1} 
        assume that $(v,t',v,t') \in \semp{s_{i-1}}{\C}$, which implies that  
        there exists $t_1 \in \{ t', t' + 2^{i-2} \}$ satisfying that $(v, t_1, v, t_1) \in \semp{s_i}{\C}$. Given that $t'$ is an integer with $i-2$ bits, $t_1$ comes from either putting $1$ or $0$ as the $(i-1)$-th bit of $t'$, which means that $t_1 < 2^{i-1}$ must hold. By induction hypothesis, this implies that $Q_i x_i \dots Q_n x_n \varphi(\sigma_{t_1}(x_1), \dots, \sigma_{t_1}(x_{i-1}), x_i, \dots, x_n)$ is valid. Since $t_1$ and $t'$ share the same first $i-2$ bits, we get that $\sigma_{t'}(x_j) = \sigma_{t_1}(x_j)$ for $j \in \{ 1, ..., i-2 \}$. Therefore, we conclude that
        $\exists x_{i-1} Q_i x_i \dots Q_n x_n \varphi(\sigma_{t'}(x_1), \dots, \sigma_{t'}(x_{i-2}), x_{i-1}, \dots,  x_n)$ is valid.
        
        To prove the direction $(\Leftarrow)$ of \eqref{eq-equiv-i-1} 
        suppose that $\exists x_{i-1} Q_i x_i \dots Q_n x_n \varphi(\sigma_{t'}(x_1), \dots, \sigma_{t'}(x_{i-2}), x_{i-1}, \dots,  x_n)$ is valid. Then 
        we know that $Q_i x_i \dots Q_n x_n \varphi(\sigma_{t'}(x_1), \dots, \sigma_{t'}(x_{i-2}), b, x_i, \dots,  x_n)$ is valid for some value $b \in \{\true,\false\}$. Define $\mathbbm{1}_b$ as $1$ if $b = \true$, and as $0$ otherwise. Notice then that by taking $t_1 = t' + 2^{i-2} \cdot \mathbbm{1}_b$, we get that $\sigma_{t_1}(x_{i-1}) = b$. Moreover, $\sigma_{t_1}(x_j) = \sigma_{t'}(x_j)$ for every $j \in \{ 1, \dots, i-2 \}$ since $t' < 2^{i-2}$ and $t_1$ shares all its first $i-2$ bits with $t'$. In consequence, this gives us that  $Q_i x_i \dots Q_n x_n \varphi(\sigma_{t_1}(x_1), \dots, \sigma_{t_1}(x_{i-1}), x_i, \dots,  x_n)$ is valid.
        By induction, this means that $(v, t_1, v, t_1) \in \semp{s_i}{\C}$. Since $t_1 = t' + 2^{i-2} \cdot \mathbbm{1}_b$, we have that either $t_1 = t'$ or $t_1 = t' + 2^{i-2}$. In any case, we get that $(v, t', v, t_1) \in \semp{\nxt[2^{i-2}, 2^{i-2}] + \nxt[0, 0] }{\C}$, so we have that $(v, t, v, t_1) \in \semp{(\nxt[2^{i-2}, 2^{i-2}] + \nxt[0, 0] ) \ / \ s_i }{\C}$. By definition of $s_{i-1}$, 
        we conclude that $(v, t', v, t') \in \semp{s_{i-1}}{\C}$, which was to be shown.
    \end{sloppypar}
    
    \item If $Q_{i-1} = \forall$, recall that
    \begin{eqnarray*}
    s_{i-1} &=& \neg ? \left( \left(\nxt [2^{i-2}, 2^{i-2}] + \nxt [0, 0] \right) \ / \ \left(\neg s_i \right) \right). \end{eqnarray*}
    \begin{sloppypar}
    Now, by definition of $s_{i-1}$,
    we have that $(v, t', v, t') \in \semp{s_{i-1}}{\C}$ if and only if 
    \begin{eqnarray*}
    (v, t') & \not\models & ? \left( \left(\nxt [2^{i-2}, 2^{i-2}] + \nxt [0, 0] \right) \ / \ \left(\neg s_i \right) \right).
    \end{eqnarray*}
    This, in turn, 
    is equivalent to 
    the fact that there is no time point $t_1 \in \Omega$ such that $(v, t', v, t_1) \in \semp{ \left(\nxt [2^{i-2}, 2^{i-2}] + \nxt [0, 0] \right) \ / \ \left(\neg s_i \right)}{\C}$.
    Hence, we know that $(v, t', v, t') \in \semp{s_{i-1}}{\C}$ if and only if, for every time point $t_1 \in \Omega$:
    \[ (v, t', v, t_1) \notin \semp{ \left(\nxt [2^{i-2}, 2^{i-2}] + \nxt [0, 0] \right) \ / \ \left(\neg s_i \right)}{\C}. \]
    This condition means that each $t_1$ satisfies $(v, t', v, t_1) \notin \semp{ \left(\nxt [2^{i-2}, 2^{i-2}] + \nxt [0, 0] \right)}{\C}$ or $(v, t_1) \not\models \left(\neg s_i \right)$. This, in turn, is equivalent to saying that if $t_1 \in \{ t', t' + 2^{i-2} \}$ then $(v, t_1) \not\models \left(\neg s_i \right)$, \textit{i.e.}, $(v, t_1) \models s_i $.
    As a consequence, $(v, t', v, t') \in \semp{s_{i-1}}{\C}$ if and only if $(v, t') \models s_i$ and $(v, t' + 2^{i-2}) \models s_i$, which is equivalent to having that $(v, t', v, t') \in \semp{s_i}{\C}$ and $(v, t' + 2^{i-2}, v, t' + 2^{i-2}) \in \semp{s_i}{\C}$.

    To prove the direction ($\Rightarrow$) of \eqref{eq-equiv-i-1} suppose that $(v, t', v, t') \in \semp{s_{i-1}}{\C}$, so we also have that $(v, t', v, t') \in \semp{s_i}{\C}$ and $(v, t' + 2^{i-2}, v, t' + 2^{i-2}) \in \semp{s_i}{\C}$. Notice then that $t' < 2^{i-2}$, so $t'$ and $t' + 2^{i-2}$ are both smaller than $2^{i-1}$. By induction hypothesis, we get then that both quantified Boolean formulae $Q_i x_i \dots Q_{n} x_n \varphi(\sigma_{t'}(x_1), \dots, \sigma_{t'}(x_{i-1}), x_i, \dots, x_n)$ and $Q_i x_i \dots Q_{n} x_n \varphi(\sigma_{t' + 2^{i-2}}(x_1), \dots, \sigma_{t' + 2^{i-1}}(x_{i-1}), x_i, \dots, x_n)$ are valid.
    Furthermore, since $t'$ is smaller than $2^{i-2}$, $\sigma_{t'}(x_{i-1}) = \false$, and since $t' + 2^{i-2}$ only differs from $t'$ in its $(i-1)$-th bit, which is $1$, we get that $\sigma_{t' + 2^{i-2}}(x_{i-1}) = \true$ and, for every $j \in \{ 1, \dots, {i-2} \}$, it holds that $\sigma_{t'}(x_j) = \sigma_{t' + 2^{i-2}}(x_j)$. Hence, both quantified Boolean formulae $Q_i x_i \dots Q_{n} x_n \varphi(\sigma_{t'}(x_1), \dots, \sigma_{t'}(x_{i-2}), \false, x_i, \dots, x_n)$ and $Q_i x_i \dots Q_{n} x_n \varphi(\sigma_{t'}(x_1), \dots, \sigma_{t'}(x_{i-2}), \true, x_i, \dots, x_n)$ are valid. Therefore, 
    we conclude that the quantified Boolean formula $\forall x_{i-1} Q_i x_i \dots Q_{n} x_n \varphi(\sigma_{t'}(x_1), \dots, \sigma_{t'}(x_{i-2}), x_{i-1}, x_i, \dots, x_n)$ is valid.
    
    To show the direction ($\Leftarrow$) of \eqref{eq-equiv-i-1} suppose that $\forall x_{i-1} Q_i x_i \dots Q_{n} x_n \varphi(\sigma_{t'}(x_1), \dots, \sigma_{t'}(x_{i-2}), x_{i-1}, x_i, \dots, x_n)$ is valid. Then
    we get that both quantified Boolean formulae $Q_i x_i \dots Q_{n} x_n \varphi(\sigma_{t'}(x_1), \dots, \sigma_{t'}(x_{i-2}), \false, x_i, \dots, x_n)$ and $Q_i x_i \dots Q_{n} x_n \varphi(\sigma_{t'}(x_1), \dots, \sigma_{t'}(x_{i-2}), \true, x_i, \dots, x_n)$ are valid.
    As before, since $t'$ is smaller than $2^{i-2}$, $\sigma_{t'}(x_{i-1}) = \false$, and since $t' + 2^{i-2}$ only differs from $t'$ in its $(i-1)$-th bit, which is 1, we get that $\sigma_{t' + 2^{i-2}}(x_{i-1}) = \true$ and for every $j \in \{ 1, \dots, {i-2} \}$, it holds that $\sigma_{t'}(x_j) = \sigma_{t' + 2^{i-2}}(x_j)$. This allows us to conclude that both $Q_i x_i \dots Q_{n} x_n \varphi(\sigma_{t'}(x_1), \dots, \sigma_{t'}(x_{i-1}), x_i, \dots, x_n)$ and $Q_i x_i \dots Q_{n} x_n \varphi(\sigma_{t' + 2^{i-2}}(x_1), \dots, \sigma_{t' + 2^{i-1}}(x_{i-1}), x_i, \dots, x_n)$ are valid.
    Finally, by induction hypothesis, this implies that $(v, t', v, t') \in \semp{s_i}{\C}$ and $(v, t' + 2^{i-2}, v, t' + 2^{i-2}) \in \semp{s_i}{\C}$, which, as shown before, holds if and only if $(v, t', v, t') \in \semp{s_{i-1}}{\C}$, which concludes the proof for this case.
    \end{sloppypar}
\end{itemize}
As we mentioned, all this together implies that \tupleeval(\ctpg, \Lthree) is \pspace-hard, since \ctpg $\C$, expression $s_1$ and tuple $(v, 0, v, 0)$ can be constructed in polynomial time in the size of $\psi$, and the problem of determining whether $\psi$ is valid 
can be reduced to the problem of verifying whether $(v, 0, v, 0) \in \semp{s_1}{\C}$.

Thus, it only remains to show that \tupleeval(\ctpg, \Lthree) is in \pspace. To do this, we will provide an algorithm in \pspace that, given an \ctpg \C, an expression $r$ in \Lthree, and a tuple $(o, t, o', t') \in \Pt(\C)$, computes whether $(o,t, o',t') \in \semp{r}{\C}$. More precisely, Algorithm $\textsc{TupleEvalSolve}\left(\C, r, (o_1, t_1, o_2, t_2)\right)$ is defined as follows.

\LinesNumbered

\begin{algorithm}
    \SetAlgoLined
    \SetKwInOut{Input}{Input}
    \SetKwInOut{Output}{Output}
    \Input{An \ctpg $\C = \left( \Omega, N, E, \rho, \lambda, \xi, \sigma \right)$, an expression $r$ in $\Lthree$ and a pair of temporal objects $(o_1, t_1, o_2, t_2) \in \Pt(\C)$}
    \Output{$\true$ if and only if $(o_1,t_1, o_2,t_2) \in \semp{r}{\C}$}
    \uIf{$r$ is a test}{
        \uIf{$(o_1, t_1) \neq (o_2, t_2)$}{
            \Return \false
        }
        \uIf{$r = \vd$}{
            \Return $(o_1 \in N)$
        }
        \uElseIf{$r = \ed$}{
            \Return $(o_1 \in E)$
        }
        \uElseIf{$r = \ell$ for some \ $\ell \in \Lab$}{
            \Return $\lambda(o_1) = \ell$
        }
        \uElseIf{$r = \propt{p}{v}$ for some $p\in \Prop$ \ \text{and} $v\in \Val$}{
            \ForEach{valued interval $(v', I) \in \sigma(o_1, p)$}{
                \uIf{$t_1 \in I$}{
                    \Return $(v' = v)$
                }
            }
        }
        \uElseIf{$r = \, < \mt$ with $\mt \in \Omega$}{
            \Return $(t_1 < \mt)$
        }
        \uElseIf{$r = \ex$}{
            \ForEach{interval $I \in \xi(o_1)$}{
                \uIf{$t_1 \in I$}{
                    \Return \true
                }
            }
        }
        \uElseIf{$r = (?r')$}{
            \ForEach{$(o', t') \in (N \cup E) \times \Omega$}{
                \uIf{$\textsc{TupleEvalSolve}\left(\C, r', (o_1, t_1, o', t') \right)$}{
                    \Return \true
                }
            }
        }
        \uElseIf{$r = (\test_1 \vee \test_2)$}{
            \Return $\textsc{TupleEvalSolve}\left(\C, \test_1, (o_1, t_1, o_1, t_1) \right) \ \textbf{or} \ \textsc{TupleEvalSolve}\left(\C, \test_2, (o_1, t_1, o_1, t_1) \right)$
        }
        \uElseIf{$r = (\test_1 \wedge \test_2)$}{
            \Return $\textsc{TupleEvalSolve}\left(\C, \test_1, (o_1, t_1, o_1, t_1) \right) \ \textbf{and} \ \textsc{TupleEvalSolve}\left(\C, \test_2, (o_1, t_1, o_1, t_1) \right)$
        }
        \uElseIf{$r  = \left( \neg r' \right)$}{
            \Return $\textbf{not} \ \textsc{TupleEvalSolve}\left(\C, r', (o_1, t_1, o_1, t_1) \right)$
        }
    }
    \uElseIf{$r = \nxt$}{
        \Return $o_1 = o_2 \ \textbf{and} \ t_2 = t_1 + 1$
    }
    \uElseIf{$r = \prv$}{
        \Return $o_1 = o_2 \ \textbf{and} \ t_2 = t_1 - 1$
    }
    \uElseIf{$r = \fw$}{
        \Return $t_1 = t_2 \ \textbf{and} \ \left(
            \left( o_1 \in E \ \textbf{and} \ o_2 = \tgt(o_1) \right)
            \ \textbf{or} \
            \left( o_2 \in E \ \textbf{and} \ o_1 = \src(o_2) \right)
        \right)$
    }
    \uElseIf{$r = \bw$}{
        \Return $t_1 = t_2 \ \textbf{and} \ \left(
            \left( o_1 \in E \ \textbf{and} \ o_2 = \src(o_1) \right)
            \ \textbf{or} \
            \left( o_2 \in E \ \textbf{and} \ o_1 = \tgt(o_2) \right)
        \right)$
    }
    \uElseIf{$r  = \left(r_1 + r_2 \right)$}{
        \Return $\textsc{TupleEvalSolve}\left(\C, r_1, (o_1, t_1, o_2, t_2) \right) \ \textbf{or} \ \textsc{TupleEvalSolve}\left(\C, r_2, (o_1, t_1, o_2, t_2) \right)$
    }
    \uElseIf{$r = \left(r_1 \ / \ r_2 \right)$}{
        \ForEach{$(o',t') \in (N \cup E) \times \Omega$}{
            \uIf{$\textsc{TupleEvalSolve}\left(\C, r_1, (o_1, t_1, o', t') \right) \textbf{and} \ \textsc{TupleEvalSolve}\left(\C, r_2, (o', t', o_2, t_2) \right)$}{
                \Return $\true$
            }
        }
    }
    \caption{$\textsc{TupleEvalSolve}\left(\C, r, (o_1, t_1, o_2, t_2)\right)$ (part I)}
\end{algorithm}

\begin{algorithm}
    \SetAlgoLined
    \setcounter{AlgoLine}{43}
    \uElseIf{$r = r' \left[ n, m \right]$}{
        \uIf{$m = n$}{
            \uIf{$n = 0$}{
                \Return $(o_1, t_1) = (o_2, t_2)$
            }
            \uElseIf{$n = 1$}{
                \Return $\textsc{TupleEvalSolve} \left( \C, r', (o_1, t_1, o_2, t_2) \right)$
            }
            $l \gets \lfloor n/2 \rfloor$ \\
            \uIf{$m $ is even}{
                \ForEach{$(o', t') \in (N \cup E) \times \Omega$}{
                    \uIf{$\textsc{TupleEvalSolve} \left( \C, r'[ l, l], (o_1, t_1, o', t') \right)
                    \textbf{and} \ 
                    \textsc{TupleEvalSolve} \left( \C, r'[ l, l], (o', t', o_2, t_2) \right)$}{
                        \Return $\true$
                    }
                }
            }
            \uElseIf{$m$ is odd}{
                \ForEach{$(o', t') \in (N \cup E) \times \Omega$}{
                    \ForEach{$(o'', t'') \in (N \cup E) \times \Omega$}{
                        \uIf{$($\\
                        $\qquad \textsc{TupleEvalSolve}\left( \C, r'[ l, l ], (o_1, t_1, o', t') \right)
                        \textbf{and} $ \\
                        $\qquad \textsc{TupleEvalSolve}\left( \C, r', (o', t', o'', t'') \right)
                        \textbf{and} $ \\
                        $\qquad \textsc{TupleEvalSolve}\left( \C, r'[ l, l ], (o', t', o_2, t_2) \right)$ \\
                        $)$}{
                            \Return $\true$
                        }
                    }
                }
            }
        }
        \uElseIf{$n = 0$}{
            \uIf{$m = 1$}{
                $\Return (o_1, t_1) = (o_2, t_2) \ \ \textbf{or} \ \ \textsc{TupleEvalSolve} \left( \C, r', (o_1, t_1, o_2, t_2) \right)$
            }
            $l \gets \lfloor m/2 \rfloor$ \\
            \uIf{$m$ is even}{
                \ForEach{$(o', t') \in (N \cup E) \times \Omega$}{
                    \uIf{$\textsc{TupleEvalSolve} \left( \C, r'[ 0, l], (o_1, t_1, o', t') \right)
                    \textbf{and} \ 
                    \textsc{TupleEvalSolve} \left( \C, r'[ 0, l], (o', t', o_2, t_2) \right)$}{
                        \Return $\true$
                    }
                }
            }
            \uElseIf{$m$ is odd}{
                \ForEach{$(o', t') \in (N \cup E) \times \Omega$}{
                    \ForEach{$(o'', t'') \in (N \cup E) \times \Omega$}{
                        \uIf{
                            $($ \\
                            $\qquad \textsc{TupleEvalSolve}\left( \C, r'[ 0, l ], (o_1, t_1, o', t') \right) \ \textbf{and} $ \\
                            $\qquad \textsc{TupleEvalSolve}\left( \C, r'[ 0, 1 ], (o', t', o'', t'') \right) \ \textbf{and}$ \\
                            $\qquad \textsc{TupleEvalSolve}\left( \C, r'[ 0, l ], (o', t', o_2, t_2) \right)$ \\
                            $)$
                        }{
                            \Return $\true$
                        }
                    }
                }
            }
        }
        \uElse{
            \ForEach{$(o', t') \in (N \cup E) \times \Omega$}{
                \uIf{$\textsc{TupleEvalSolve}\left( \C, r'[ n, n ], (o_1, t_1, o', t') \right) \textbf{and} \  \textsc{TupleEvalSolve}\left( \C, r'[ 0, m-n ], (o', t', o_2, t_2) \right)$}{
                    \Return $\true$
                }
            }
        }
    }
    \uElseIf{$r = r' \left[ n, \_ \right]$}{
        $m \gets n + (\left| \Omega \right| \cdot \left| N \cup E \right|)^2$ \\
        \Return $\textsc{TupleEvalSolve}\left(\C, r'[n, m], (o_1, t_1, o_2, t_2) \right)$
    }
    \Return \false
    \caption{$\textsc{TupleEvalSolve}\left(\C, r, (o_1, t_1, o_2, t_2)\right)$ (part II)}
\end{algorithm}

Next we show that
for every \ctpg $\C = \left( \Omega, N, E, \rho, \lambda, \xi, \sigma \right)$, expression $r$ in $\Lthree$ and tuple $(o_1, t_1, o_2, t_2) \in \Pt(\C)$, it holds that $(o_1, t_1, o_2, t_2) \in \semp{r}{\C}$ if and only if $\textsc{TupleEvalSolve} \left(\C, r, (o_1, t_1, o_2, t_2) \right)$ returns $\true$. Besides, we will prove that 
$\textsc{TupleEvalSolve}$ works in polynomial space in the size of the input.

Notice firstly that the algorithm is recursive, and that the depth of the recursion is polynomial, since at every step on which the algorithm is called, the size of the path expression strictly decreases. There is one exception, that happens when $r$ is of the form $\pt[n, \_]$. Notice that here this expression is treated as if it was $\pt[n, m]$, where $m = n + \left| \Omega \right| \cdot \left| N \cup E \right|$ (a term with polynomial size with respect to the input). Thus, the whole expression $r$ can be thought as an equivalent expression $r'$ where all terms of the form $\pt[n, \_]$ are replaced with similar ones, in a manner that makes the whole input remain polynomial to the original. Although it might seem that $\pt[n, m]$ could reach an exponential number of recursive calls, notice that this expression is always parsed as two expressions, $\pt[n, n]$ and $\pt[0, m-n]$, and then each of those is solved in a way similar to that of exponentiation by squaring, which allows to always get rid of the numerical occurrence indicator $[n, m]$ after at most $O(\log m)$ recursive calls, a number that is polynomial in the size of the input. Hence the recursion tree has polynomial height.

Secondly, assume that conjunctions ($\wedge$) and disjunctions ($\vee$) are computed from left to right, \textit{i.e.}, $A(x) \wedge A(y)$ first computes $A(x)$ and then computes $A(y)$. Hence, at the most, we will need to have in memory as many calls to the algorithm as the recursion tree height. This number is polynomial, and since every non-recursive step is either a non-deterministic guess or clearly in polynomial time in the size of the input, we get that the whole algorithm gives an answer in \pspace.

Now, let  $\C = \left( \Omega, N, E, \rho, \lambda, \xi, \sigma \right)$ be an \ctpg, let $r$ be an expression in $\Lthree$ and let $(o_1, t_1, o_2, t_2) \in \Pt(\C)$ be a tuple concatenating two temporal objects. First suppose that $(o_1, t_1, o_2, t_2) \in \semp{r}{\C}$. We will show, by induction on the recursion level of the recursion tree, that there exists an execution of algorithm $\textsc{TupleEvalSolve} \left( \C, r, (o_1, t_1, o_2, t_2) \right) $ that returns $\true$. Notice that the base case is given for those cases where there is no recursion.
The base test cases, \textit{i.e.}, when $r$ is equal to either $\vd$, $\ed$, $\ell$, $\propt{p}{v}$, $<\mt$, or $\ex$, are easily checked, since all the algorithm does is checking their definitions over the temporal object $(o_1, t_1)$ after checking that it is equal to the temporal object $(o_2, t_2)$. Notice that for property-value checking and existence checking, the default value is $\false$, returned at the end of the algorithm.
The base navigation operators $\nxt$, $\prv$, $\fw$ and $\bw$ are also easily checked by their definitions. For time navigation, we check that the objects are the same and that their associated times are consecutive, whereas for spatial navigation, we check that the times are equal, and that the respective objects are consecutive, by looking at the functions $\tgt$ and $\src$, as defined by the operators.

As for the recursive cases, assume that the property holds up to recursion level $n$ and we want to prove that it holds at recursion level $n-1$ (one level higher in the recursion tree).
Firstly, a path expression matching any of the regular expressions $(\test \wedge \test)$, $(\test \vee \test)$ or $(\neg \test)$ can also be checked quite straightforwardly by definition, and since the flow is deterministic, we will omit further formal proofs.
Secondly, if the path expression $r$ is of the form $(?r')$, then we know that $(o_1, t_1, o_2, t_2) \in \semp{r}{\C}$ if and only $(o_1, t_1) = (o_2, t_2)$ and there exists a temporal object $(o', t') \in \Pt(\C)$ such that $(o_1, t_1, o', t') \in \semp{r'}{\C}$. In such case, the algorithm iterates one by one over the possible temporal objects to find one that satisfies the condition. If such temporal object exists, the call to $\textsc{TupleEvalSolve}$ returns $\true$, since we then know that the tuple $(o_1, t_1, o', t')$ satisfies $r'$ if and only if there exist an execution of the call that returns $\true$. Conversely, if no such temporal object exists, all recursive calls to $\textsc{TupleEvalSolve}$ will return $\false$ by induction hypothesis. In this case, the algorithm will finish the loop without returning and it will then reach the last line (in part II), returning $\false$.

As for regular path expressions $r$ of the form $(r_1 + r_2)$ where $r_1$ and $r_2$ are also regular path expressions, we know that by definition $\semp{(r_1 + r_2)}{\C} = \semp{r_1}{\C} \cup \semp{r_2}{\C}$. Hence, $(o_1, t_1, o_2, t_2) \in \semp{r}{\C}$ if and only if $(o_1, t_1, o_2, t_2) \in \semp{r_1}{\C}$ or $(o_1, t_1, o_2, t_2) \in \semp{r_2}{\C}$. By induction hypothesis, this means that $(o_1, t_1, o_2, t_2) \in \semp{r}{\C}$ if and only if either $\textsc{TupleEvalSolve}(\C, r_1, (o_1, t_1, o_2, t_2))$ returns $\true$ or $\textsc{TupleEvalSolve}(\C, r_2, (o_1, t_1, o_2, t_2))$ returns $\true$. As the algorithm returns the disjunction of this two results, we get that $(o_1, t_1, o_2, t_2) \in \semp{r}{\C}$ if and only if $\textsc{TupleEvalSolve}(\C, r, (o_1, t_1, o_2, t_2))$ returns $\true$

For regular path expressions $r$ of the form $(r_1 \ / \ r_2)$ where $r_1$ and $r_2$ are also TRPQs, we know that if $(o_1, t_1, o_2, t_2) \in \semp{r}{\C}$, then there must exist a temporal object $(o',t') \in \Pt(\C)$ such that $(o_1, t_1, o', t') \in \semp{r_1}{\C}$ and $(o', t', o_2, t_2) \in \semp{r_2}{\C}$. Thus, by iterating over all temporal object $(o',t')$, when we reach that exact temporal object, both $\textsc{TupleEvalSolve}(\C, r_1, (o_1, t_1, o', t'))$ and $\textsc{TupleEvalSolve}(\C, r_2, (o', t', o_2, t_2))$ will be $\true$, so the algorithm will return $\true \ \textbf{and} \ \true = \true$. On the other hand, if $(o_1, t_1, o_2, t_2) \notin \semp{r}{\C}$, then no matter what temporal object $(o', t')$ is being considered, we will either have that $(o_1, t_1, o', t') \notin \semp{r_1}{\C}$ or $(o', t', o_2, t_2) \notin \semp{r_2}{\C}$. By induction hypothesis, this means that, for every execution, either the first call will be $\false$ or the second will, so the condition that makes the algorithm return $\true$ will not be met. Hence, the last line is reached and the algorithm returns $\false$.

For expressions $r$ matching regular expressions with numerical occurrence indicators of the form $r'[n, m]$, recall that, by definition, $\semp{r'[n, m]}{\C} = \bigcup_{k=n}^m \semp{r'^k}{\C}$. This implies that $(o_1, t_1, o_2, t_2) \in \semp{r'[n, m]}{\C}$ if and only if there exists an integer $k \in [n, m]$ such that $(o_1, t_1, o_2, t_2) \in \semp{r'^k}{\C}$. We split this case into three cases.

\begin{enumerate}
    \item When $n = m$, then $(o_1, t_1, o_2, t_2) \in \semp{r'[n, m]}{\C}$ if and only if $(o_1, t_1, o_2, t_2) \in \semp{r'^n}{\C}$. Recall that the concatenation operator is associative, and that $\semp{r'^n}{\C} = \semp{r' \ / \ r'^{n-1}}{\C} = \semp{r' \ / \ \dots \ / \ r'}{\C}$ ($n$ repetitions). Hence, if we define $l = \lfloor n/2 \rfloor$, then if $n$ is even, $\semp{r'^n}{\C} = \semp{(r'^l \ / \ r'^l)}{\C}$, whereas if $n$ is odd, $\semp{r'^n}{\C} = \semp{(r'^l \ / \ r' \ / \ r'^l)}{\C}$.

    In the first case then, by the definition of concatenation, $(o_1, t_1, o_2, t_2) \in \semp{(r'^l \ / \ r'^l)}{\C}$ if and only if there exists a temporal object $(o', t')$ such that $(o_1, t_1, o', t') \in \semp{r'^l}{\C}$ and $(o', t', o_2, t_2) \in \semp{r'^l}{\C}$. By induction hypothesis this is equivalent to having that there exists a temporal object $(o', t')$ such that both $\textsc{TupleEvalSolve}(\C, r'[l, l], (o_1, t_1, o', t'))$ and $\textsc{TupleEvalSolve}(\C, r'[l, l], (o', t', o_2, t_2))$ return $\true$, since $\semp{r'^l}{\C} = \semp{r'[l, l]}{\C}$. Since the algorithm iterates over all possible temporal objects to check this condition, if such temporal object exists, it will return $\true$, if it does not, then it will reach the last line and return $\false$.

    The second case is similar, except that now we need two temporal objects as there are two concatenations, which means that $(o_1, t_1, o_2, t_2) \in \semp{(r'^l \ / \ r' \ / \ r'^l)}{\C}$ if and only if there exist two temporal objects $(o', t')$ and $(o'', t'')$ such that $(o_1, t_1, o', t') \in \semp{r'^l}{\C}$, $(o', t', o'', t'') \in \semp{r'}{\C}$ and $(o'', t'', o_2, t_2) \in \semp{r'^l}{\C}$. By induction hypothesis, and recalling again that $\semp{r'^l}{\C} = \semp{r'[l, l]}{\C}$, that means that $(o_1, t_1, o_2, t_2) \in \semp{r}{\C}$ if and only if there exist two temporal objects $(o', t')$ and $(o'', t'')$ such that the calls $\textsc{TupleEvalSolve}(\C, r'[l, l], (o_1, t_1, o', t'))$, $\textsc{TupleEvalSolve}(\C, r', (o', t', o'', t''))$ and $\textsc{TupleEvalSolve}(\C, r'[l, l], (o'', t'', o_2, t_2))$ return $\true$. Again, since the algorithm iterates over all possible pairs of temporal objects $(o', t')$ and $(o'', t'')$ to check this condition, if such temporal objects exist, it will return $\true$, if it does not, then it will reach the last line and return $\false$.
    
    Since this recursion must stop at some point, the base case $n = 1$ is included, in which case $r$ is $r'[1, 1]$, which is equivalent to $r'$, since in such case we know that $(o_1, o_2, t_1, t_2) \in \semp{r}{\C}$ if and only if $(o_1, t_1, o_2, t_2) \in \semp{r'}{C}$. By hypothesis induction, this happens if and only if $\textsc{TupleEvalSolve}(\C, r', (o_1, t_1, o_2, t_2))$ returns $\true$, which is why the algorithm returns that result.
    
    Finally, the recursion works for $n \geq 2$. The case when $n = 1$ is covered as a base case, so it only remains to look for the case when $n = 0$. For such case, recall that $r'^0 = (\exists \vee \neg \exists)$, \textit{i.e.}, a test that is a tautology. Hence, $(o_1, t_1, o_2, t_2) \in \semp{r}{\C}$ if and only if $(o_1, t_1) = (o_2, t_2)$, which is what the algorithm tests.
    
    \item When $n = 0$, the base case will be slightly different. We can assume that $n \neq m$ since the case where $n = m$ was already covered. Hence, the base case only needs to consider the value $m = 1$. In this case, we have that $\semp{r}{\C} = \semp{r'[0, 1]}{\C} = \semp{r'^0}{\C} \cup \semp{r'^1}{\C}$. As we already discussed, checking whether $(o_1, t_1, o_2, t_2) \in \semp{r'^0}{\C}$ comes down to checking whether $(o_1, t_1) = (o_2, t_2)$, and also $\semp{r'^1}{\C} = \semp{r'}{\C}$. By induction, we know that $(o_1, t_1, o_2, t_2) \in \semp{r'}{\C}$ if and only if $\textsc{TupleEvalSolve}(\C, r', (o_1, t_1, o_2, t_2))$ returns $\true$. Since the algorithm returns $\true$ if either of these conditions hold, this case is correctly covered.
    
    As for the recursive case, it is very similar to the previous one. If we define again $l = \lfloor m/2 \rfloor$, we can notice that, if $m$ is even, then $\semp{r'[0, m]}{\C} = \semp{(r'[0, l] \ / \ r'[0,l])}{\C}$, whereas if $m$ is odd, then $\semp{r'[0, m]}{\C} = \semp{(r'[0, l] \ / \ r' \ / \ r'[0,l])}{\C}$.
    
    To show the part when $m$ is even, notice that $(o_1, t_1, o_2, t_2) \in \semp{r}{\C}$ if and only if $(o_1, t_1, o_2, t_2) \in \semp{r'^i}{\C}$ for some $i \in [0, m]$. For every $i \in [0, m]$ there exist two integers, $j_i := \lfloor i/2 \rfloor$ and $k_i := \lceil i/2 \rceil$, both in the interval $[0, l]$, that satisfy that $j_i + k_i = i$. Since the concatenation operator is associative, this means that $\semp{r'_i}{\C} = \semp{(r'^{j_i} \ / \ r'^{k_i})}{\C}$, which in turn implies that $\semp{r}{\C} = \bigcup_{i=0}^m \semp{r'_i}{\C} = \bigcup_{i=0}^m \semp{(r'^{j_i} \ / \ r'^{k_i})}{\C}$.
    
    Then, by definition of the concatenation operator, $(o_1, t_1, o_2, t_2) \in \bigcup_{i=0}^m \semp{(r'^{j_i} \ / \ r'^{k_i})}{\C}$ if and only if there exists a temporal object $(o', t')$ such that $(o_1, t_1, o', t') \in \semp{r'^{j_i}}{\C}$ and $(o', t', o_2, t_2) \in \semp{r^{k_i}}{\C}$.
    
    Since both $j_i$ and $k_i$ are bounded by $l$, $\semp{r'^{j_i}}{\C} \subseteq \semp{\bigcup_{i=0}^l r'^i}{\C}$, and $\semp{r'^{k_i}}{\C} \subseteq \semp{\bigcup_{i=0}^l r'^i}{\C}$, so any tuple $(o, t, o', t')$ in $\semp{r'^{k_i}}{\C}$ or $\semp{r'^{j_i}}{\C}$ will also be in $\semp{\bigcup_{i=0}^l r'^i}{\C} = \semp{r'[0, l]}{\C}$. Hence, $(o_1, t_1, o', t') \in \semp{r'^{j_i}}{\C}$ implies that $(o_1, t_1, o', t') \in \semp{r'[0, l]}{\C}$ and $(o', t', o_2, t_2) \in \semp{r'^{k_i}}{\C}$ implies that $(o', t', o_2, t_2) \in \semp{r'[0, l]}{\C}$. It can then be inferred that having that $(o_1, t_1, o_2, t_2) \in \bigcup_{i=0}^m \semp{(r'^{j_i} \ / \ r'^{k_i})}{\C}$ can only hold if there exists a temporal object $(o', t')$ satisfying that $(o_1, t_1, o', t') \in \semp{r'^[0, l]}{\C}$ and $(o', t', o_2, t_2) \in \semp{r'[0, l]}{\C}$. As a result, $(o_1, t_1, o_2, t_2) \in \bigcup_{i=0}^n \semp{(r'^{j_i} \ / \ r'^{k_i})}{\C}$ implicates that $(o_1, t_1, o_2, t_2) \in \semp{(r'[0, l] \ / \ r'[0, l])}{\C}$. In consequence, we get that $\bigcup_{i=0}^n \semp{(r'^{j_i} \ / \ r'^{k_i})}{\C} \subseteq \semp{(r'[0, l] \ / \ r'[0, l])}{\C}$.
    
    For the inverse inclusion, notice that if $(o_1, t_1, o_2, t_2) \in \semp{(r'[0, l] \ / \ r'[0, l])}{\C}$, then there must exist a temporal object $(o', t')$ and two integers $j$ and $k$ in $[0, l]$ such that $(o_1, t_1, o', t') \in \semp{r'^j}{\C}$ and $(o', t', o_2, t_2) \in \semp{r'^k}{\C}$, which implies that $(o_1, t_1, o_2, t_2) \in \semp{(r'^j \ / \ r'^k)}{\C}$. Again, since concatenation is associative, $\semp{(r'^i \ / \ r'^j)}{\C} = \semp{r'^{i+j}}{\C}$, and because $i + j$ is at most $n$, $(o_1, t_1, o_2, t_2) \in \bigcup_{i=0}^n \semp{r'^i}{\C}$, \textit{i.e.}, $(o_1, t_1, o_2, t_2) \in \semp{r}{\C}$. As a result, we get that $\semp{(r'[0, l] \ / \ r'[0, l])}{\C} \subseteq \bigcup_{i=0}^n \semp{(r'^{j_i} \ / \ r'^{k_i})}{\C}$.
    
    Combining these two inclusions with the equality $\semp{r}{\C} = \bigcup_{i=0}^n \semp{(r'^{j_i} \ / \ r'^{k_i})}{\C}$, we get that $\semp{r}{\C} = \semp{(r'[0, l] \ / \ r'[0, l])}{\C}$.
    
    To show the part where $n$ is odd, notice that $(o_1, t_1, o_2, t_2) \in \semp{r}{\C}$ if and only if there exists an integer $i \in [0, m]$ such that $(o_1, t_1, o_2, t_2) \in \semp{r'^i}{\C}$. Notice then that $i$ can be written as the sum of three integers, $k_i := \lfloor i/2 \rfloor$ ($\in [0, l]$), $j_i =: \lfloor i/2 \rfloor$ ($\in [0, l]$) and $s_i := (i mod 2)$ ($\in [0, 1]$). Since the concatenation operator is associative, this means that $\semp{r'_i}{\C} = \semp{(r'^{j_i} \ / \ r'^{s_i} \ / \ r'^{k_i})}{\C}$, which in turn implies that $\semp{r}{\C} = \bigcup_{i=0}^m \semp{r'_i}{\C} = \bigcup_{i=0}^m \semp{(r'^{j_i} \ / \ r'^{s_i} \ / \ r'^{k_i})}{\C}$.
    
    Then, by definition of the concatenation operator, $(o_1, t_1, o_2, t_2) \in \bigcup_{i=0}^m \semp{(r'^{j_i} \ / \ r'^{s_1}\ / \ r'^{k_i})}{\C}$ if and only if there exist two temporal objects $(o', t')$ and $(o'', t'')$ such that $(o_1, t_1, o', t') \in \semp{r'^{j_i}}{\C}$, $(o', t', o'', t'') \in \semp{r'^{s_i}}{\C}$ and $(o'', t'', o_2, t_2) \in \semp{r'^{k_i}}{\C}$.
    
    Since both $j_i$ and $k_i$ are bounded by $l$, $\semp{r'^{j_i}}{\C} \subseteq \semp{\bigcup_{i=0}^l r'^i}{\C}$, and $\semp{r'^{k_i}}{\C} \subseteq \semp{\bigcup_{i=0}^l r'^i}{\C}$, so any tuple $(o, t, o', t')$ in $\semp{r'^{k_i}}{\C}$ or $\semp{r'^{j_i}}{\C}$ will also be in $\semp{\bigcup_{i=0}^l r'^i}{\C} = \semp{r'[0, l]}{\C}$. Similarly, any tuple in $\semp{r'^{s_1}}{\C}$ will also be in $\semp{r'[0,1]}{\C}$. Hence, $(o_1, t_1, o', t') \in \semp{r'^{j_i}}{\C}$ implies that $(o_1, t_1, o', t') \in \semp{r'[0, l]}{\C}$, $(o', t', o'', t'') \in \semp{r'^{s_i}}{\C}$ implies that $(o', t', o'', t'') \in \semp{r'[0, 1]}{\C}$ and $(o'', t'', o_2, t_2) \in \semp{r^{k_i}}{\C}$ implies that $(o'', t'', o_2, t_2) \in \semp{r[0, l]}{\C}$. It can then be inferred that having that $(o_1, t_1, o_2, t_2) \in \bigcup_{i=0}^m \semp{(r'^{j_i} \ / \ r'^{s_i} \ / \ r'^{k_i})}{\C}$ can only hold if there exist two temporal objects $(o', t')$ and $(o'', t'')$ satisfying that $(o_1, t_1, o', t') \in \semp{r'^[0, l]}{\C}$, $(o', t', o'', t'') \in \semp{r'[0, 1]}{\C}$ and $(o'', t'', o_2, t_2) \in \semp{r[0, l]}{\C}$. As a result, $(o_1, t_1, o_2, t_2) \in \bigcup_{i=0}^n \semp{(r'^{j_i} \ / \ r'^{k_i})}{\C}$ implicates that $(o_1, t_1, o_2, t_2) \in \semp{(r'[0, l] \ / \ r'[0, 1] \ / \ r'[0, l])}{\C}$. In consequence, we get that $\bigcup_{i=0}^n \semp{(r'^{j_i} \ / \ r'^{k_i})}{\C} \subseteq \semp{(r'[0, l] \ / \ r'[0, 1] \ / \ r'[0, l])}{\C}$.
    
    For the inverse inclusion, notice that if $(o_1, t_1, o_2, t_2) \in \semp{(r'[0, l] \ / \ r'[0, 1] \ / \ r'[0, l])}{\C}$, then there must exist two temporal objects $(o', t')$ and $(o'', t'')$, and three integers $j \in [0, l]$, $s \in [0, 1]$ and $k \in [0, l]$ such that $(o_1, t_1, o', t') \in \semp{r'^j}{\C}$, $(o'', t'') \in \semp{r'^s}{\C}$ and $(o'', t'', o_2, t_2) \in \semp{r'^k}{\C}$, which implies that $(o_1, t_1, o_2, t_2) \in \semp{(r'^j \ / \ r'^s \ / \ r'^k)}{\C}$. Again, since concatenation is associative, $\semp{(r'^i \ / \ r'^s \ / \ r'^j)}{\C} = \semp{r'^{i+ s + j}}{\C}$, and because $i + s + j$ is at most $n$, $(o_1, t_1, o_2, t_2) \in \bigcup_{i=0}^n \semp{r'^i}{\C}$, \textit{i.e.}, $(o_1, t_1, o_2, t_2) \in \semp{r}{\C}$. As a result, we get that $\semp{(r'[0, l] \ / \ r'[0, 1] \ / \ r'[0, l])}{\C} \subseteq \bigcup_{i=0}^n \semp{(r'^{j_i} \ / \ r'^{k_i})}{\C}$.

\begin{sloppypar}    
    As before, combining these two inclusions with the equality $\semp{r}{\C} = \bigcup_{i=0}^n \semp{(r'^{j_i} \ / \ r'^{l_i} \ / \ r'^{k_i})}{\C}$, we get that $\semp{r}{\C} = \semp{(r'[0, l] \ / \ r'[0, 1] \ / \ r'[0, l])}{\C}$.
    With these two results in mind then, \textit{i.e.}, that $\semp{r'[0, m]}{\C} = \semp{(r'[0, l] \ / \ r'[0,l])}{\C}$ when $m$ is even, whereas if $m$ is odd, then $\semp{r'[0, m]}{\C} = \semp{(r'[0, l] \ / \ r' \ / \ r'[0,l])}{\C}$, we know that $(o_1, t_1, o_2, t_2) \in \semp{r'[0, m]}{\C}$ if and only if (i) $m$ is even and there exists a temporal object $(o', t')$ such that $(o_1, t_1, o', t') \in \semp{r'[0,l]}{\C}$ and $(o', t', o_2, t_2) \in \semp{r'[0, l]}{\C}$, or (ii) $m$ is odd and there exist two temporal objects $(o', t')$ and $(o'', t'')$ such that $(o_1, t_1, o', t') \in \semp{r'[0, l]}{\C}$, $(o', t', o'', t'') \in \semp{r'[0, 1]}{\C}$ and $(o'', t'', o_2, t_2) \in \semp{r'[0, l]}{\C}$.
    
    By induction, (i) holds if and only if there exists a temporal object $(o', t')$ such that both $\textsc{TupleEvalSolve}(\C, r'[0, l], (o_1, t_1, o', t'))$ and $\textsc{TupleEvalSolve}(\C, r'[0, l], (o', t', o_2, t_2))$ return $\true$. Since the algorithm iterates over all temporal objects $(o', t')$ for this case, and then checks that both those conditions are met to return $\true$, it will return $\true$ if $(i)$ holds, and it will reach the last line and return $\false$ if no such pair existed.
    
    Also by induction, (ii) holds if and only if there exist two temporal objects $(o', t')$ and $(o'', t'')$ such that the three calls $\textsc{TupleEvalSolve}(\C, r'[0, l], (o_1, t_1, o', t'))$, $\textsc{TupleEvalSolve}(\C, r'[0, l], (o', t', o'', t''))$ and $\textsc{TupleEvalSolve}(\C, r'[0, l], (o'', t'', o_2, t_2))$ return $\true$. 
    Here again, since the algorithm iterates over all pairs of temporal objects $(o', t')$ and $(o'', t'')$ and sees if these three conditions are met to return $\true$, it will return $\true$ if (ii) holds, and it will reach the last line and return $\false$ otherwise.
    
    Hence, when $n = 0$, the algorithm also returns $\true$ if and only if $(o_1, t_1, o_2, t_2) \in \semp{r}{\C}$
    
    \item When $m \neq n$ and $n \neq 0$, then $(o_1, t_1, o_2, t_2) \semp{r'[n, m]}{\C}$ if and only if there exists $i \in [n, m]$ such that $(o_1, t_1, o_2, t_2) \semp{r'^i}{\C}$. In this case, $i = n + d$ for some $d \in [0, m-n]$, and since the concatenation operator is associative, $(o_1, t_1, o_2, t_2) \semp{r'^i}{\C}$ if and only if there exists $(o', t')$ such that $(o_1, t_1, o', t') \in \semp{r'^n}{\C}$ and $(o', t', o_2, t_2) \in \semp{r'^d}{\C}$. 
    By induction, and since $\semp{r'^n}{\C} = \semp{r'[n, n]}{\C}$, $(o_1, t_1, o', t') \in \semp{r'^n}{\C}$ if and only if $\textsc{TupleEvalSolve}(\C, r'[n, n], (o_1, t_1, o', t'))$. Similarly, $(o', t', o_2, t_2) \in \semp{r'^d}{\C}$ for some $d \in [0, m-n]$ if and only if $(o', t', o_2, t_2) \in \semp{r'[0, m-n]}{\C}$, which by induction holds if and only if $\textsc{TupleEvalSolve}(\C, r'[0, m-n], (o', t', o_2, t_2))$.

    Together, this means that $(o_1, t_1, o_2, t_2) \in \semp{r}{\C}$ if and only if there exists a temporal object $(o', t')$ such that $\textsc{TupleEvalSolve}(\C, r'[n, n], (o_1, t_1, o', t')$ and $\textsc{TupleEvalSolve}(\C, r'[0, m-n], (o', t', o_2, t_2))$. Since the algorithm iterates over all temporal objects $(o', t')$ and checks if these conditions are met to return $\true$, it will return $\true$ if $(o_1, t_1, o_2, t_2) \in \semp{r}{\C}$ and it will reach the last line and return $\false$ otherwise.
    \end{sloppypar}
\end{enumerate}
Finally, for an expression $r$ matching a regular expressions with numerical occurrence indicators of the form $r'[n, \_]$, recall that we showed in Section \ref{sec-app-tpg-pol} that $\semp{r'[n, \_]}{\C} = \semp{r'[n, m]}{\C}$, where the expression $m := n + (|\Omega| + |V \cup E|)^2$ is polynomial in the size of the original input.
By induction, $(o_1, t_1, o_2, t_2) \in \semp{r'[n, m]}{\C}$ if and only if $\textsc{TupleEvalSolve}(\C, r'[n, m], (o_1, t_1, o_2, t_2))$ returns $\true$, which is what the algorithm returns for this case.
Altogether, we proved that $\textsc{TupleEvalSolve}$ works in polynomial space, and that $\textsc{TupleEvalSolve}(\C, r, (o_1, t_1, o_2, t_2))$ returns $\true$ if and only if $(o_1, t_1, o_2, t_2) \in \semp{r}{\C}$. 
This concludes the proof of the theorem.

\section{Allowing numerical occurrence indicators only in the axes: additional complexity results}

A natural question is whether there is a restriction on $\Lonep$ that can reduce the complexity of the evaluation problem but is still expressive enough to represent some useful queries. At this point, a restriction used in the study of XPath comes to the rescue~\cite{M05}. 
In what follows, we show that the complexity of the evaluation problem is lower if numerical occurrence indicators
are only allowed in the axes. 

\begin{theorem}\label{th:LoneppTupleNP}
{\em \tupleeval(\ctpg, \Lonepp)} is \np-complete.
\end{theorem}

\begin{proof}
To show \np-hardness, consider the following decision problem called Subset Sum (\sss), which is known to be \np-complete \cite{SubsetSum1990}:

\begin{center}
	\framebox{
	    \begin{tabular}{p{1.6cm} p{10cm}}
			\textbf{Problem:} & $\sss$ \\
		    \textbf{Input:} & A finite set of integers $A \subsetneq \mathbb{N}$, and a positive integer $S \in \mathbb{N}$ \\
		    \textbf{Output:} & $\true$ if there exists a subset $A' \subseteq A$ of $A$ such that $\sum_{a \in A'} a = S$.
		\end{tabular}
	}
\end{center}
Given a set $A \subsetneq \mathbb{N}$, and an integer $S \in \mathbb{N}$, the goal is to provide a polynomial-time algorithm that returns an \ctpg $\C$, a tuple $(o_1, t_1, o_2, t_2)$, and an expression $r$ in $\Lonepp$ such that $(o_1, t_1, o_2, t_2) \in \semp{r}{\C}$ if and only if there exists $A' \subseteq A$ such that $\sum_{a \in A'} a = S$.
More specifically,
$\C$ will be the \ctpg $\left( \Omega, N, E, \rho, \lambda, \xi, \sigma \right)$ where $\Omega = [0, S]$, $N = \lbrace v \rbrace$, $E = \varnothing$, $\rho$ is an empty function, $\lambda(v) = l$, $\xi(v)=\left\lbrace [0, S] \right\rbrace$ and $\sigma$ is an empty function. In other words, $\C$ is an \ctpg consisting of only one node existing from time $0$ to time $S$, with no edges or properties. The tuple $(o_1, t_1, o_2, t_2)$ in our reduction will be given by $(v, 0, v, S)$. Moreover, assuming that $A = \{ a_1, \dots, a_n \}$, expression $r$ is defined as follows:
\begin{eqnarray*} r &=& \left( \nxt[ a_1, a_1 ] + \nxt[ 0, 0 ] \right) \ / \ \cdots \ / \ \left( \nxt[ a_n, a_n ] + \nxt[ 0, 0 ] \right) 
\end{eqnarray*}
Notice that \ctpg $\C$, expression $r$ in $\Lonepp$ and tuple $(v,0,v,S)$ can be computed in polynomial time in the sizes of $A$ and $S$. Besides, it is straightforward to prove that $(v, 0, v, S) \in \semp{r}{\C}$ if and only if there exists $A' \subseteq A$ such that $\sum_{a \in A'} a = S$. This concludes the of $\np$-hardness of \tupleeval(\ctpg, \Lonepp).

To show that this problem is \np-complete, it only remains to show that the problem is also in \np. We present a nondeterministic algorithm that works in polynomial time, $\textsc{TupleEvalSolve\_ANOI}$, that, given an \ctpg $\C$, an expression $r$ in $\Lonepp$ and a pair of temporal objects $(o_1, t_1, o_2, t_2)$, has a run that returns $\true$ if and only if $(o, t, o', t') \in \semp{r}{\C}$. This procedure is presented in Algorithm \ref{algo:ANOI}.

\LinesNumbered

\begin{algorithm}
    \SetAlgoLined
    \SetKwInOut{Input}{Input}
    \SetKwInOut{Output}{Output}
    \SetKwInOut{Initialization}{Initialization}
    \Input{\ An \ctpg $\C = \left( \Omega, N, E, \rho, \lambda, \xi, \sigma \right)$, an expression $r$ in $\Lonepp$ and a pair of temporal objects $\left( o_1, t_1, o_2, t_2 \right)$}
    \Output{\ $\true$ if $\left( o, t, o', t' \right) \in \semp{r}{\C}$}
    \uIf{$r$ is a $\test$}{
        \uIf{$(o_1, t_1) \neq (o_2, t_2)$}{
            \Return $\false$
        }
        \uElseIf{$r = \vd$}{
            \Return $(o_1 \in N)$
        }
        \uElseIf{$r = \ed$}{
            \Return $(o_1 \in E)$
        }
        \uElseIf{$r = \ell$ for some \ $\ell \in \Lab$}{
            \Return $(\lambda(o_1) = \ell)$
        }
        \uElseIf{$r = \propt{p}{v}$ for some $p\in \Prop$ \ \text{and} $v\in \Val$}{
            \ForEach{valued interval $(v', I) \in \sigma(o_1, p)$}{
                \uIf{$t_1 \in I$}{
                    \Return $v' = v$
                }
            }
            \Return $\false$
        }
        \uElseIf{$r \, = \, < \mt$ with $\mt \in \Omega$}{
            \Return $(t_1 < \mt)$
        }
        \uElseIf{$r = \ex$}{
            \ForEach{interval $I \in \xi(o_1)$}{
                \uIf{$t_1 \in I$}{
                    \Return $\true$
                }
            }
            \Return $\false$
        }
        \uElseIf{$r = (\test_1 \vee \test_2)$}{
            \Return $\textsc{TupleEvalSolve}\left(\C,  (o_1, t_1, o_1, t_1), \test_1 \right)$ \textbf{or} $\textsc{TupleEvalSolve}\left(\C, (o_1, t_1, o_1, t_1), \test_2 \right)$
        }
        \uElseIf{$r   = (\test_1 \wedge \test_2)$}{
            \Return $\textsc{TupleEvalSolve}\left(\C, (o_1, t_1, o_1, t_1), \test_1 \right)$ \textbf{and} $\textsc{TupleEvalSolve}\left(\C, (o_1, t_1, o_1, t_1), \test_2 \right)$
        }
        \uElseIf{$r  = \left( \neg r' \right)$}{
            \Return \textbf{not} $ \textsc{TupleEvalSolve}\left(\C, (o_1, t_1, o_1, t_1), r' \right)$
        }
    }
    \uElseIf{$r = \nxt$}{
        \Return $\left(
            o_1 = o_2 \ \textbf{and} \ t_2 = t_1 + 1
        \right)$
    }
    \uElseIf{$r = \prv$}{
        \Return $\left(
            o_1 = o_2 \ \textbf{and} \ t_2 = t_1 - 1
        \right)$
    }
    \uElseIf{$r = \fw$}{
        \Return $\left(
            t_1 = t_2 \ \textbf{and} \ \left(
                \left( o_1 \in E \ \textbf{and} \ o_2 = \tgt\left( o_1 \right) \right)
                \ \textbf{or} \
                \left( o_2 \in E \ \textbf{and} \ o_1 = \src\left( o_2 \right) \right)
            \right)
        \right)$
    }
    \uElseIf{$r = \bw$}{
        \Return $\left(
            t_1 = t_2 \ \textbf{and} \ \left(
                (o_1 \in E \ \textbf{and} \ o_2 = \src(o_1))
                \ \textbf{or} \
                (o_2 \in E \ \textbf{and} \ o_1 = \tgt(o_2))
            \right)
        \right)$
    }
    \caption{$\textsc{TupleEvalSolve\_ANOI}(\C, (o_1, t_1, o_2, t_2), r)$ (part I)}
    \label{algo:ANOI}
\end{algorithm}

$\textsc{TupleEvalSolve\_ANOI}$
is very similar to $\textsc{TupleEvalSolve}$, so we will not discuss in detail what it does.
Instead, we give an intuition of what the differences are that allow to return the right answer in non-deterministic polynomial time, instead of polynomial space.
First, notice that if $r$ is a test, then the algorithm works by solving basic tests efficiently,
and then conjunctions, disjunctions and negations of tests are solved just by using directly the definition of these Boolean connectives. Hence,
unlike what happens in the presence of path conditions, where we can have nested expressions with existential conditions and negations of existential conditions, sub-expressions for tests are efficiently solved by $\textsc{TupleEvalSolve\_ANOI}$.
Second, notice that for spatial navigation, we write the problem in terms of the reachability problem for graphs in a number of steps in a set $\{n, \ldots, m\}$.
This problem can be efficiently solved by using exponentiation by squaring on the adjacency matrix. Besides, notice that for spatial navigation expressions in $\Lonepp$, we need to consider as many new objects as there are in $V \cup E$ since the time is fixed, which is why expressions $\fw[n, \_]$ and $\bw[n, \_]$ are equivalent to $\fw[n, m]$ and $\bw[n, m]$, respectively, with $m = n + |N \cup E| = n + |N| + |E|$.
Finally, polynomial time executions are ensured by the non-deterministic guess for $(o', t')$ in Line $64$, and the fact that the depth of the recursion tree is linear 
with respect to the size of the input expression $r$.
In particular, we do a single non-deterministic guess in Line $64$,
instead of an exponential number of attempts (with respect to the size of the representation of $\Omega$) that would be necessary to find the right pair $(o', t')$ in a deterministic algorithm.




\begin{algorithm}
    \SetAlgoLined
    \setcounter{AlgoLine}{35}
    \uElseIf{$r = \nxt[n, m]$}{
        \Return $\left(
            o_1 = o_2 \ \textbf{and} \ n \leq (t_2 - t_1) \leq m
        \right)$
    }
    \uElseIf{$r = \prv[n, m]$}{
        \Return $\left(
            o_1 = o_2 \ \textbf{and} \ n \leq (t_1 - t_2) \leq m
        \right)$
    }
    \uElseIf{$r = \fw[n, m]$}{
        \uIf{$t_1 \neq t_2$}{
            \Return $\false$
        }
        \uElse{
            Let $G = (N', E')$ be the graph where: \\
            \begin{itemize}
                \item $N' = (N \cup E)$
                \item $E' = \{
                    (v, e) \in N \times E \ \vert \ \src(e) = v
                \} \cup \{ 
                    (e, v) \in E \times N \ \vert \ \tgt(e) = v
                \}$
            \end{itemize}
            \Return $o_2$ is reachable from $o_1$ in $k$ steps in $G$, where $k \in \{n, \ldots, m\}$
        }
    }
    \uElseIf{$r = \bw[n, m]$}{
        \uIf{$t_1 \neq t_2$}{
            \Return $\false$
        }
        \uElse{
            Let $G = (N', E')$ be the graph where: \\
            \begin{itemize}
                \item $N' = (N \cup E)$
                \item $E' = \{
                    (v, e) \in N \times E \ \vert \ \tgt(e) = v
                \} \cup \{ 
                    (e, v) \in E \times N \ \vert \ \src(e) = v
                \}$
            \end{itemize}
            \Return $o_2$ is reachable from $o_1$ in $k$ steps in $G$, where $k \in \{n, \ldots, m\}$
        }
    }
    \uElseIf{$r = \nxt[n, \_]$}{
        \Return $\left(
            o_1 = o_2 \ \textbf{and} \ n \leq (t_2 - t_1)
        \right)$
    }
    \uElseIf{$r = \prv[n, \_]$}{
        \Return $\left(
            o_1 = o_2 \ \textbf{and} \ n \leq (t_1 - t_2)
        \right)$
    }
    \uElseIf{$r = \fw[n, \_]$}{
            $m \gets n + |N| + |E|$ \\
            \Return $\textsc{TupleEvalSolve\_ANOI}(\C, (o_1, t_1, o_2, t_2), \fw[n,m])$
    }
    \uElseIf{$r = \bw[n, \_]$}{
            $m \gets n + |N| + |E|$ \\
            \Return          $\textsc{TupleEvalSolve\_ANOI}(\C, (o_1, t_1, o_2, t_2), \bw[n,m])$
    }
    \uElseIf{$r = (r_1 + r_2)$}{
        Guess $i \in \{ 1, 2 \}$ \\
        \Return $\textsc{TupleEvalSolve\_ANOI}(\C, ( o_1, t_1, o_2, t_2), r_i)$
    }
    \uElseIf{$r = (r_1 \ / \ r_2)$}{
        Guess $(o', t') \in (N \cup E) \times \Omega$ \\
        \Return $\textsc{TupleEvalSolve\_ANOI}(\C, (o_1, t_1, o', t'), r_1)$ \textbf{and}  $\textsc{TupleEvalSolve\_ANOI}(\C, (o', t', o_2, t_2), r_2)$ \\
    }
    \caption{$\textsc{TupleEvalSolve\_ANOI}(\C, (o_1, t_1, o_2, t_2), r)$ (part II)}
\end{algorithm}

\end{proof}

We have that \tupleeval(\ctpg, \Ltwo) can be solved in polynomial time by Theorem \ref{th:main-summary}, and we know that \tupleeval(\ctpg, \Ltwo) is \np-complete by Theorem \ref{th:LoneppTupleNP}. A natural question 
then is whether the complexity remains the same if these functionalities are combined. Notice that \tupleeval(\ctpg, \Lthree) is $\pspace$-complete, so a positive answer to this question means a significant decrease in the complexity of the query evaluation problem. Unfortunately,  we show that the complexity of the entire language does not decrease by restricting numerical occurrence indicators to occur only in the~axes.
\begin{theorem}\label{th:LthreepTupleNP}
{\em \tupleeval(\ctpg, \Lthreep)} is \pspace-complete.
\end{theorem}

\begin{proof}
Notice that every expression in $\Lthreep$ is also an expression in $\Lthree$, so \pspace-membership follows immediately from Theorem \ref{th:main-summary}.
Hence, we only need to prove \pspace-hardness for $\Lthreep$.

To show this, we replace test expressions $r_i$ in the proof in Section \ref{sec-proof-itpg-pspace} 
by an 
expression in $\Lthreep$ that will be 
denoted by $q_i$. 
Expression $q_i$ is defined in such a way that,
for every time $t$, it holds that $(v, t, v, t) \in \semp{q_i}{\C}$ if and only if $(v, t, v, t) \in \semp{r_i}{\C}$, \textit{i.e.}, if and only if $\bit(i, t)$ is $\true$, where $\bit(i, t)$ holds if the $i$-th bit of time $t$ (from right to left when written in its binary representation) is 1. More precisely,
expression $q_i$ 
is defined as follow:
\begin{eqnarray*}
    q_i &  = & ?\left(
        \left(
            ( \prv[0, 0] + \prv[2^n, 2^n] ) \ / \
            \ldots \ / \
            ( \prv[0, 0] + \prv[2^i, 2^i] )
        \right) \ / \ \left(
            < 2^i \wedge \neg < 2^{i-1}
        \right)
    \right)
\end{eqnarray*}
Notice that the length of the representation $2^k$ is $k$, so the whole expression $q_i$ has length $O(n^2)$, which is polynomial with respect to the size of $\psi$. Also, notice that as before, we only need a polynomial number of these expressions for the reduction, and no further nesting of numerical occurrence indicators is required for the proof. Hence, we only need to prove that $(v, t, v, t) \in \semp{q_i}{\C}$ if and only if $\bit(i, t)$ is $\true$.
Recall that for this reduction, $\C$ is an \ctpg consisting of only one node $v$, existing from time $0$ to time $2^n-1$, with no edges or properties, so 
any temporal object considered will be of the form $(v,t)$.

First, notice that $q_i$ is a path test, so $(v, t, v, t) \in \semp{q_i}{\C}$ if and only if there exists a time point $t'$ such that
\begin{eqnarray*}
    (v, t, v, t') &  \in & \semp{\left(
        ( \prv[0, 0] + \prv[2^n, 2^n] ) \ / \
        \ldots \ / \
        ( \prv[0, 0] + \prv[2^i, 2^i] )
    \right) \ / \ \left(
        < 2^i \wedge \neg < 2^{i-1}
    \right)}{\C}
\end{eqnarray*}
As in \textbf{Step 1} of Section \ref{sec-proof-itpg-pspace}, since the last part of the expression is a test, this is equivalent to the existence of a time point $t'$ such that $(v, t') \models \left( < 2^i \wedge \neg < 2^{i-1} \right)$, \textit{i.e.}, the $i$-th bit of $t'$ is $1$ and
\begin{eqnarray}\label{eq-qi-int}
    (v, t, v, t') &  \in & \semp{
        ( \prv[0, 0] + \prv[2^n, 2^n] ) \ / \
        \ldots \ / \ ( \prv[0, 0] + \prv[2^i, 2^i] )
    }{\C}
\end{eqnarray}
We now prove that $(v, t, v, t) \in \semp{q_i}{\C}$ if and only if $\bit(i, t)$ is $\true$.
To show direction ($\Leftarrow$), suppose that $\bit(i, t)$ is $\true$. Notice then that $(v, t_1, v, t_2) \in \semp{( \prv[0, 0] + \prv[2^k, 2^k])}{\C}$, if and only if $t_2 = t_1$ or $t_2 = t_1 - 2^k$. In particular, if the $(k+1)$-th bit of $t_1$ is $1$, then $t_2 = t_1 - 2^k$ has a binary representation that is equal to that of $t_1$ except on the $(k+1)$-th bit,
and $(v, t_1, v, t_2) \in \semp{( \prv[0, 0] + \prv[2^k, 2^k] )}{\C}$. Similarly, if the $(k+1)$-th bit of $t_1$ is $0$, then $t_2 = t_1$ has a binary representation that is equal to that of $t_1$, 
and also $(v, t_1, v, t_2) \in \semp{( \prv[0, 0] + \prv[2^k, 2^k])}{\C}$.
As in Section \ref{sec-proof-itpg-pspace}, given $b \in \{\true, \false\}$, let $\mathbbm{1}_b$ be $1$ if $b = \true$, and be $0$ otherwise. Moreover, define the sequence of time points $t_{n+1}, \dots, t_i$ such that $t_{n+1} = t$ and $t_k = t_{k + 1} - \mathbbm{1}_{\bit(k + 1, t)} \cdot 2^k$ for $k \in \{ i, \dots, n \}$. 
Then for every $k \in \{ i, \dots, n \}$, it holds that 
$(v, t_k, v, t_{k+1}) \in \semp{( \prv[0, 0] + \prv[2^k, 2^k] )}{\C}$, and in particular, $t' = t - \sum_{k=i}^n \mathbbm{1}_{\bit(k + 1, t)} \cdot 2^k$ satisfies \eqref{eq-qi-int}.
%
Therefore, 
if the $i$-th bit of $t$ is $1$, then the $i$-th bit of $t'$ will be 1 as well.
Hence, given $\bit(i, t)$ is $\true$, we conclude that $(v, t, v, t) \in \semp{q_i}{\C}$, since for $t' = t - \sum_{k=i}^n \mathbbm{1}_{\bit(k + 1, t)} \cdot 2^k$, equation \eqref{eq-qi-int} holds and $(v,t') \models (< 2^i \wedge \neg < 2^{i-1})$.

To show direction ($\Rightarrow$), suppose that $(v, t, v, t) \in \semp{q_i}{\C}$. Then there exists a time point $t'$ such that \eqref{eq-qi-int} holds, which only holds if there exists a sequence of time points $t_{n+1}, \dots, t_i$ where $t_{n+1} = t$ and either $t_k = t_{k + 1}$ or $t_k = t _{k + 1} - 2^k$ for $k \in \{ i, \dots, n \}$, and $t_i = t'$. Notice that for such values for $k$, $2^k$ is a multiple of $2^i$, so $t' = t + d \cdot 2^i$ for some integer $d$. 
We conclude that the $i$-th bit of $t$ is equal to $1$ if and only if the $i$-th bit of $t'$ is equal to $1$. Moreover, $(v, t') \models ( < 2^i \wedge \neg < 2^{i-1})$, so the $i$-th bit of $t'$ is indeed equal to $1$, so $\bit(i, t)$ must be equal to $\true$.

From the previous paragraphs, we conclude that $(v, t, v, t) \in \semp{q_i}{\C}$ if and only if $\bit(i, t)$ is $\true$. Hence, by replacing $r_i$ with $q_i$ in the proof of Section \ref{sec-proof-itpg-pspace},
we deduce that $\Lthreep$ is also \pspace-hard, which was to be shown.
\end{proof}

\section{Supplementary Experimental Result}
\label{sec-app-epxr}
\begin{figure*}[!htb]
\centering
       \includegraphics[width=0.75\textwidth]{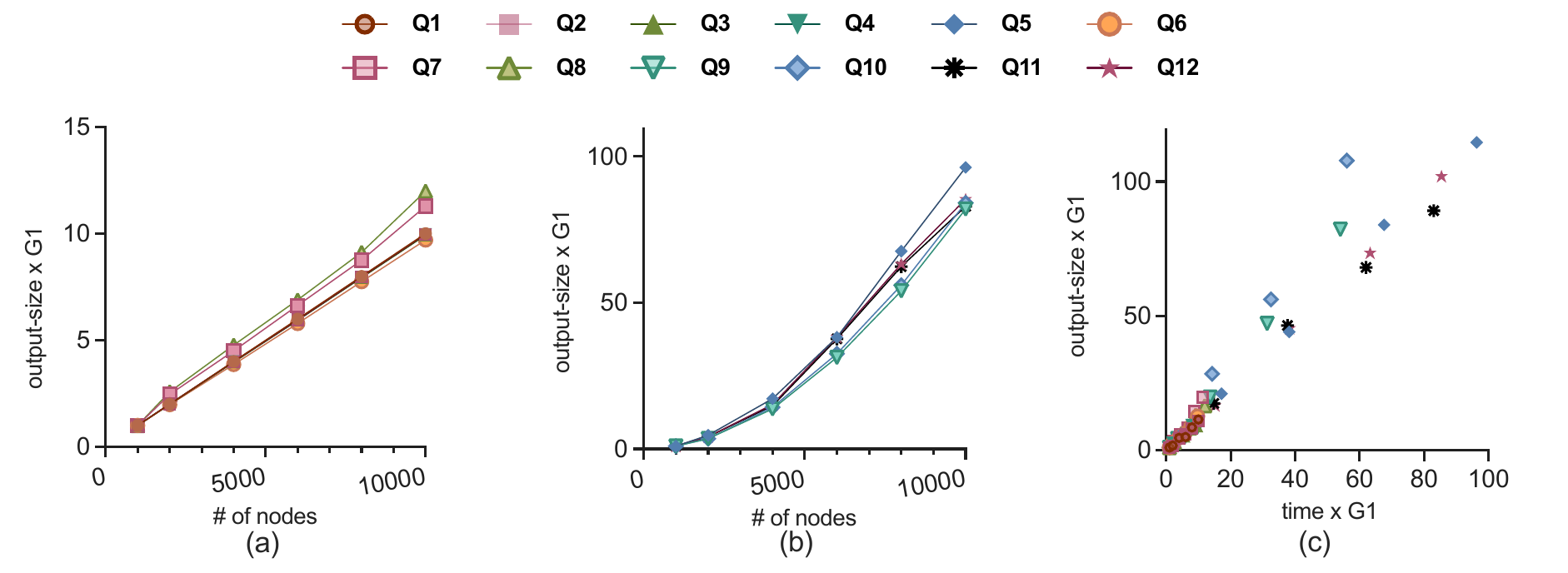}
\vspace{-0.3cm}       
\caption{Relationship between input size, output size, and query execution time for all queries.  Execution time and output size are computed for graphs G2-G6 (with between 2,000 and 10,000 nodes) in  Table~\ref{tbg:ds-desc}, in proportion to these quantities for graph G1 (with 1,000 nodes). }
\label{fig:exp-output-size}
\end{figure*}

In section~\ref{sec-experiments-data} we discussed the impact of \tgs size on query execution, and pointed out that the trends presented in Figure~\ref{fig:exp-nodes} can be explained by the size of output. To study this, we computed the increase in output size for graphs G2-G6 (with between 2,000 and 10,000 nodes, as summarized in Table~\ref{tbg:ds-desc})  relative to the size of the output for G1 (with 1,000 nodes), for each query.  Figures~\ref{fig:exp-output-size} (a) and (b) show this result. In these figures, the $x$-axis shows the number of nodes in each graph, and the $y$-axis shows the relative size of output bindings table, in comparison to the output of the same query over G1. Similarly to Figure~\ref{fig:exp-nodes}, it can be observed that the output size for all queries except Q5, Q9, Q10, Q11, and Q12 follows a linear trend. For Q5, Q9, Q10, Q11 and Q12, the output size increases quadratically. 

Figures~\ref{fig:exp-output-size} (c) gives another presentation of these results.  Here, in addition to computing the output size relative to G1 for each query (shown on the $y$-axis), we also computed the  execution time relative to G1 (shown on the $x$-axis).  This plot show that relative query execution time and relative increase in output size are highly correlated for all queries, and for the majority of our queries we have perfect correlation.

\end{document}